\documentclass[aps,prl,preprint,groupedaddress]{revtex4-2}
\usepackage[utf8]{inputenc}
\usepackage[T1]{fontenc}
\usepackage[english]{babel}
\usepackage{amsmath, amssymb, amsbsy}
\usepackage{multirow}
\usepackage{upgreek} 
\usepackage{cancel} 
\usepackage{mathdots} 
\usepackage{mathrsfs} 
\usepackage{stackrel} 
\usepackage{subcaption}
\usepackage{xurl} 
\usepackage[breaklinks=true,colorlinks=true,linkcolor=blue,urlcolor=blue]{hyperref}
\usepackage{adjustbox} 
\usepackage{graphicx}

\begin{document}


\title{Observation of Pseudogap in $\mathrm{Cr_{1-x}Y_xN}$ magnetic alloy and its impact on the Seebeck coefficient by ab-initio calculations}



\author{Luis Felipe León-Pinzón}
\affiliation{Universidad Nacional de Colombia sede Manizales}
\affiliation{Centro de Investigación en Materiales Avanzados, CIMAV}
\email{lufleonpi@unal.edu.co}

\author{Elisabeth Restrepo Parra}
\affiliation{Universidad Nacional de Colombia sede Manizales}
\email{erestrepopa@unal.edu.co}

\author{Andrés Manuel Garay-Tapia}
\affiliation{Centro de Investigación en Materiales Avanzados, CIMAV}
\email[Corresponding author]{andres.garay@cimav.edu.mx}



\date{\today}

\begin{abstract}
Thermoelectric materials require high electronic conductivity and low thermal conductivity. CrN has been shown to have low phononic thermal conductivity, making it a potential candidate for thermoelectric applications. In addition, similarities have been observed between YN and ScN suggesting that the CrYN alloy may have interesting thermoelectric properties. As CrYN has not been studied in detail at the level of thermoelectric properties, the first study on CrYN alloy of Seebeck coefficient and zT figure of merit is proposed in this study. For this purpose, cubic special quasirandom structures were constructed at values of x = 0.25, 0.5 and 0.75 in the alloy $\mathrm{Cr_{1-x}Y_xN}$ starting from different magnetic structures. After analyzing lattice parameters, Cr magnetic moments, octahedron deformation, second neighbors distribution around metals, density of states and band structures, it was concluded that to obtain high values of Seebeck coefficient and zT, it is necessary the presence of a pseudo gap in both spin channels and it is also necessary that the Fermi level is on a steep decreasing slope of number of states, since due to Mott's approximation, the value of this slope is proportional to the Seebeck coefficient. Density of states of all the structures shows a metallic behavior. In structures with x=0.5, the presence of small indirect energy gaps is observed. Although no structure retains the initial magnetic configuration, there is a possible influence of this on the electronic structure. Considerable deformations in octahedra can suppress thermoelectric properties.
\end{abstract}


\maketitle

\section{Introduction}
Transition metal nitrides (TMNs) belonging to groups IV-VI, are an amazing class of binary alloys due to their unique combination of covalent, ionic and metallic bonds \cite{Nita2016}. They generally have a base state with cubic symmetry and have an interesting combination of mechanical, electrical, and chemical properties that make them extremely hard and have very high melting points \cite{Nagao2006}. All these features have received great theoretical and experimental attention in recent years for potential applications such as hard coatings and thin films in semiconductors, cutting tools, optoelectronics, plasmonics, photovoltaic industries, information storage, high power energy, spintronics, among others \cite{Gupta2014,FernondezGuillermet1992,Nita2016,Nagao2006,papaconstantopoulos1985,Shimizu1997,Filippetti1999,Stampfl2001,Isaev2005,Holec2012,Pierson1996}.          

Initially, chromium nitride, CrN, was recognized for its impressive physical properties, including high hardness, oxidation resistance, corrosion resistance, and wear resistance \cite{Alam2017,Zhou2014}. But among TMNs, it is the only compound that in its ground state does not have cubic symmetry but orthorhombic symmetry, which makes it particularly interesting \cite{Ebad-Allah2016}. In addition, their practical applications have attracted attention for their peculiar mechanical, magnetic, optical, electronic and, especially, thermoelectric (TE) properties in its crystalline form, both in volume and epitaxial. \cite{Mozafari2015,Rojas2017,Rojas2018}.

One of the key factors behind TE materials is their low lattice electrical conductivity, which is fundamental to defining the dimensionless TE figure of merit (zT) \cite{Quintela2015}. CrN with space group $\mathrm{Pnma}$ shows intrinsic lattice instabilities that suppress its thermal conductivity. By ab-initio calculations, it was determined in \cite{Quintela2015} that the origin of these instabilities is similar to that observed in compounds IV-VI with resonant bond states. According to \cite{Lee2014}, these states are the main reason for the low thermal conductivity. Through the fabrication of high quality epitaxial (001) CrN thin films, in \cite{Quintela2015} reported an increase of 250 in zT at room temperature compared to bulk CrN. These results, together with its high thermal stability, corrosion resistance and exceptional mechanical properties, make CrN a promising material for high-temperature TE applications.

On the other hand, yttrium nitride (YN), still has few experimental verification in the literature \cite{Cherchab2008,Villars1985,Morris1985} and its theoretical studies are very scarce \cite{Rovere2010,Zhou2013}. The most recent investigations \cite{Salguero2003,Cherchab2008} have shown that YN has many similarities with ScN, as a reference, such as both crystallize in cubic phase and have a local second minimum in the wurzite structure and under high pressures are expected to crystallize in the CsCL structure. When both are in cubic phase, they have semiconducting behavior and their lattice parameters are quite close: 4.877 $\mathring{A}$ for YN and 4.54--4.651 $\mathring{A}$ for ScN \cite{Salguero2003,Stampfl2001}.

In addition, scandium nitride (ScN) has also shown a set of properties suitable for TE applications \cite{Stampfl2001,Cherchab2008,Louhadj2009,Kerdsongpanya2016,Sukkabot2019}. According to \cite{Kerdsongpanya2016}, these two materials constitute a well-defined model systems for investigating their mixing thermodynamics, phase stability and band structure. They demonstrated by a theoretical and experimentally proven approach that it is possible to improve their thermoelectric properties by solid solutions ($\mathrm{Sc_{1-x}Cr_{x}N}$). In \cite{Sukkabot2019} it was shown that doping ScN with Cr in a substitutional solid solution exhibits half-metallic behavior. They demonstrated that this alloy is potentially applicable as magnetically dilute semiconductors, which enables a range of interesting applications in thermoelectricity, spintronics, piezoelectricity, among others.

In this study, first-principles calculations were carried out on the $\mathrm{Cr_{1-x}Y_{x}N}$ solid solutions. The results obtained of electronic properties and their relation to magnetic properties are presented. First, the results of energy and lattice parameters obtained for the different compositions and magnetic structures are reported. The band structures of each system with their respective densities of states (DOS) are then discussed. Finally, the obtained Seebeck coefficient calculations ($S$), the figure of merit zT and the conclusions obtained from the thermoelectric properties of these alloys are shown.

\begin{figure}[htbp] 
\centering
\begin{subfigure}[t]{0.3\linewidth}
    \includegraphics[width=\linewidth]{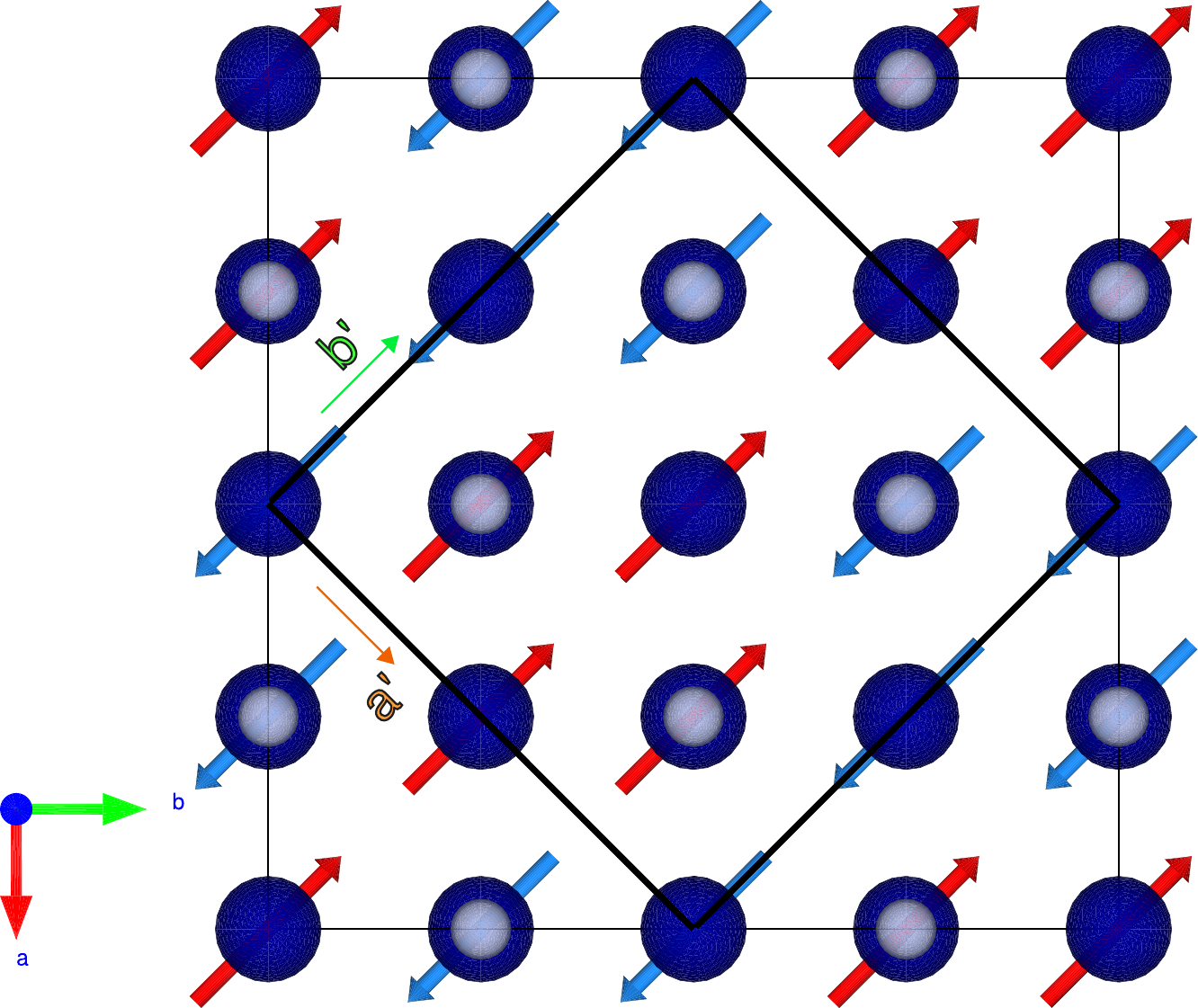}
    \caption{CrN $\mathrm{AFM_{[100]}^{1}}$}
    \label{fig:AFM1}
\end{subfigure}%
\hfill 
\begin{subfigure}[t]{0.3\linewidth}
    \includegraphics[width=\linewidth]{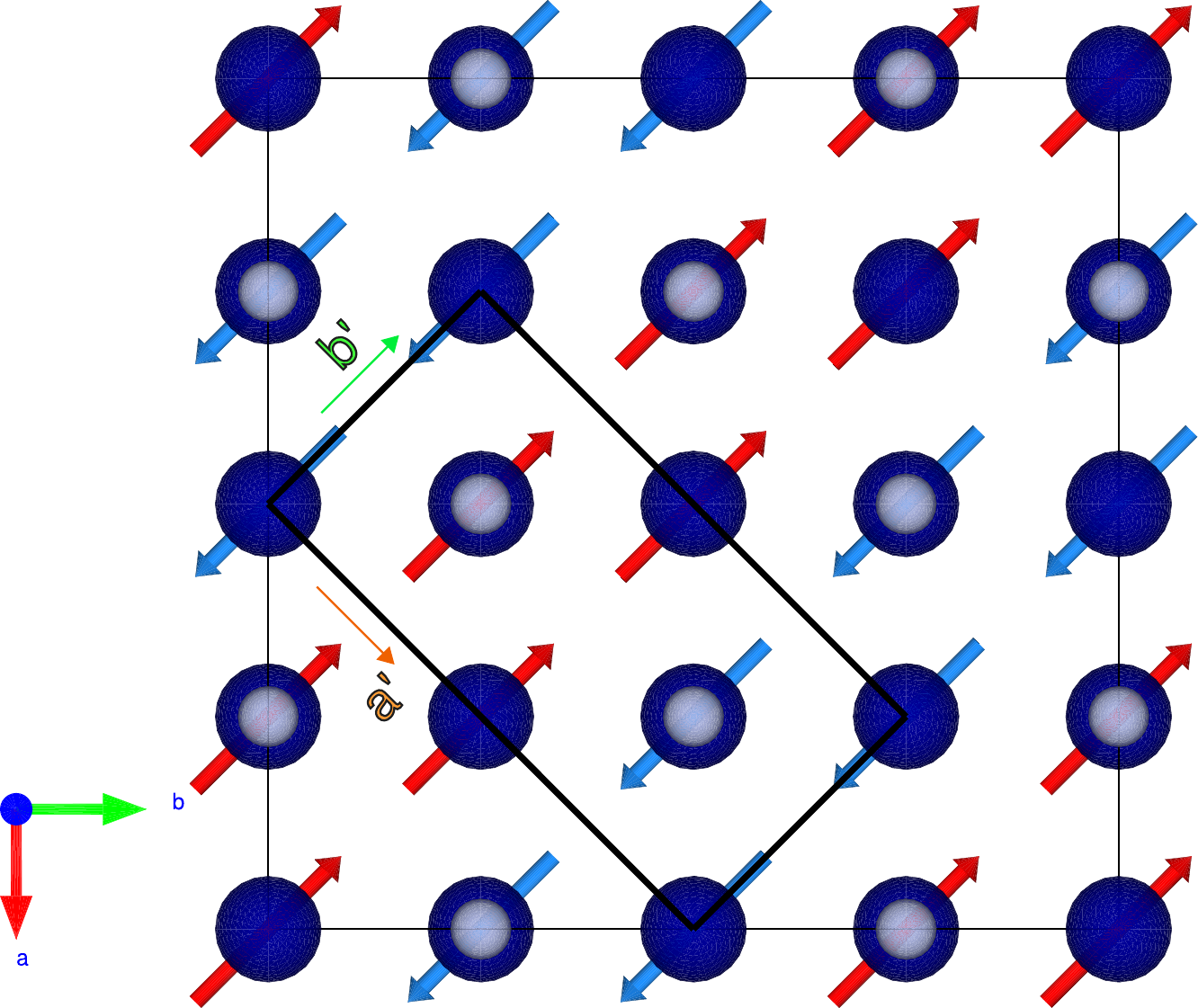}
    \caption{CrN $\mathrm{AFM_{[110]}^{2}}$}
    \label{fig:AFM2}
\end{subfigure}%
\hfill
\begin{subfigure}[t]{0.3\linewidth}
    \includegraphics[width=\linewidth]{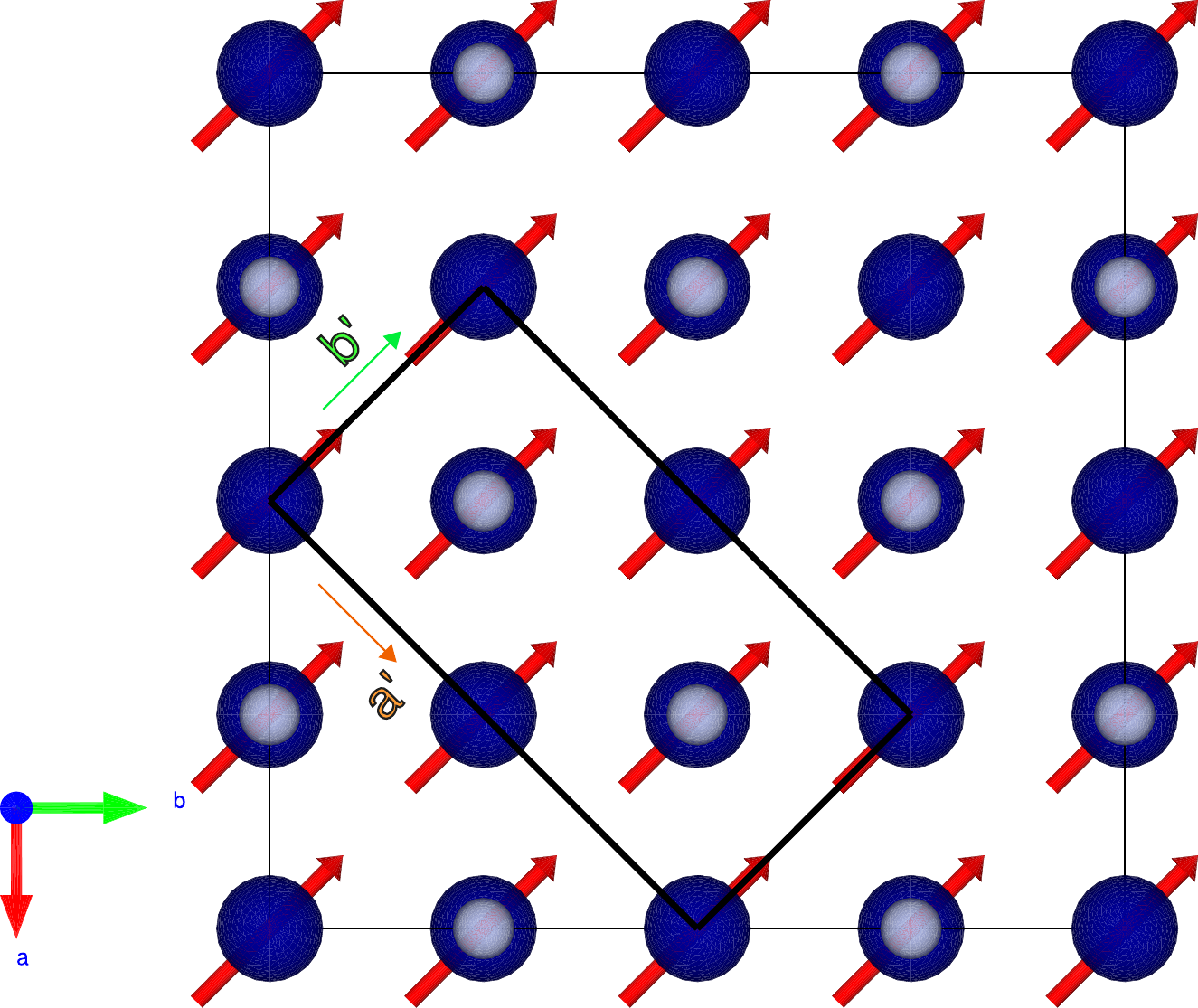}
    \caption{CrN $\mathrm{FM}$}
    \label{fig:FM}
\end{subfigure}

\begin{subfigure}[t]{0.3\linewidth}
    \includegraphics[width=\linewidth]{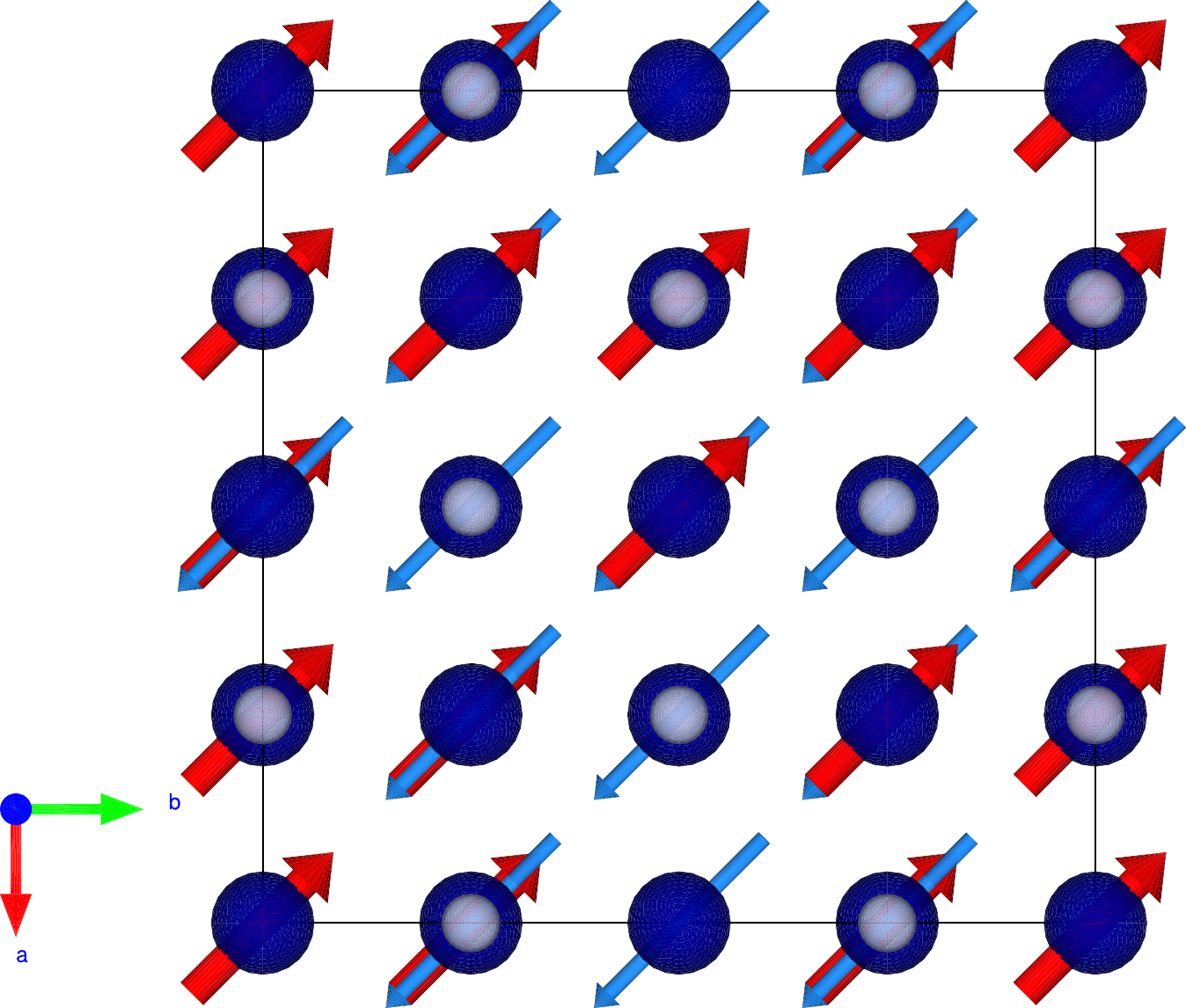}
    \caption{CrN $\mathrm{PM}$}
    \label{fig:PM}
\end{subfigure}%
\hfill
\begin{subfigure}[t]{0.3\linewidth}
    \includegraphics[width=\linewidth]{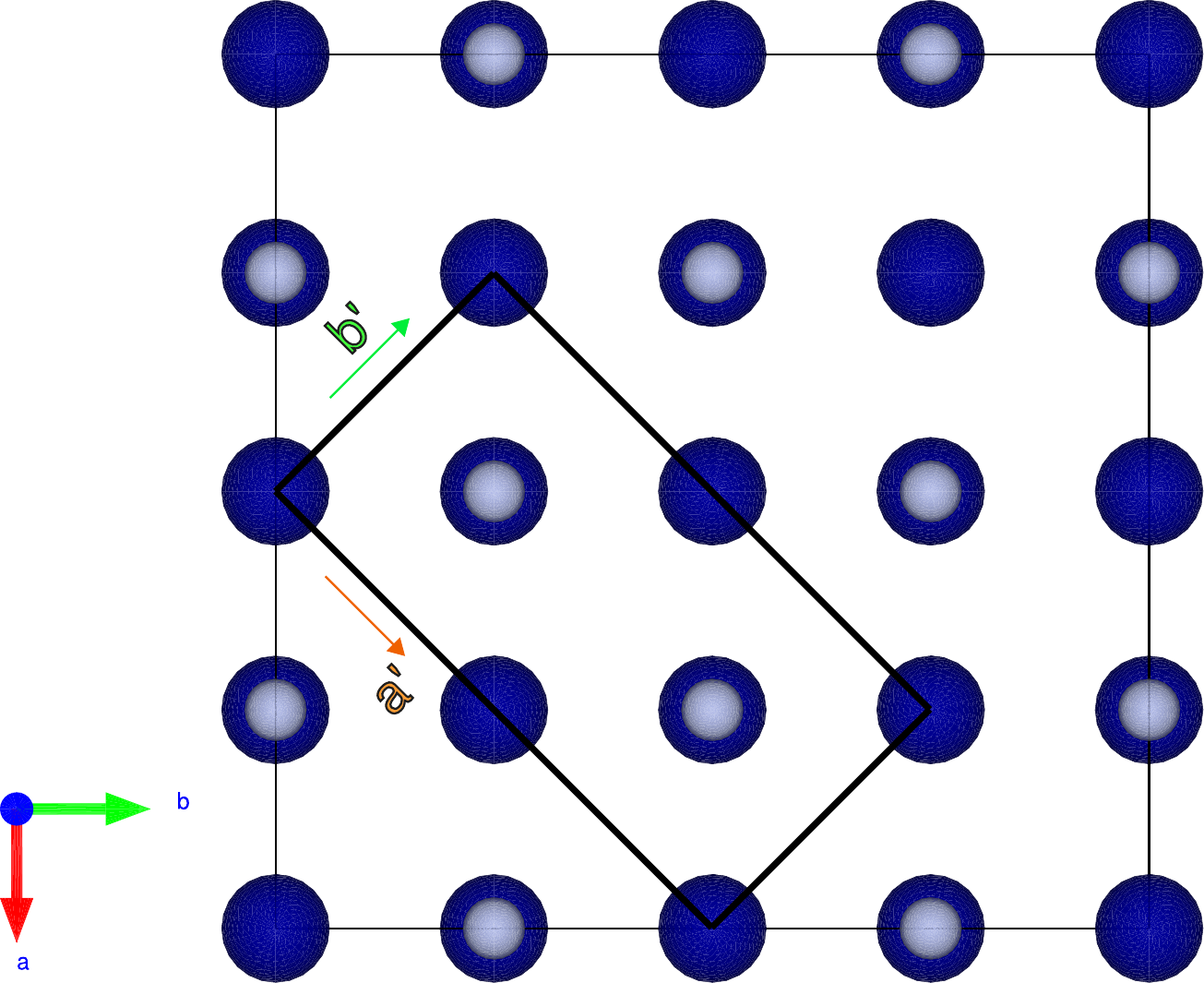}
    \caption{CrN $\mathrm{NM}$}
    \label{fig:NM}
\end{subfigure}%
\hfill
\begin{subfigure}[t]{0.3\linewidth}
    \includegraphics[width=\linewidth]{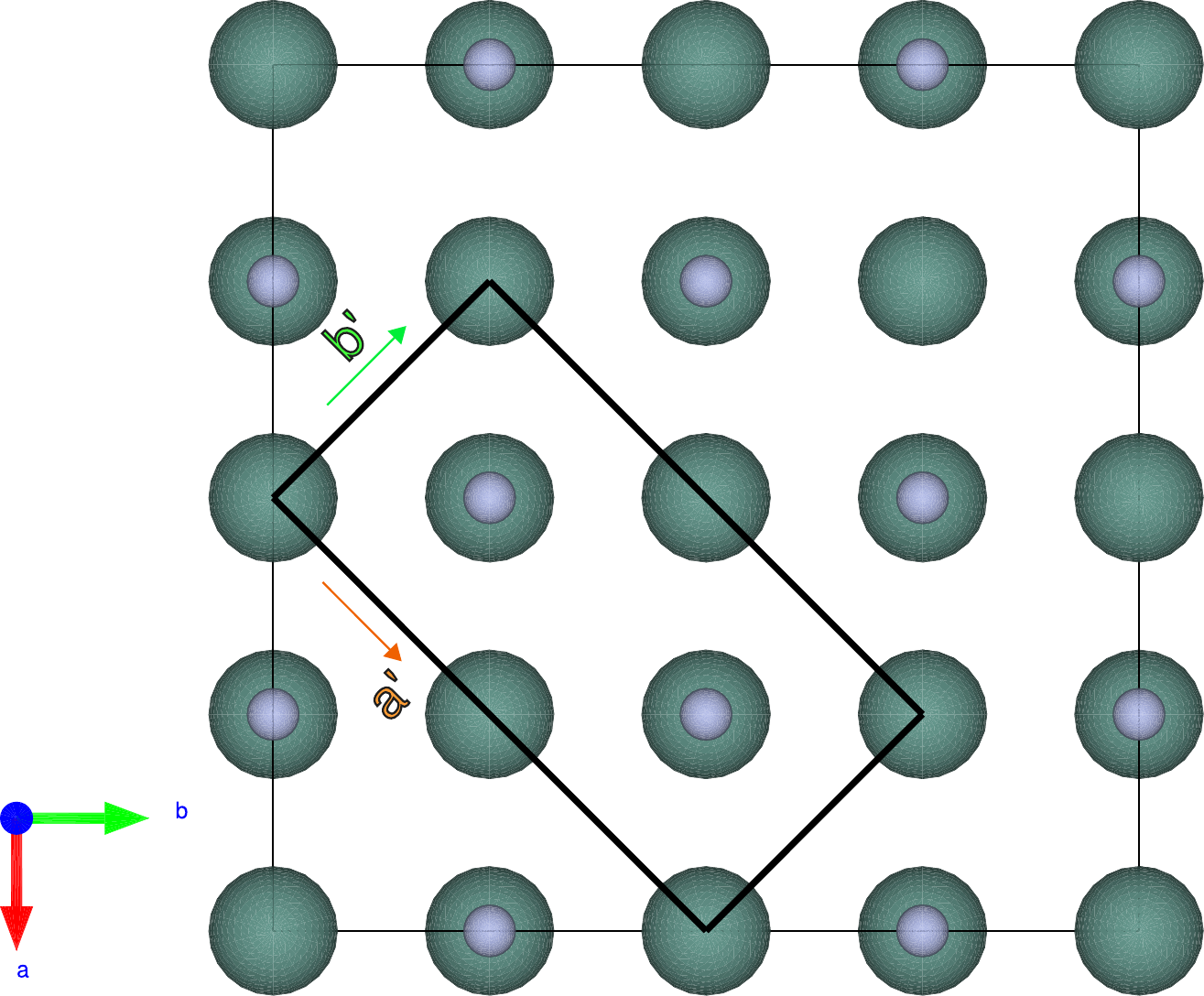}
    \caption{YN}
    \label{fig:YN}
\end{subfigure}
\caption{Model magnetic structures of CrN and YN in FCC phase obtained from \cite{unal_85410}}
\label{fig:estructuras}
\end{figure}

To determine structural and electronic properties of CrN, YN and CrYN alloy, calculations based on density functional theory (DFT) were developed using the Vienna Ab initio Simulation Package (VASP) software \cite{VASP}, which employs the projector augmented wave (PAW) method \cite{Blochl1994}. 

For the solution of the correlation and exchange functional, the strongly constrained and appropriately normed (SCAN) functional was used \cite{Sun2015}. According to \cite{Sun2016}, there are indications that this functional is superior to most gradient-corrected functionals and in this study it has been found, compared to another previous theoretical study, that SCAN fits better the lattice parameters and band structures with respect the experiments, compared to PBE, PBEsol and LDA without U \cite{Filippetti1999,unal_85410,Cheiwchanchamnangij2020,Herwadkar2009}. 

For the cubic and orthorrombic structures of CrN and YN (see figures \ref{fig:estructuras}), 2x2x2 supercells of 64 atoms (32 metal atoms and 32 N atoms) were constructed starting from the orthorhombic cell. For CrN, the three magnetic structures proposed in \cite{Corliss1960} and named $\mathrm{AFM^{1}_{[100]}}$, $\mathrm{AFM^{2}_{110}}$ and $\mathrm{FM}$ were constructed, including the nonmagnetic structure (NM). To define the reference primitive crystalline cell, it must be taken into account that it can reproduce at the same time the magnetic cell in a periodic way. The orthorhombic cell of 8 atoms each cell (4 metal atoms and 4 N atoms) is chosen over the original primitive cell (4 atoms each cell), because the primitive magnetic cell is orthorhombic for CrN ($\mathrm{AFM^{2}_{110}}$, $\mathrm{FM}$) and YN structures. For the $\mathrm{AFM^{1}_{[100]}}$ CrN structure, a 1x2x1 expansion must be performed to reproduce the magnetic structure in a periodic pattern. The lattice parameters and atomic positions of the cubic cell ($\mathbf{a,b,c}$) can be described in terms of the orthorhombic cell ($\mathbf{a',b',c'}$) by applying the following transformation to the orthorhombic cell:

\begin{equation} 
(\mathbf{a,b,c})=
\begin{pmatrix} 
-1 & 1 & 0 \\ 
2 & 2 & 0 \\ 
0 & 0 & 2 \\ 
\end{pmatrix} 
(\mathbf{a',b',c'})
\end{equation}
In the case of the $\mathrm{AFM^{1}_{[100]}}$ CrN structure, the following transformation is performed on the 1x2x1 orthorhombic cell:

\begin{equation} 
(\mathbf{a,b,c})=
\begin{pmatrix} 
-1 & 1 & 0 \\ 
1 & 1 & 0 \\ 
0 & 0 & 2 \\ 
\end{pmatrix} 
(\mathbf{a',b',c'})
\end{equation}
The initial lattice parameters of all CrN orthorhombic cells were: $a=5.757 \: \mathrm{\mathring{A}}$, $b=2.964 \: \mathrm{\mathring{A}}$ and $c=4.134 \: \mathrm{\mathring{A}}$, extracted from the values measured by neutron diffraction in \cite{Corliss1960}. The initial values of the orthorhombic cell of YN were taken from the cubic cell measured by X-ray diffraction in \cite{Morris1985} ($a=4.894 \: \mathrm{\mathring{A}}$).

\begin{figure}[htbp]
\centering
\begin{subfigure}[t]{0.38\linewidth}
    \includegraphics[width=\linewidth]{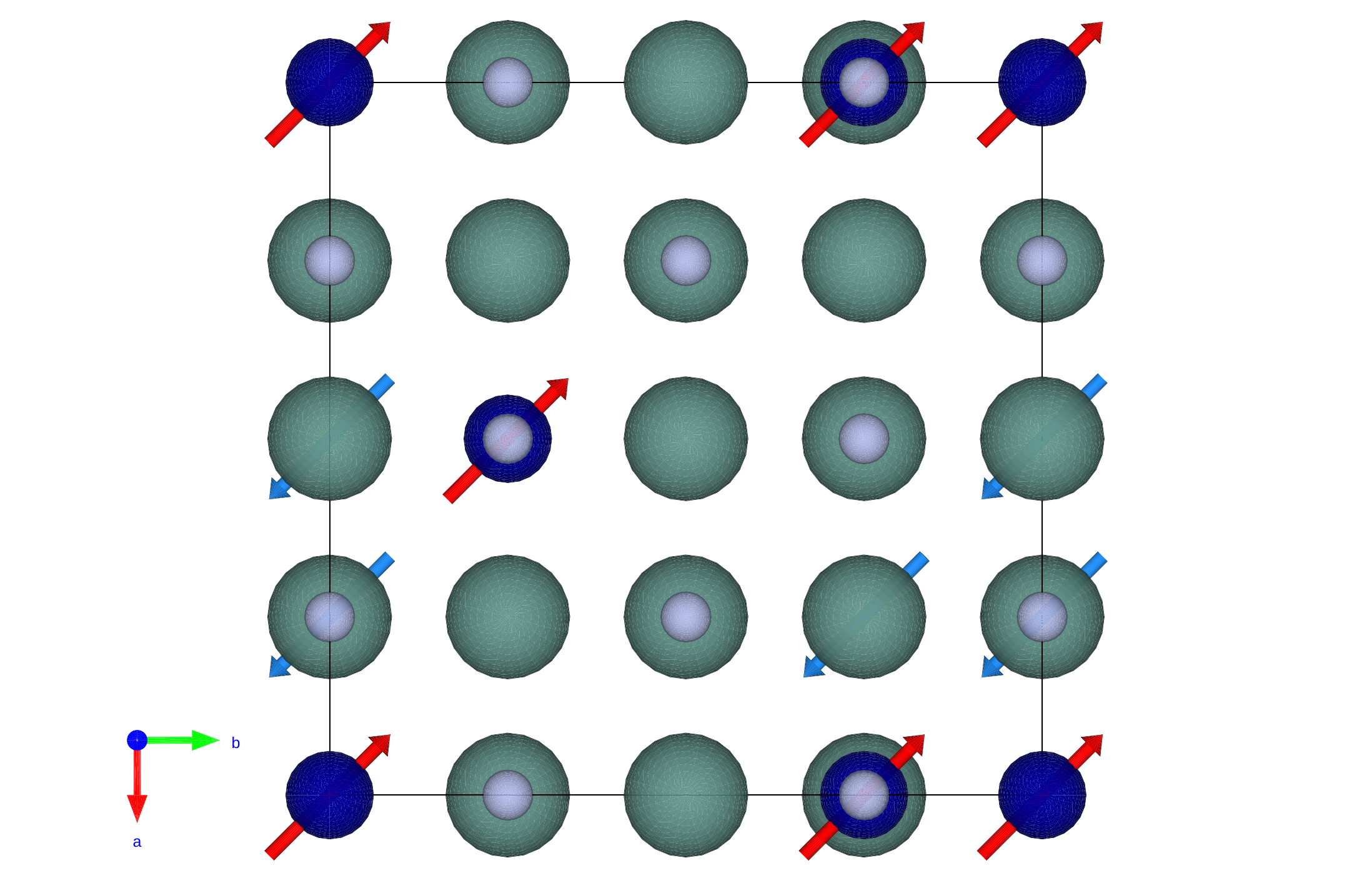}
    \caption{$\mathrm{AFM_{[100]}^{1}}$}
    \label{fig:25AFM1}
\end{subfigure}%
\hspace{-1.3cm}%
\begin{subfigure}[t]{0.38\linewidth}
    \includegraphics[width=\linewidth]{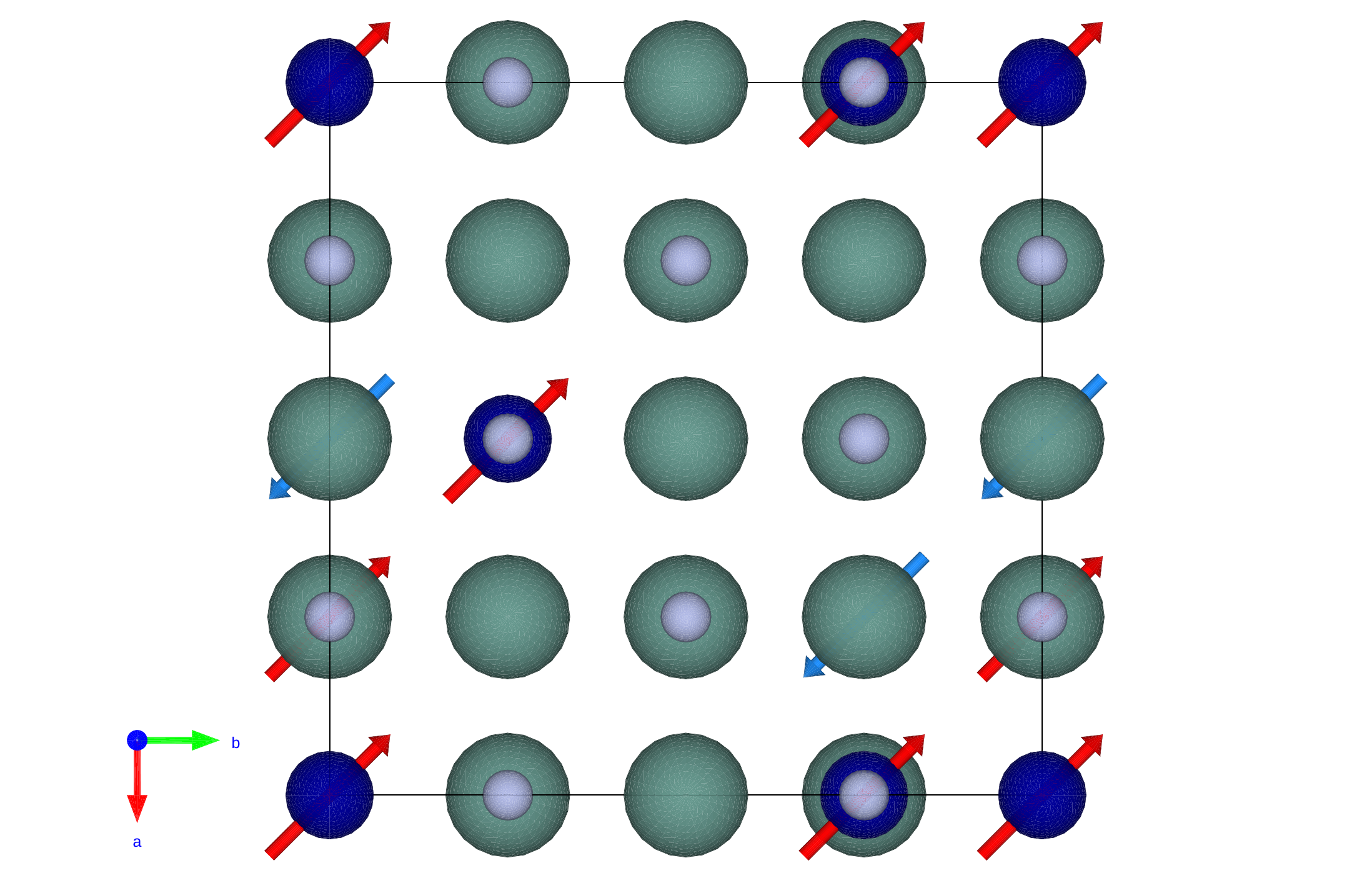}
    \caption{$\mathrm{AFM_{[110]}^{2}}$}
    \label{fig:25AFM2}
\end{subfigure}%
\hspace{-1.3cm}%
\begin{subfigure}[t]{0.38\linewidth}
    \includegraphics[width=\linewidth]{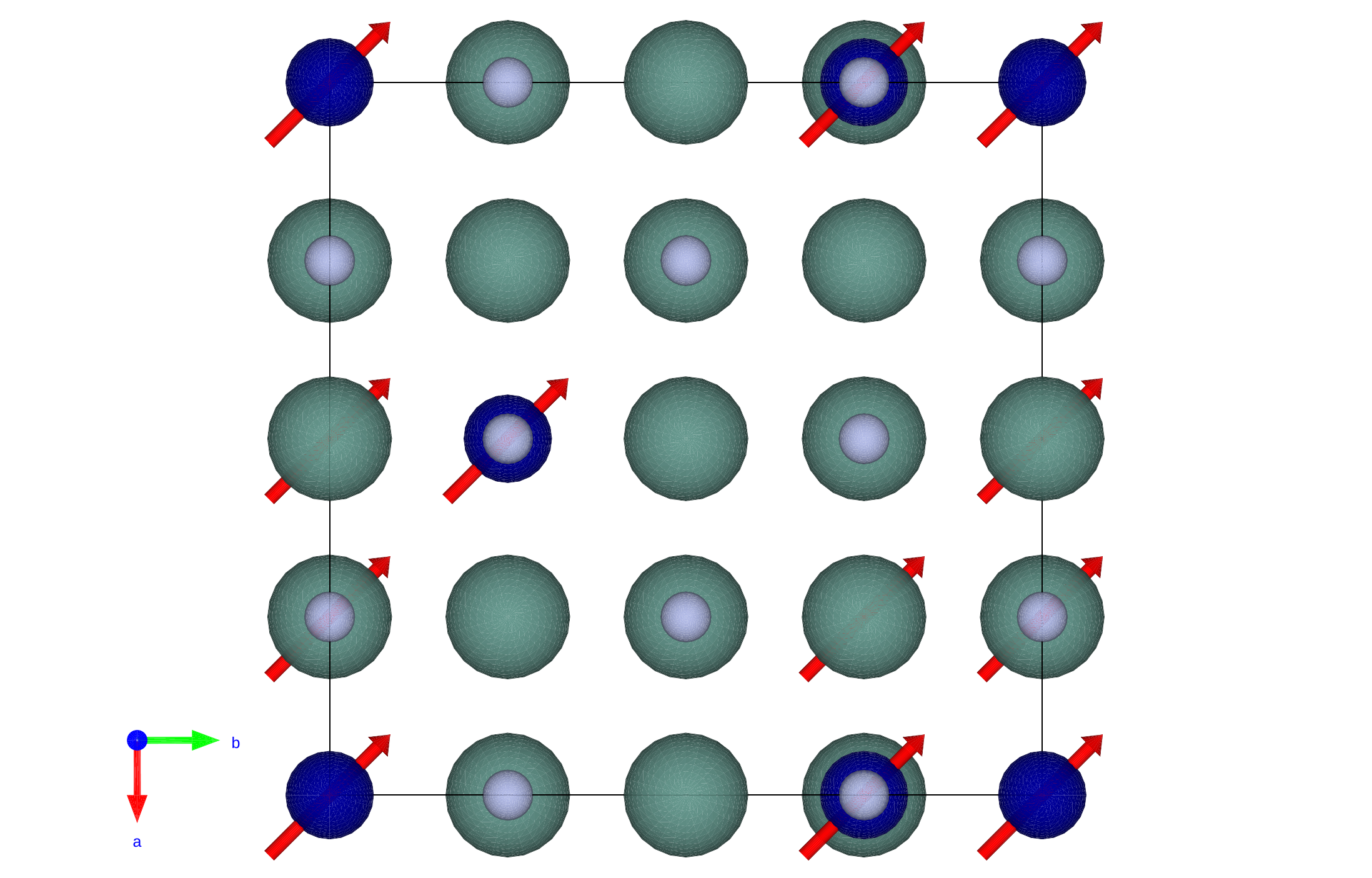}
    \caption{$\mathrm{FM}$}
    \label{fig:25FM}
\end{subfigure}        
\caption{$\mathrm{Cr_{0.25}Y_{0.75}N}$}
\label{fig:25CrYN}
\end{figure}

\begin{figure}[htbp]
\centering
\begin{subfigure}[t]{0.38\linewidth}
    \includegraphics[width=\linewidth]{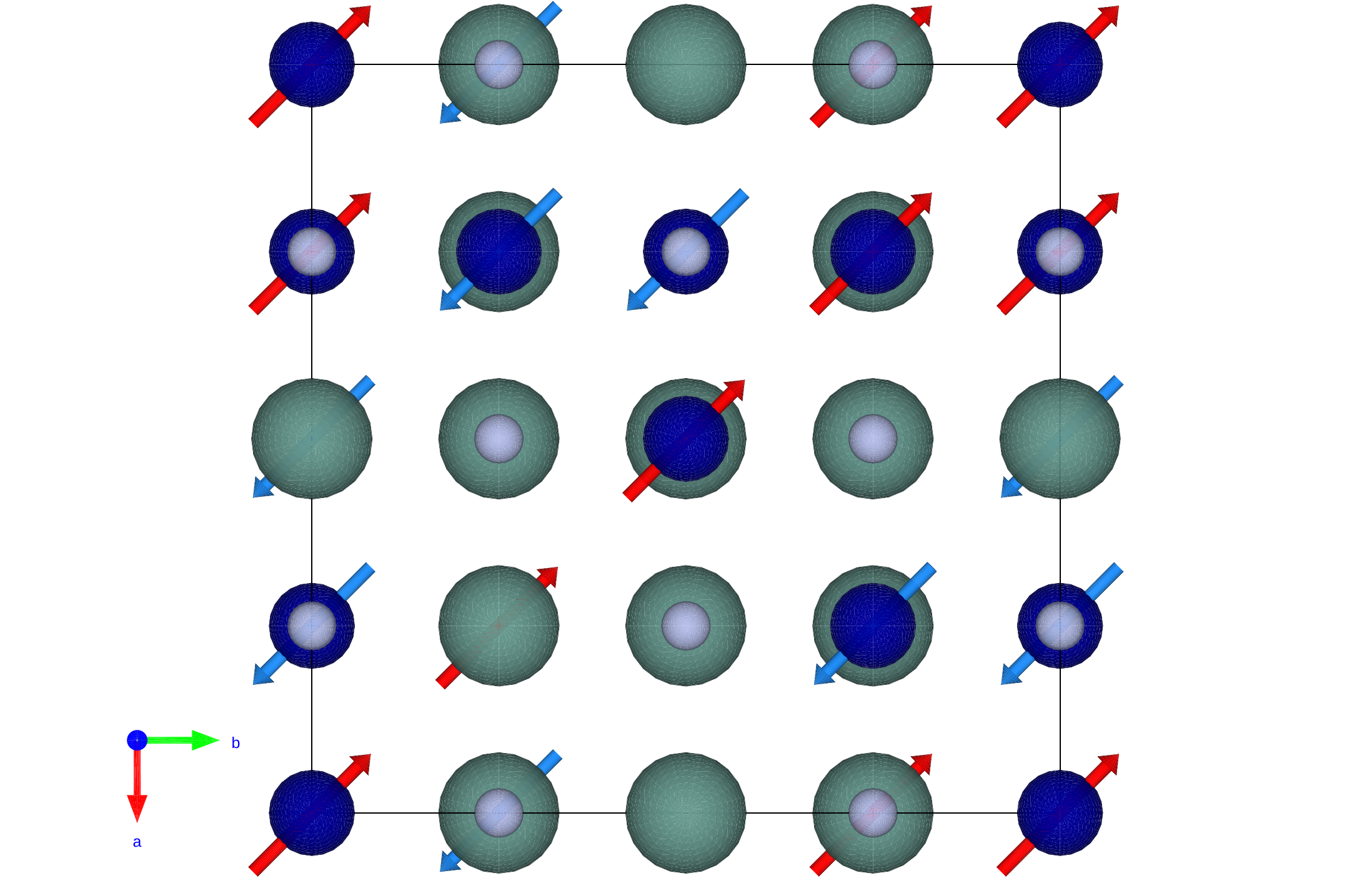}
    \caption{$\mathrm{AFM_{[100]}^{1}}$}
    \label{fig:50AFM1}
\end{subfigure}%
\hspace{-1.3cm}%
\begin{subfigure}[t]{0.38\linewidth}
    \includegraphics[width=\linewidth]{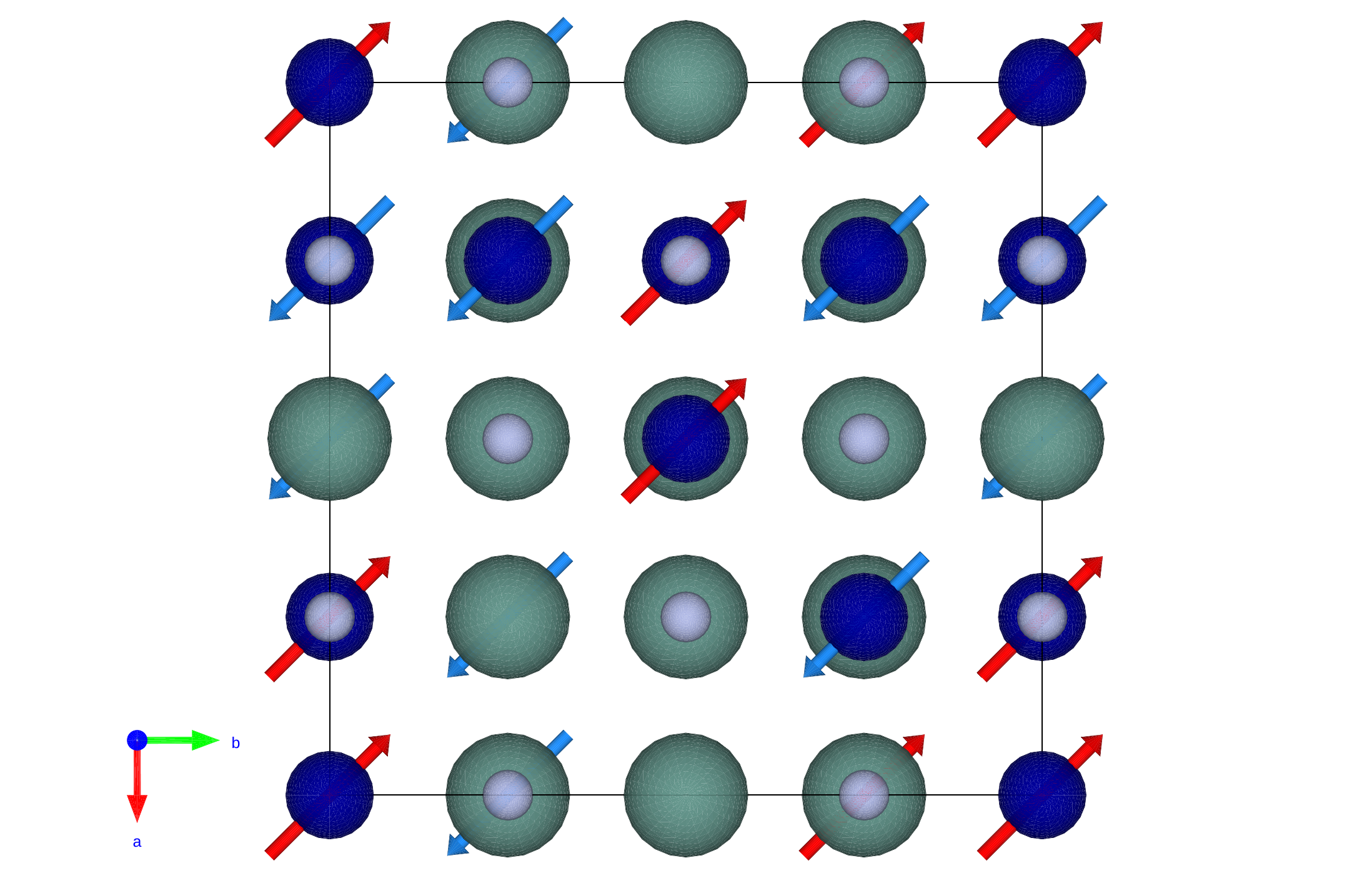}
    \caption{$\mathrm{AFM_{[110]}^{2}}$}
    \label{fig:50AFM2}
\end{subfigure}
\hspace{-1.3cm}
\begin{subfigure}[t]{0.38\linewidth}
    \includegraphics[width=\linewidth]{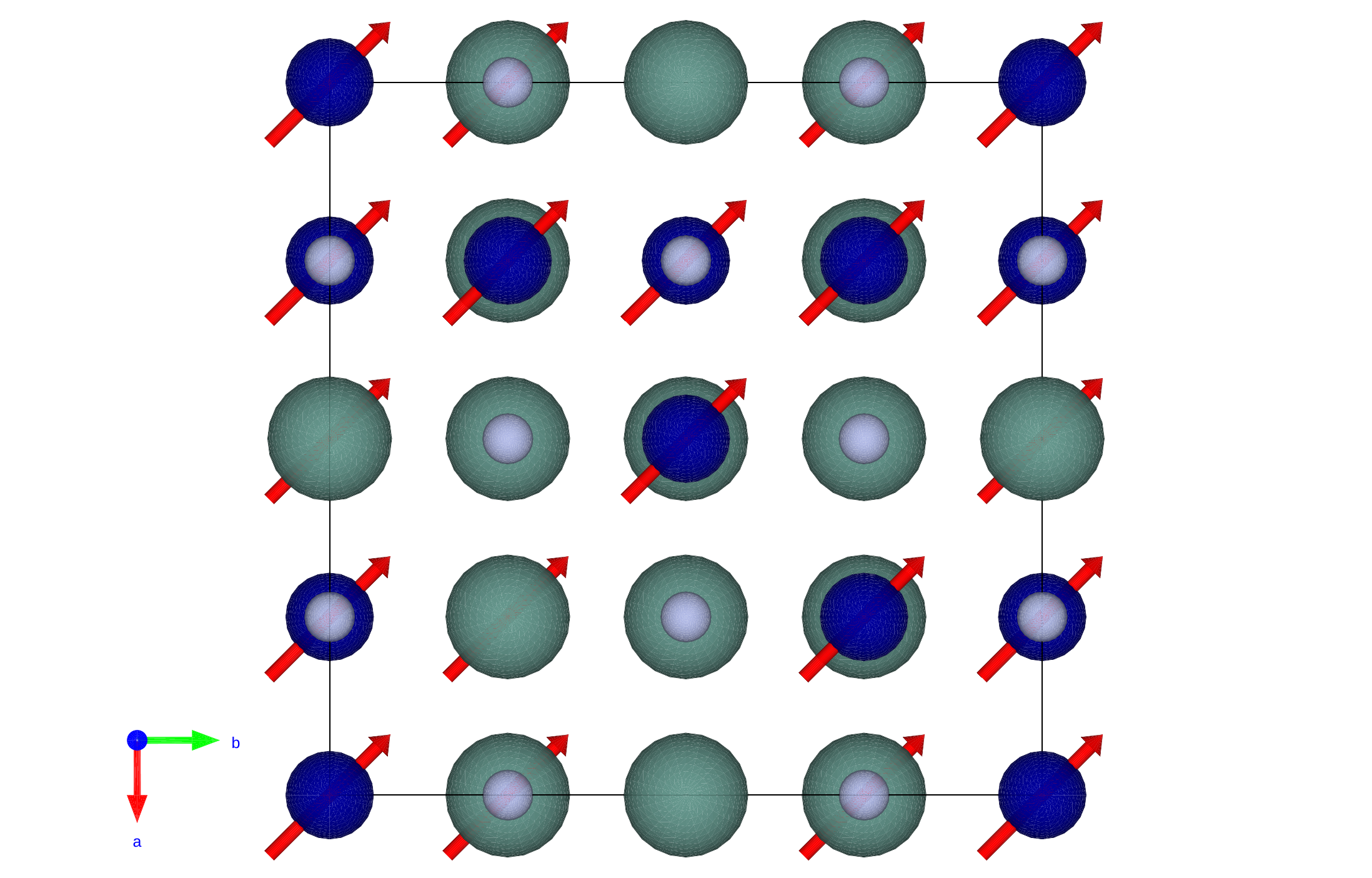}
    \caption{$\mathrm{FM}$}
    \label{fig:50FM}
\end{subfigure}        
\caption{$\mathrm{Cr_{0.5}Y_{0.5}N}$}
\label{fig:50CrYN}
\end{figure}

\begin{figure}[htbp]
\centering
\begin{subfigure}[t]{0.28\linewidth}
    \includegraphics[width=\linewidth]{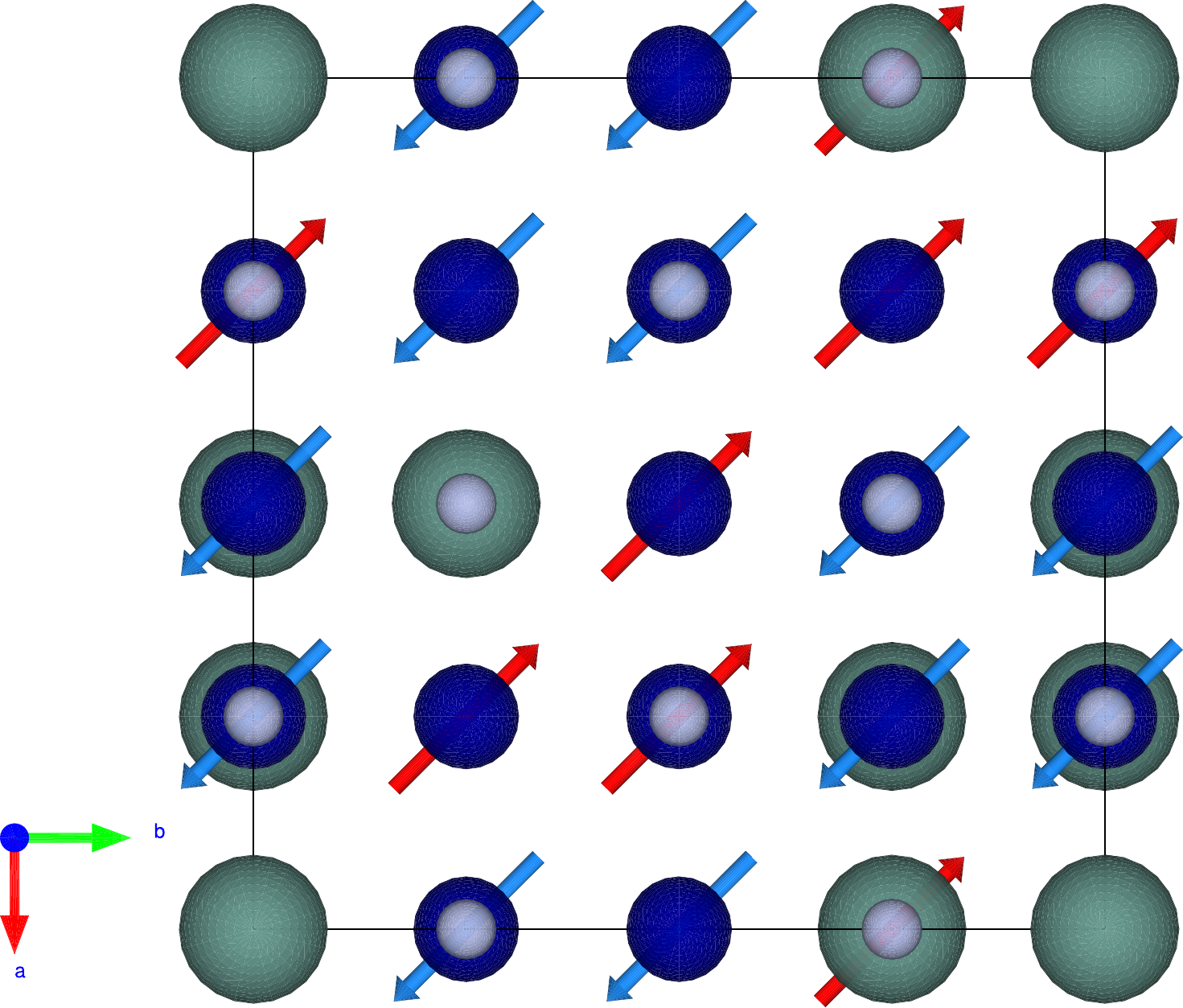}
    \caption{$\mathrm{AFM_{[100]}^{1}}$}
    \label{fig:75AFM1}
\end{subfigure}
\hspace{.5cm}
\begin{subfigure}[t]{0.28\linewidth}
    \includegraphics[width=\linewidth]{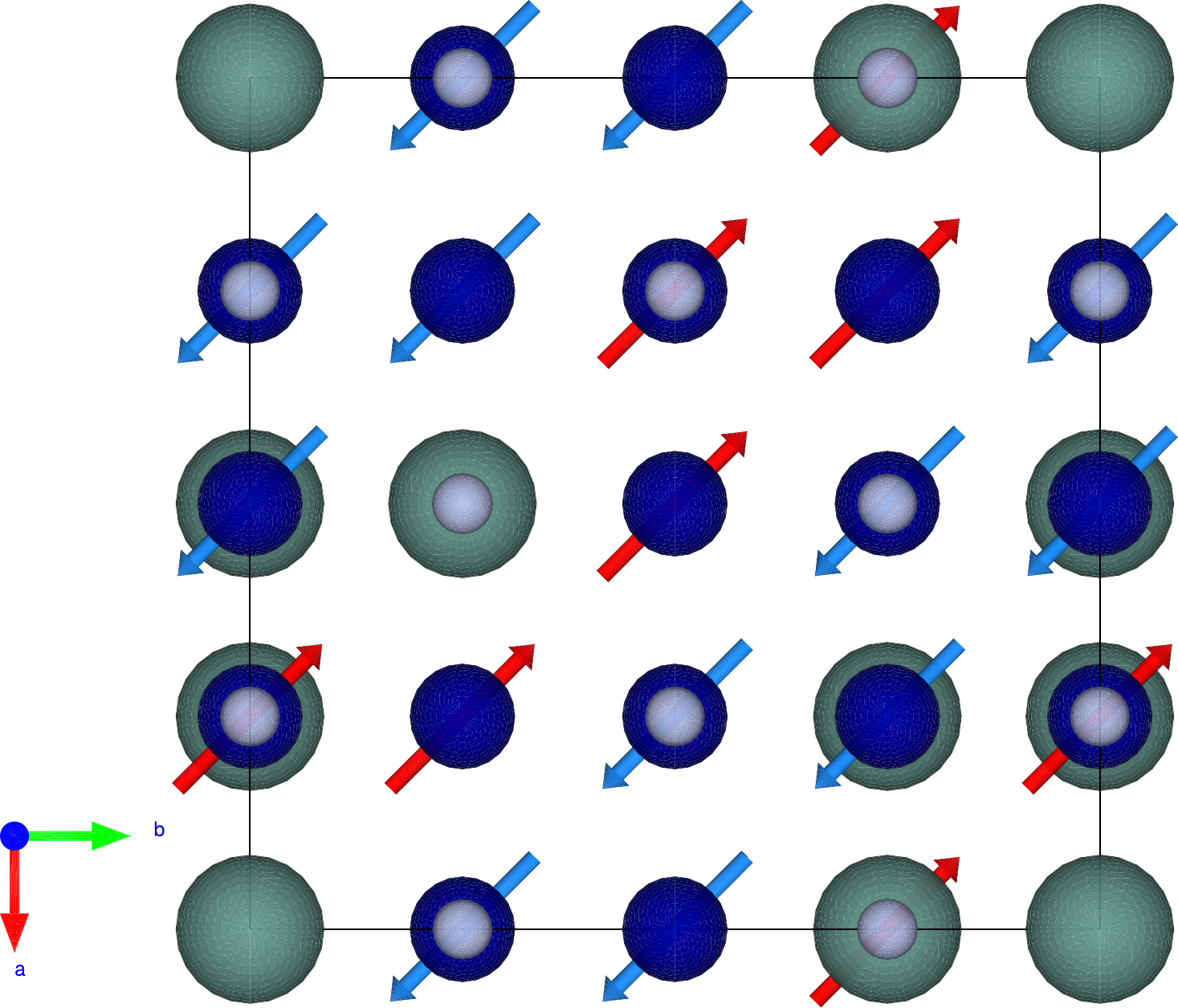}
    \caption{$\mathrm{AFM_{[110]}^{2}}$}
    \label{fig:75AFM2}
\end{subfigure}
\hspace{.5cm}
\begin{subfigure}[t]{0.28\linewidth}
    \includegraphics[width=\linewidth]{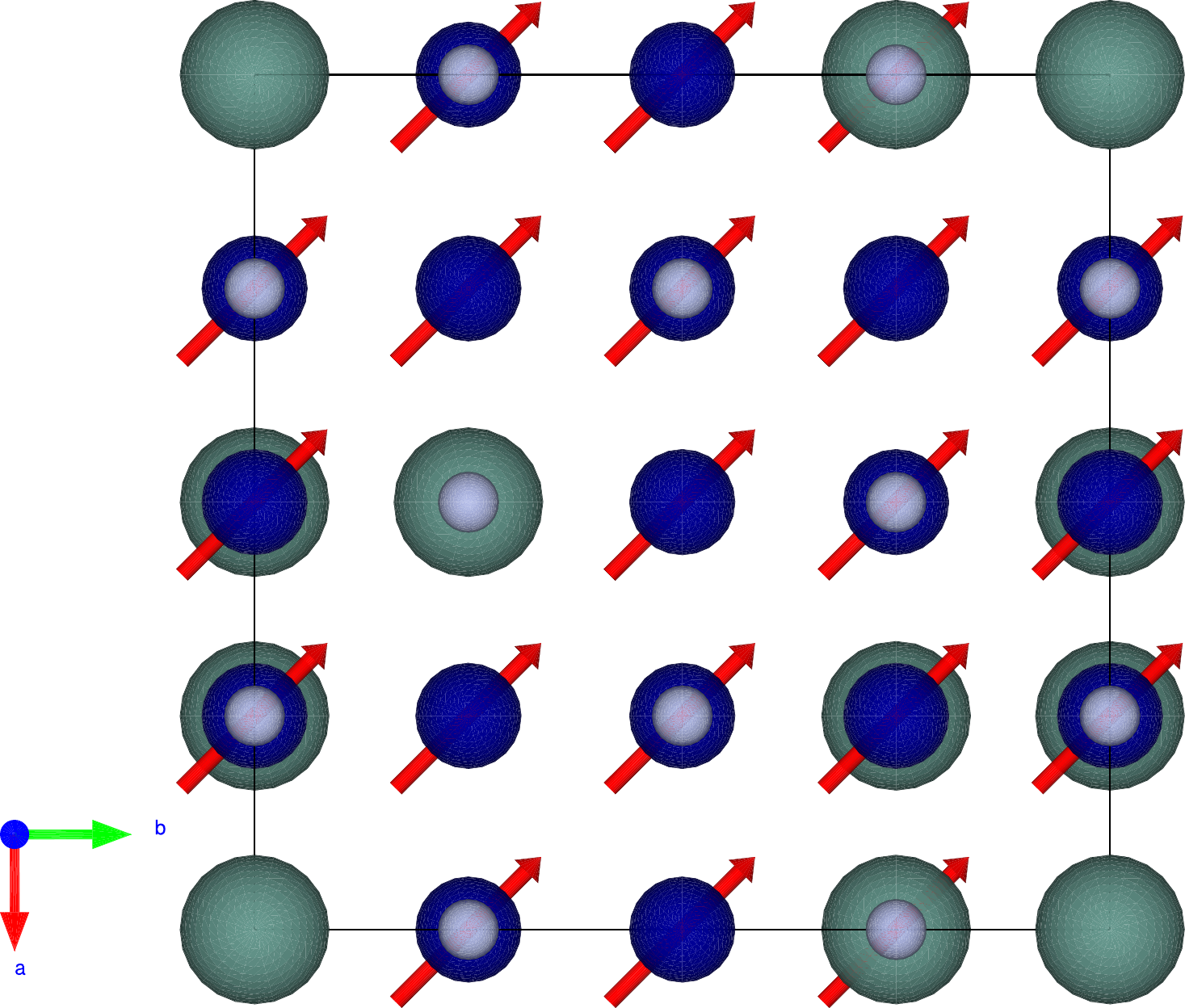}
    \caption{$\mathrm{FM}$}
    \label{fig:75FM}
\end{subfigure}        
\caption{$\mathrm{Cr_{0.75}Y_{0.75}N}$}
\label{fig:75CrYN}
\end{figure}

For (Cr,Y)N  alloys, the same cubic supercells of 64 atoms were also used, but in three different compositions: $\mathrm{Cr_{0.25}Y_{0.75}N}$, $\mathrm{Cr_{0.5}Y_{0.5}N}$ and $\mathrm{Cr_{0.75}Y_{0.25}N}$ (see figures \ref{fig:25CrYN}, \ref{fig:50CrYN} and \ref{fig:75CrYN}). The proposed magnetic structures for CrN were built in the Cr atoms available at each concentration (see figures \ref{fig:25CrYN}, \ref{fig:50CrYN}, and \ref{fig:75CrYN}). To simulate a short-range random distribution of Cr and Y atoms at all concentrations within cubic structures, the method of quasi-random structures (SQS) \cite{Zunger1990} was used using the ATAT software \cite{atat} for the construction of the correlation functions (for details of these functions, see the study \cite{unal_85410}). This method has been shown in different theoretical studies on ternary CrN alloys with transition metals to be a reliable thermodynamic model for disordered phases and provides a considerably simpler and computationally efficient approach compared to other techniques, as demonstrated in \cite{Zhou2013}. Thus, first-principles calculations on SQS are a good combination for accurately reproducing the structural, electronic, and thermodynamic properties of disordered ternary alloys. The lattice parameter values obtained after relaxing all CrN magnetic structures were used as initial values for each SQS labeled with the template magnetic structure. Using the SQS for $\mathrm{Cr_{0.5}Y_{0.5}N}$, a paramagnetic cell (PM) was constructed in order to build an SQS for $\mathrm{Cr_{0.5\uparrow}Cr_{0.5\downarrow}N}$ (see figure \ref{fig:PM}). To achieve this, all $\mathrm{Cr_{\uparrow}}$ atoms were placed in the Cr positions and all $\mathrm{Cr_{\downarrow}}$ atoms were placed in the Y positions.

For structural and magnetic properties, all structures were fully relaxed with a cut-off kinetic energy of 400 eV and a Monkhorst Pack mesh of 4x4x4, and initial magnetic moments in $\mathrm{5\:\mu_B}$ in all the cases. For electronic density of states (DOS), self-consistent calculations were performed with a finner Monkhorst-Pack mesh of 6x6x6 and an energy difference criterion of $\mathrm{10^{-7}\: eV}$. Since the distribution of Cr and Y atoms in the alloys is random, as are the spins in the Cr PM supercell, it is not possible to construct a primitive orthorhombic cell, the cubic supercell being the primitive cell for these cases. For that reason, the band structure unfolding was made with help of \cite{Zhu2024}. From these results the Seebeck coefficient and zT were calculated using BoltaTrap2 \cite{BoltzTraP2} software. For these calculations, 100 points were used in the interpolation of the band structure obtained by DFT.

\section{Results and discussion}

\subsection{Structural and magnetic properties}

\begin{table}[htbp]
    \centering
    \caption{Lattice parameters and magnetic moments of different magnetic structures}
    \label{t:param_magmom_part1}
    \resizebox{\textwidth}{!}{%
    \begin{tabular}{|c|c|c|c|c|c|c|c|c|c|}
    \hline 
    Structure &  
    $\mathrm{a \, (\mathring{A})}$ & 
    $\mathrm{b \, (\mathring{A})}$ & 
    $\mathrm{c \, (\mathring{A})}$ & 
    $\alpha \, (^{\circ})$ & 
    $\mathrm{m\!\uparrow} \, (\mu_B)$ & 
    $\mathrm{N \, (m\!\uparrow)}$ & 
    $\mathrm{m\!\downarrow} \, (\mu_B)$ & 
    $\mathrm{N \, (m\!\downarrow)}$ & 
    $\mathrm{M} \, (\mu_B)$ \\ \hline \hline
    
    $\mathrm{YN}$ &  4.865 & -- & -- & 90 & -- & -- & -- & -- & -- \\ \hline
    $\mathrm{CrN}$-$\mathrm{NM}$ &  4.022 & -- & -- & 90 & -- & -- & -- & -- & -- \\ \hline
    $\mathrm{CrN}$-$\mathrm{PM}$ &  4.132 & 4.131 & 4.133 & 90 & 2.56 & 16 & 2.54 & 16 & 0.00 \\ \hline
    \multicolumn{10}{|l|}{$\mathrm{AFM^{1}_{[100]}}$} \\ \hline 
    $\mathrm{Cr_{0.25}Y_{0.75}N}$ &  4.633 & 4.621 & 4.650 & 90.02 & 2.63 & 4 & 2.62 & 4 & 0.01 \\ \hline 
    $\mathrm{Cr_{0.50}Y_{0.50}N}$ &  4.482 & 4.479 & 4.489 & 90.18 & 2.70 & 10 & 2.52 & 6 & 12.00 \\ \hline 
    $\mathrm{Cr_{0.75}Y_{0.25}N^{\ast}}$ &  4.309 & 4.297 & 4.331 & 89.79 & $2.58\,(0.27)$ & $13\,(2)$ & $2.59\,(0.48)$ & $11\,(2)$ & 6.00 \\ \hline 
    $\mathrm{CrN}$ &  4.126 & 4.126 & 4.138 & 90 & 2.57 & 16 & 2.57 & 16 & 0.00 \\ \hline
    \multicolumn{10}{|l|}{$\mathrm{AFM^{2}_{[110]}}$} \\ \hline
    $\mathrm{Cr_{0.25}Y_{0.75}N}$ & 4.635 & 4.623 & 4.646 & 90.01 & 2.56 & 5 & 2.78 & 3 & 6.00 \\ \hline 
    $\mathrm{Cr_{0.50}Y_{0.50}N}$ & 4.486 & 4.476 & 4.487 & 90.25 & 2.64 & 7 & 2.66 & 9 & 6.01 \\ \hline 
    $\mathrm{Cr_{0.75}Y_{0.25}N^{\ast}}$ &  4.309 & 4.301 & 4.332 & 89.78 & $2.70\,(0.27)$ & $9\,(2)$ & $2.51\,(0.48)$ & $11\,(2)$ & 6.00 \\ \hline 
    $\mathrm{CrN}$ & 4.122 & 4.122 & 4.113 & 88.31 & 2.52 & 16 & 2.52 & 16 & 0.00 \\ \hline 
    \multicolumn{10}{|l|}{$\mathrm{FM}$} \\ \hline
    $\mathrm{Cr_{0.25}Y_{0.75}N}$ &  4.631 & 4.621 & 4.650 & 89.98 & 2.65 & 5 & 2.51 & 3 & 6.00 \\ \hline 
    $\mathrm{Cr_{0.50}Y_{0.50}N}$ &  4.479 & 4.481 & 4.484 & 90.11 & 2.36 & 5 & 2.75 & 11 & 18.00 \\ \hline 
    $\mathrm{Cr_{0.75}Y_{0.25}N^{\ast}}$ &  4.310 & 4.301 & 4.327 & 89.93 & $2.56\,(0.56)$ & 12 & $2.61\,(0.56)$ & $11\,(1)$ & 2.00 \\ \hline 
    $\mathrm{CrN}$ &  4.155 & -- & -- & 90 & 2.76 & 32 & -- & 0 & 88.32 \\ \hline 
    \multicolumn{10}{|l|}{$\mathrm{Theoretic}$} \\ \hline
    $\mathrm{Cr_{0.875}Y_{0.125}N}$-$\mathrm{AFM^{2}_{[110]}}$ \cite{Matas2025} & 4.266 & -- & -- & -- & 2.87 & -- & -- & -- & -- \\ \hline
    $\mathrm{Cr_{0.875}Y_{0.125}N}$-$\mathrm{PM}$ \cite{Matas2025} & 4.266 & -- & -- & -- & 2.908 & -- & -- & -- & -- \\ \hline
    $\mathrm{CrN}$-$\mathrm{AFM^{2}_{[110]}}$ \cite{Matas2025} & 4.266 & -- & -- & -- & 2.882 & -- & -- & -- & -- \\ \hline
    $\mathrm{CrN}$-$\mathrm{PM}$ \cite{Matas2025} & 4.266 & -- & -- & -- & 2.89 & -- & -- & -- & -- \\ \hline
    $\mathrm{Cr_{0.25}Y_{0.75}N}$-$\mathrm{PM}$ \cite{Zhou2013} & 4.33 & -- & -- & -- & -- & -- & -- & -- & -- \\ \hline
    $\mathrm{Cr_{0.50}Y_{0.50}N}$-$\mathrm{PM}$ \cite{Zhou2013} & 4.6 & -- & -- & -- & -- & -- & -- & -- & -- \\ \hline
    $\mathrm{Cr_{0.75}Y_{0.25}N}$-$\mathrm{PM}$ \cite{Zhou2013} & 4.75 & -- & -- & -- & -- & -- & -- & -- & -- \\ \hline
    $\mathrm{CrN}$-$\mathrm{PM}$ \cite{Zhou2013} & 4.146 & -- & -- & -- & -- & -- & -- & -- & -- \\ \hline
    $\mathrm{YN}$ \cite{Zhou2013} & 4.91 & -- & -- & -- & -- & -- & -- & -- & -- \\ \hline
    \end{tabular}}
\end{table}

\begin{table}[htbp]
    \centering
    \label{t:param_magmom_part2}
    \resizebox{\textwidth}{!}{%
    \begin{tabular}{|c|c|c|c|c|c|c|c|c|c|}
    \hline 
    Structure &  
    $\mathrm{a \, (\mathring{A})}$ & 
    $\mathrm{b \, (\mathring{A})}$ & 
    $\mathrm{c \, (\mathring{A})}$ & 
    $\alpha \, (^{\circ})$ & 
    $\mathrm{m\!\uparrow} \, (\mu_B)$ & 
    $\mathrm{N \, (m\!\uparrow)}$ & 
    $\mathrm{m\!\downarrow} \, (\mu_B)$ & 
    $\mathrm{N \, (m\!\downarrow)}$ & 
    $\mathrm{M} \, (\mu_B)$ \\ \hline \hline

	$\mathrm{CrN}$-$\mathrm{PM}$ \cite{Shulumba2014} & 4.14 & -- & -- & -- & -- & -- & -- & -- & -- \\ \hline
    $\mathrm{CrN}$-$\mathrm{PM}$ \cite{Siegel2003} & 4.058 & -- & -- & -- & -- & -- & -- & -- & -- \\ \hline
    $\mathrm{CrN}$-$\mathrm{PM}$ \cite{Kerdsongpanya2016} & 4.128 & -- & -- & -- & 2.81 & -- & -- & -- & -- \\ \hline
    $\mathrm{CrN}$-$\mathrm{PM}$ \cite{Zhang2013} & 4.146 & -- & -- & -- & -- & -- & -- & -- & -- \\ \hline
    $\mathrm{CrN}$-$\mathrm{NM}$ \cite{Zhang2013} & 4.05 & -- & -- & -- & -- & -- & -- & -- & -- \\ \hline
    $\mathrm{CrN}$-$\mathrm{NM}$ \cite{Alling2010a} & 4.055 & -- & -- & -- & -- & -- & -- & -- & -- \\ \hline
    $\mathrm{CrN}$-$\mathrm{PM}$ \cite{Alling2010a} & 4.149 & -- & -- & -- & -- & -- & -- & -- & -- \\ \hline
    $\mathrm{CrN}$-$\mathrm{AFM^{2}_{[110]}}$ \cite{Alling2007} & 4.196 & -- & -- & -- & 2.20 & -- & -- & -- & -- \\ \hline
    $\mathrm{CrN}$-$\mathrm{PM}$ \cite{Alling2007} & 4.206 & -- & -- & -- & 2.32 & -- & -- & -- & -- \\ \hline
    $\mathrm{CrN}$-$\mathrm{FM}$ \cite{Alling2007} & 4.22 & -- & -- & -- & 2.19 & -- & -- & -- & -- \\ \hline
    $\mathrm{YN}$ \cite{Holec2012} & 4.917 & -- & -- & -- & -- & -- & -- & -- & -- \\ \hline
    $\mathrm{YN}$ \cite{Amrani2007} & 4.915 & -- & -- & -- & -- & -- & -- & -- & -- \\ \hline
    $\mathrm{YN}$ \cite{Salguero2003} & 4.93 & -- & -- & -- & -- & -- & -- & -- & -- \\ \hline
    $\mathrm{YN}$ \cite{Stampfl2001} & 4.85 & -- & -- & -- & -- & -- & -- & -- & -- \\ \hline
    $\mathrm{CrN}$-$\mathrm{AFM^{1}_{[100]}}$ \cite{Alling2010} & 4.128 & -- & -- & -- & 2.72 & -- & -- & -- & -- \\ \hline
    $\mathrm{CrN}$-$\mathrm{AFM^{2}_{[110]}}$ \cite{Alling2010} & 4.117 & -- & -- & -- & 2.81 & -- & -- & -- & -- \\ \hline
    $\mathrm{CrN}$-$\mathrm{FM}$ \cite{Alling2010} & 4.158 & -- & -- & -- & 2.96 & -- & -- & -- & -- \\ \hline
    $\mathrm{CrN}$-$\mathrm{PM}$ \cite{Alling2010} & 4.133 & -- & -- & -- & 2.82 & -- & -- & -- & -- \\ \hline
    \multicolumn{10}{|l|}{$\mathrm{Experimental}$} \\ \hline
    $\mathrm{CrN}$-$\mathrm{AFM_{[110]}^{2}}$ \cite{Corliss1960} & 4.148 & -- & -- & 88.23 & 2.36 & -- & -- & -- & -- \\ \hline
    $\mathrm{CrN}$-$\mathrm{AFM_{[110]}^{2}}$ \cite{Gall2002} & 4.162 & -- & -- & -- & -- & -- & -- & -- & -- \\ \hline
    $\mathrm{CrN}$-$\mathrm{PM}$ \cite{Rivadulla2009} & 4.131 & -- & -- & -- & 2.5 & -- & -- & -- & -- \\ \hline
    $\mathrm{YN}$ \cite{Villars1985} & 4.877 & -- & -- & -- & -- & -- & -- & -- & -- \\ \hline   
    $\mathrm{YN}$ \cite{Morris1985} & 4.894 & -- & -- & -- & -- & -- & -- & -- & -- \\ \hline
    $\mathrm{YN}$ \cite{Cherchab2008} & 4.84 & -- & -- & -- & -- & -- & -- & -- & -- \\ \hline
    \end{tabular}}
\end{table}

\begin{figure}[htbp]
\centering
  \includegraphics[width=0.5\textwidth]{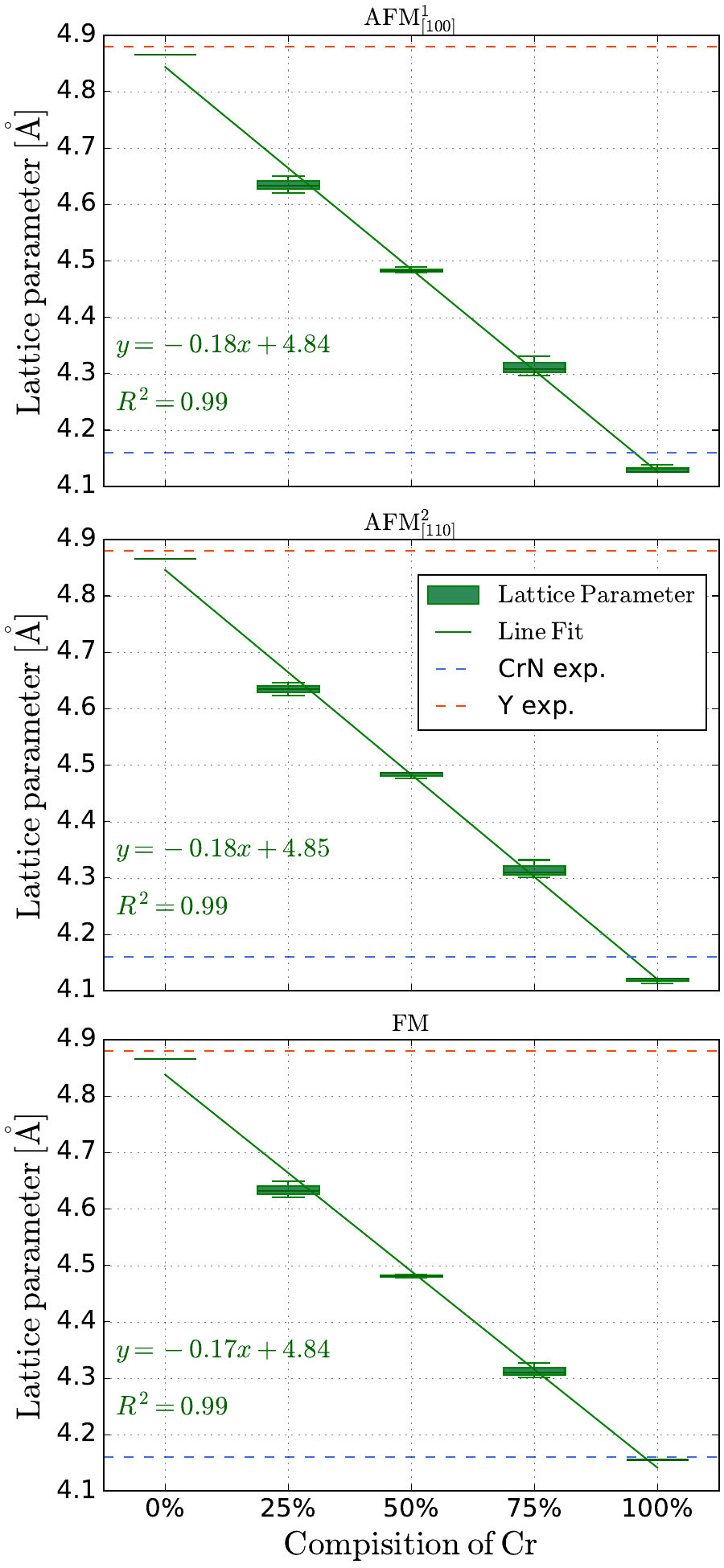}
  \caption{Lattice parameter with respect to the percentage of Cr composition for the different magnetic structures. The box-and-whisker plot represents the variation of the lattice parameter for each structure}
  \label{f:PvsC}
\end{figure}

After having relaxed all structures completely, the results of lattice parameters, average magnetic moments per spin direction, number atoms per spin and magnetization per cell are given in Table \ref{t:param_magmom}. The figure \ref{f:PvsC} plots the relationship between lattice parameter and composition, and divides the figure with respect to the 3 magnetic structures presented in \cite{Corliss1960}. In each plot, the fitting equation is exposed, where the intercept with the y-axis corresponds to the value of the YN lattice parameter and the squared correlation coefficient ($R^{2}$). Also noted is the value of the experimental lattice parameters against which the CrN and YN results are compared (\cite{Corliss1960} and \cite{Morris1985} respectively). In all 3 cases it can be seen that the behavior of the lattice parameter with respect to the composition has a linear decreasing behavior as the concentration of Cr atoms increases, with a high $R^{2}$, indicating a good linear fit. This behavior is similar to that calculated by \cite{Zhou2013} at different concentrations for CrYN. This linear behavior indicates that vegard's law is satisfied \cite{Vegard1921}, i.e. the properties of the alloys can be predicted based on the properties of their constituent elements and their concentrations. This behavior in the lattice parameters is attributed to the fact that the atomic radius of Y is larger than that of Cr, therefore, the more Cr atoms, the smaller the size of the lattice parameter. 

On this same figure it can be seen from the bar and whisker diagrams that the dispersion of the lattice parameters after total relaxations is quite small in all cases, i.e., the distance between the maximum and minimum values that can be adopted in all cases is small (starts to vary from the second decimal) and the interquartile distance is even smaller. This indicates that the structures tend to maintain the cubic structure with deformations small enough to prevent the formation of new bonds and thus undesired phase transitions. In the three magnetic structures of the alloys it can be seen that in the $\mathrm{Cr_{0.5}Y_{0.5}N}$ configurations there is less variation between the three lattice parameters. This feature suggests that a uniform distribution of Cr and Y atoms within the cubic lattice favors the symmetry of the structure, i.e., apparently the lattice distortion is lower, possibly because of a more equal distribution in the strengths of the N-bonds. To verify this, an analysis of the deformation of the octahedra by composition is necessary and will be carried out later. 

For the pure structures, the cubic character of their cells can be better appreciated as the variation in the second decimal is not greater than $10^{-2} \: \mathrm{\mathring{A}}$. This variation can be understood because the way the magnetic moment arrangement is defined affects the way the cell is distorted, since the distribution of forces in the octahedra formed around the metal atoms can change slightly to compensate for the AFM, as is the case for the  $\mathrm{CrN}$-$\mathrm{AFM_{[110]}^{2}}$ structure, which presents a small distortion angle (change from 90$^{\circ}$ to 88.31$^{\circ}$, more than 1$^{\circ}$), as observed in \cite{Corliss1960} and \cite{unal_85410}, and accompanied by a compression of the parameter c, showing an orthorhombic distortion. This deformation is known as Jhan-Teller type deformation \cite{H.A.JahnandE.Teller1937} and is in agreement with the observations made by \cite{Filippetti2000}. In the case of $\mathrm{CrN}$-$\mathrm{AFM_{[100]}^{1}}$ and $\mathrm{CrN}$-$\mathrm{FM}$, the magnetic moments, on the contrary, seek to stabilize the perfect cubic symmetry, tending to stabilize the octahedra within the cell, minimizing the distortion angle. Regarding the $\mathrm{CrN}$-$\mathrm{AFM_{[100]}^{1}}$, a lengthening of the parameter c occurs to stabilize the crystal structure, as mentioned by \cite{Filippetti2000}. This correspond to a special kind of Jhan-Teller type deformation \cite{H.A.JahnandE.Teller1937} named tetragonal distortion. In the $\mathrm{AFM_{[110]}^{2}}$ and $\mathrm{AFM_{[100]}^{1}}$ configurations, the value of the slopes is the same, while the slope of the FM configuration is slightly smaller. This indicates that as the Cr composition increases, the variation in the lattice parameter of the FM configuration is slightly slower, while in the others, this variation is practically the same. In the case of YN, the cubic cell remains perfect without deformation.

\begin{figure}[htbp]
\centering
  \includegraphics[width=.5\textwidth]{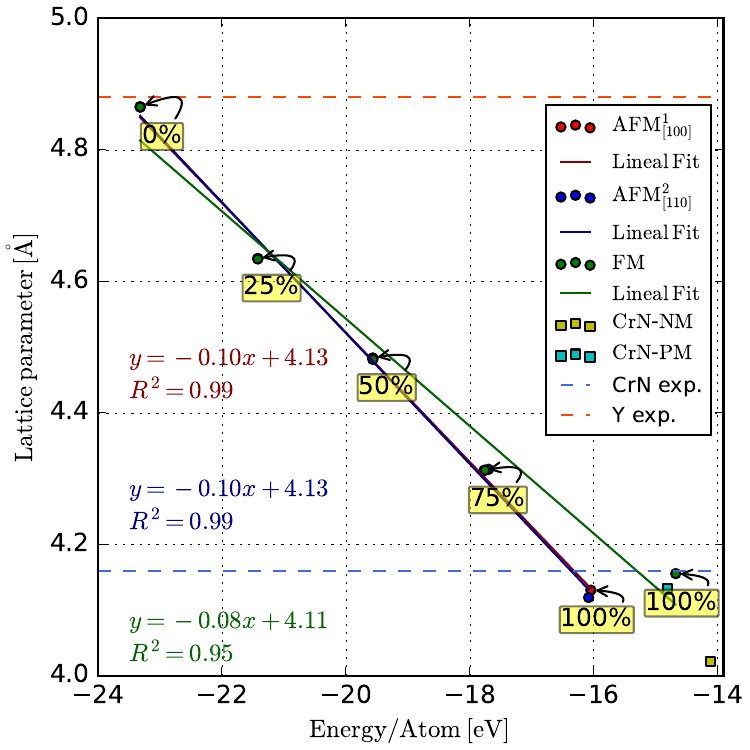}
  \caption{Average lattice parameter with respect to the free energy of each magnetic structure. The yellow box indicates the percentage of Cr composition}
  \label{f:PvsE}
\end{figure}

Figure \ref{f:PvsE} presents the average lattice parameter versus free energy of each magnetic structure and the experimental values of CrN \cite{Corliss1960} and YN \cite{Morris1985} are indicated. The corresponding values for the CrN-NM and CrN-PM cell are also presented. Above are the fitting equations, where the value of the intercept with the y-axis indicates the value of the ideal lattice parameter of each CrN magnetic structure. The behavior of the lattice parameters with respect to free energy shows a linear decreasing trend. When looking at the energy per magnetic structure, it is found that something similar happens as in the plots of lattice parameter versus composition with respect to slope. However, the $R^{2}$ of the fit line of the FM configuration is smaller than that of the other configurations, which indicates that in this case, the linear fit is of lower precision, so it is a little less accurate to make predictions regarding free energy. This variation may be due to the resulting magnetic configurations within the magnetic structures of the alloys. 

When analyzing the results by composition (indicated by the yellow boxes), it can be seen that the energy differences in the three magnetic configurations are minimal in each composition. This result seems to indicate that the resulting magnetic structures are very similar to each other after relaxation. To understand these results in detail, it is necessary to perform an analysis of magnetic moments in the alloy structures and verify whether the initial magnetic configurations were retained (see figures \ref{fig:25CrYN}, \ref{fig:50CrYN} and \ref{fig:75CrYN}). From the energy of the magnetic configurations of pure CrN, it can be clearly seen that the CrN-AFM$^{2}$ structure is the minimum energy structure.  Interestingly, the ideal lattice parameter value marked on the AFM structure fit lines coincides with the experimental value measured for CrN-PM: $a=4.131 \: \mathrm{\mathring{A}}$ \cite{Rivadulla2009}.

\begin{figure}[htbp]
\centering
  \includegraphics[width=.5\textwidth]{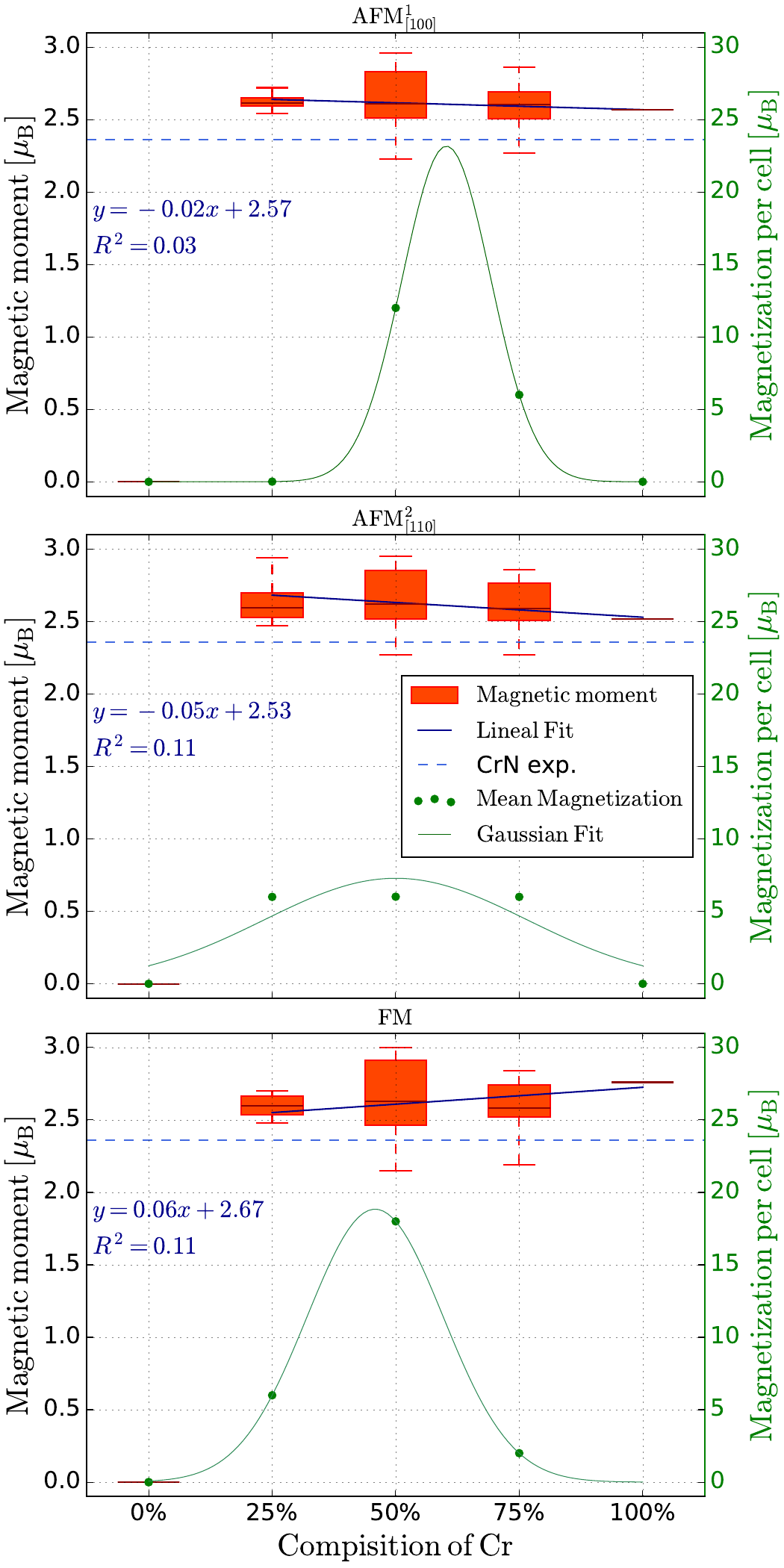}
  \caption{Magnetic moment with respect to the percentage of Cr composition for the different magnetic structures. The box-and-whisker plot represents the variation of the magnetic moment in each structure}\label{f:MvsC}
\end{figure}

Regarding the magnetic moments, in the figure \ref{f:MvsC} it can be seen the magnetic moment results divided into three plots. Each one corresponds to one of the three proposed magnetic configurations. Each plot shows the dispersion of magnetic moments in box and whisker plots, the linear fitting equation (where the cut with the y-axis represents the value of the fitting magnetic moment of pure CrN in its respective magnetic configuration), the value of the experimental magnetic moment measured on CrN in \cite{Corliss1960}, the magnetization values per cell and a Gaussian fitting curve over these values. In all the magnetic configurations, that for the three different SQS, the box and whiskers diagram shows a dispersion of the magnetic moments, i.e., the distribution of the magnitudes of the magnetic moments is not uniform, unlike pure CrN which does show a uniform distribution of magnetic moments.

By observing the fitting equations of the magnetic moments in the three magnetic configurations, it can be inferred that the linear fit is not very good over the distribution of the points, since the values of $R^2$ are almost zero. This means that with respect to the alloys, one cannot expect a linear behavior of the magnetic moments with respect to the composition. The reason for this behavior is due to the fact that in each of the SQS, in each magnetic configuration, for each Cr composition value there is a significant magnetic moment dispersion, which gives signs that the chemical environment in each composition surrounding the Cr atoms has an important effect on the Cr magnetic moment values. The values predicted by the fitted line for the magnetic moment of pure CrN ($2.57$ y $2.53 \: \mathrm{\mu B}$) coincides almost exactly with the values calculated for the $\mathrm{CrN}$-$\mathrm{AFM_{[100]}^{1}}$ and $\mathrm{CrN}$-$\mathrm{AFM_{[110]}^{2}}$ structures ($2.57$ y $2.52 \: \mathrm{\mu B}$). In the case of the $\mathrm{CrN}$-FM structure ($2.67 \: \mathrm{\mu B}$) this prediction differs significantly from the calculated magnetic moment value ($2.76 \: \mathrm{\mu B}$). 

Another important observation is that, in all cases, the mean values in the box and whisker plots are above the experimental value, which seems to indicate an overestimation of the value of the magnetic moments. The reason for this behavior may be due to two things: First, it has already been analyzed in a previous study \cite{Tran2020} that metaGGA functionals, such as SCAN, tend to overestimate these values in magnetic structures. Second, it may be that within the alloy, Y has an effect on the magnetic moments that causes them to rise. This could be due to the dispersion of the magnetic moments within the alloys, and also that in certain cases, some outliers appear that seem to indicate a suppression of the magnetic moment of some Cr atoms (see Table 1). This second assumption can be discarded, since in the study of \cite{Matas2025} it could be observed that in spite of alloying CrN with different metallic atoms, no significant change in the magnetic moment values of CrN is observed. A common feature of the three graphs is that the dispersion of magnetic moments in the 50\% Cr composition is the highest with respect to the other compositions. This coincides with what was seen in the lattice parameters, where at this composition the dispersion of lattice parameters is the lowest. This seems to indicate that although the number of Cr and Y atoms is the same in the cell, the distribution of these is not uniform, which seems to create different chemical environments experienced by the Cr atoms within the cell, which although they balance the distribution of forces within the lattice, the magnitude of the magnetic moments of Cr is affected in different ways, possibly due to the random distribution. Possibly, in regions of low Cr, the atoms are mostly surrounded by Y atoms, and depending on their proximity to other Cr atoms, the level of delocalization of the wave functions can vary depending on the surrounding neighbors. Since the number of Cr atoms is so low or high, the magnetic behavior is expected to be non-uniform, as each Cr atom interacts in markedly different ways with its surroundings. To clarify this situation, an analysis of second neighbors will be made later. 

Speaking now of the magnetization per cell, it can be clearly appreciated that only the $\mathrm{Cr_{0.25}Y_{0.75}N}$-$\mathrm{AFM_{[100]}^{1}}$ structure remained AFM, although this does not mean that it retained its initial magnetic configuration. The other structures, although with different magnitudes, presented a remanent magnetization. Another important annotation is that in the $\mathrm{AFM^{2}}$ and FM magnetic structures of $\mathrm{Cr_{0.75}Y_{0.25}N}$ a charge suppression occurred in some Cr atoms (see Table \ref{t:param_magmom}). From this it can be determined that none of the proposed magnetic structures retained their initial magnetic configuration. The magnetic reconfiguration of the Cr atoms is due precisely to the introduction of Y atoms within the structure, which are responsible for modifying the 'd' orbitals of Cr, even leading to charge suppression of some of these Cr atoms, since the Y also possesses 'd' orbitals which have a low electron occupancy where these wave functions could be delocalized. These effects will be addressed later in the respective band structure analysis. 

With the exception of the $\mathrm{Cr_{0.25}Y_{0.75}N}$-$\mathrm{AFM_{[100]}^{1}}$ structure, it can then be said that the magnetic reconfiguration of the supercells results in a ferrimagnetic (fM) configuration. It cannot be said that they are FM because evidently there are spin up and spin down in opposition, but with sufficient unbalance of magnetic moments for the remanent magnetization to occur. Regarding the $\mathrm{Cr_{0.25}Y_{0.75}N}$-$\mathrm{AFM_{[100]}^{1}}$ structure, since there is no internal magnetic symmetry, it can be said that its resulting magnetic state is PM. 

Focusing now on the Gaussian fits for each magnetic configuration, several things can be appreciated: For the $\mathrm{AFM^{1}}$ structures, the Gaussian curve predicts that from a composition of 25\% Cr up to 100\% Cr a remanent magnetization can be found in the cells, reaching a maximum of magnetization with a concentration close to 60\% Cr content. In the FM structures, a good Gaussian fit is also observed (it should be noted that the magnetization point corresponding to pure CrN was not taken into account because being FM, the magnetization is considerably high with respect to those of the alloys). In this case, the cells of the different alloys will present a remanent magnetization throughout all the compositions. This behavior indicates that structures built with the initial FM configuration will not retain this configuration anywhere in the composition range. Finally, in the structures built with the initial $\mathrm{AFM^{2}}$ magnetic configuration, it can be seen that the Gaussian fit is not good, therefore it is not useful to make predictions.  However, it is clearly seen that in the three alloy structures at different compositions, the magnetization per cell is 6 $\mathrm{\mu B}$ in each of them, i.e., possibly, over the whole compositional range, with this initial magnetic configuration similar magnetization values per cell will be obtained as a result. With these results it could be inferred that the initial magnetic configuration could influence the magnetization per cell remaining after relaxation of the structure.

\begin{figure}[htbp]
\centering
  \includegraphics[width=.5\textwidth]{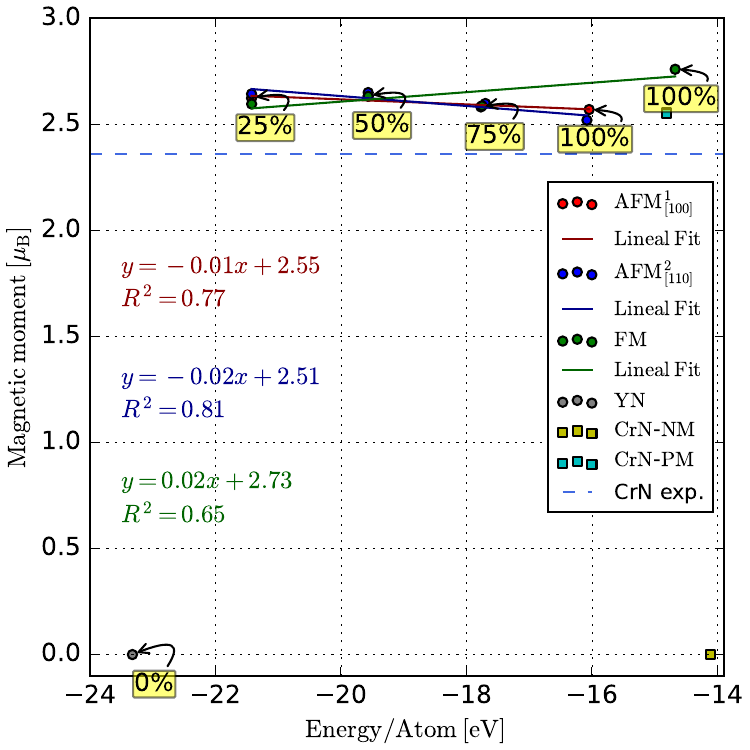}
  \caption{Average magnetic moment with respect to the free energy of each magnetic structure. The yellow box indicates the percentage of Cr composition}\label{f:MvsE}
\end{figure}

Figure \ref{f:MvsE} shows the average magnetic moments per cell versus free energy per cell, the experimental magnetic moment value measured in \cite{Corliss1960} and the linear fitting equations (the corresponding YN, CrN-PM and CrN-NM values are also plotted). In contrast to figure \ref{f:MvsE}, the $R^2$ values show a better linear fit. However, ideally these values should be as close to 1 as possible for the free energy predictions to be more reliable. The value predicted by the fitting curve for pure CrN is improved for the FM structure (2.73) with respect to the calculated one (2.76). The linear behavior of the AFM$^{1}$ and AFM$^{1}$ structures is decreasing, while that of the FM structure is increasing. Knowing the result of the magnetic moments within each cell, it can be determined that the inconsistencies with respect to the structures that started with an initial FM configuration are due to the fact that in the alloys there is no FM co-position, making these structures tend to have similar characteristics to those of the other magnetic configurations. As shown in Figure \ref{f:PvsE}, the CrN-AFM$^{2}$ structure is also the lowest energy one, coinciding with the neutron diffraction measurements made by \cite{Corliss1960}.

Analyzing the case of CrN-PM, it can be seen that the average magnetic moment value is similar to those obtained in AFM$^{1}$ and AFM$^{2}$ structures. In addition, despite having a higher free energy than the latter mentioned, it is lower than that of the NM structure. This coincides with the study of \cite{Matas2025}, where they propose that to study the PM case of CrN, it is a better approximation to use the SQS than a hypothetical non-magnetic structure. This statement is further verified in the study of band structures.

\subsection{Phase stability}

\begin{table}[htbp]
    \centering
    \begin{tabular}{|c|c|c|c|} \hline 
         Estructure&  $\mathrm{E \: (eV)}$&  $\mathrm{H_{f} \: (eV)}$&  $\mathrm{H_{mix} \: (eV)}$\\ \hline \hline
         $\mathrm{YN}$& -23.3206 &  -1.9318&  0.0\\ \hline
         $\mathrm{CrN}$-$\mathrm{NM}$&-14.9676 & 1.7912 &  0.0\\ \hline
         $\mathrm{CrN}$-$\mathrm{PM}$& -16.0396& 0.7192 &0.0\\ \hline
         $\mathrm{Y}$-$\mathrm{HCP}$& -27.8235& - &-\\ \hline
         $\mathrm{Cr}$-$\mathrm{BCC}$-$\mathrm{AFM}$& -18.5637& - &-\\ \hline
         $\mathrm{N_2}$& -14.9540& - &-\\ \hline
         \multicolumn{4}{|l|}{$\mathrm{AFM^{1}_{[100]}}$} \\ \hline
         $\mathrm{Cr_{0.25}Y_{0.75}N }$&-21.4224  &  -1.1911&  0.0787\\ \hline 
         $\mathrm{Cr_{0.50}Y_{0.50}N}$& -19.5633 & -0.4895 &  0.1184\\ \hline 
         $\mathrm{Cr_{0.75}Y_{0.25}N}$&-17.7680 &  0.1483&  0.0942\\ \hline 
         $\mathrm{CrN }$&  -16.0428&  0.7160&  0.0\\ \hline 
         \multicolumn{4}{|l|}{$\mathrm{AFM^{2}_{[110]}}$} \\ \hline
         $\mathrm{Cr_{0.25}Y_{0.75}N}$&  -21.4219&  -1.1906&  0.0881\\ \hline 
         $\mathrm{Cr_{0.50}Y_{0.50}N}$&  -19.5641&-0.4903 &  0.1354\\ \hline 
         $\mathrm{Cr_{0.75}Y_{0.25}N}$&  -17.6958&  0.2205&  0.1932\\ \hline 
         $\mathrm{CrN}$&  -16.0784&  0.6804&  0.0\\ \hline 
         \multicolumn{4}{|l|}{$\mathrm{FM}$} \\ \hline
         $\mathrm{Cr_{0.25}Y_{0.75}N}$&  -21.4209& -1.1896& -0.2626\\ \hline 
         $\mathrm{Cr_{0.50}Y_{50}N}$&  -19.5633&  -0.4895& -0.5672\\ \hline 
         $\mathrm{Cr_{0.75}Y_{0.25}N}$&  -17.7506&  0.1657& -0.9167\\ \hline 
         $\mathrm{CrN}$& -14.6718&  2.0870&  0.0\\ \hline 
         \multicolumn{4}{|l|}{$\mathrm{Theoric}$}\\ \hline
         $\mathrm{Cr_{0.25}Y_{0.75}N}$-$\mathrm{PM}$&  -& -& 0.125\\ \hline 
         $\mathrm{Cr_{0.50}Y_{50}N}$-$\mathrm{PM}$&  -&  -& 0.15\\ \hline 
         $\mathrm{Cr_{0.75}Y_{0.25}N}$-$\mathrm{PM}$&  -&  &0.13\\ \hline 
         $\mathrm{YN}$ \cite{Holec2012}&  -& -2.737 &-\\ \hline    
    \end{tabular}
    \caption{Energy per atom ($\mathrm{E}$), formation enthalpy ($\mathrm{H_f}$) and mixing enthalpy ($\mathrm{H_{mix}}$) of the different magnetic structures}
    \label{t:phase_stab}
\end{table}

To analyze the stability of the magnetic phases in the three compositions studied, the SQS method used by several theoretical studies will be followed \cite{Alling2007,Zaukauskaite2021}. All results of enthalpy of formation and mixing enthalpy are presented in the table \ref{t:phase_stab}. It was first necessary to calculate the enthalpies of formation of the three magnetic phases proposed for CrYN, CrN and YN from the equations \ref{ec:form}. In this equation, the subscripts indicate that the energies of these compounds are calculated at different phases. To do this, the energy of the nitrides in their most stable phase and also that of the different individual atoms in their most stable phases must be calculated. In the case of CrYN, YN,$\mathrm{CrN}$-$\mathrm{AFM^{1}}$ and $\mathrm{CrN}$-FM, the most stable phase is cubic FCC. That of $\mathrm{CrN}$-$\mathrm{AFM^{2}}$ is orthorhombic, but it will be done with the FCC phase to maintain uniformity with the other magnetic structures, since it still gives lower energy anyway. That of the Cr atom is BCC and that of the Y atom is HCP. Similarly, for the N atom a diatomic molecule must be constructed and must be confined in a 10x10x10 A box.  The results of these enthalpies are given in Table \ref{t:phase_stab}. 

\begin{align}
\centering
\mathrm{H}_{f\mathrm{CrN}}&= \mathrm{E}_{\mathrm{CrN\text -FCC}}-[0.5\mathrm{E}_{\mathrm{Cr\text -BCC}}+0.5\mathrm{E}_{\mathrm{N_2}}] \\
\mathrm{H}_{f\mathrm{YN}}&= \mathrm{E}_{\mathrm{YN\text -FCC}}-[0.5\mathrm{E}_{\mathrm{Y\text -HCP}}+0.5\mathrm{E}_{\mathrm{N_2}}]\\
\mathrm{H}_{f\mathrm{Cr_{1-x}Y_{x}N}}&= \mathrm{E}_{\mathrm{Cr_{1-x}Y_{x}N\text -FCC}}\\&-0.5[x\mathrm{E}_{\mathrm{Cr\text -BCC}}+(1-x)\mathrm{E}_{\mathrm{Y\text -HCP}}+\mathrm{E}_{\mathrm{N_2}}]
\label{ec:form}
\end{align}

From the results for formation energy, it should be clarified that negative values mean that the compound is more stable than its individual components, while positive values mean that the individual components are more stable than the compound, and to hold together they require additional energy. It can be seen then that YN is the compound with the lowest formation energy, and together with $\mathrm{Cr_{0.25}Y_{0.75}N}$ and $\mathrm{Cr_{0.5}Y_{0.5}N}$ have negative formation energies. On the other hand, CrN has the highest formation energy and together with $\mathrm{Cr_{0.75}Y_{0.25}N}$ have positive values. Now, when looking at the formation energies by magnetic structure, no difference is seen with respect to the different compositions, except for pure CrN.  As the Cr composition increases, the formation energy begins to increase. This increase has a linear trend for the $\mathrm{AFM^{1}}$ and $\mathrm{AFM^{2}}$ magnetic structures, and a quadratic trend for the FM. However, with respect to the latter it can be seen that without taking into account the CrN value, its trend, like the other magnetic structures, appears to be linear. As mentioned above, since the SQS of the alloys do not retain the FM configuration, they tend to behave the same as those starting with AFM configuration. Reviewing the results by magnetic structure, it can be seen that in all the alloys the formation energy is practically the same. However, in pure CrN there are notable differences and it can be clearly determined that the $\mathrm{AFM^{2}}$ structure has the lowest formation energy.

\begin{figure}[htbp]
\centering
  \includegraphics[width=.5\textwidth]{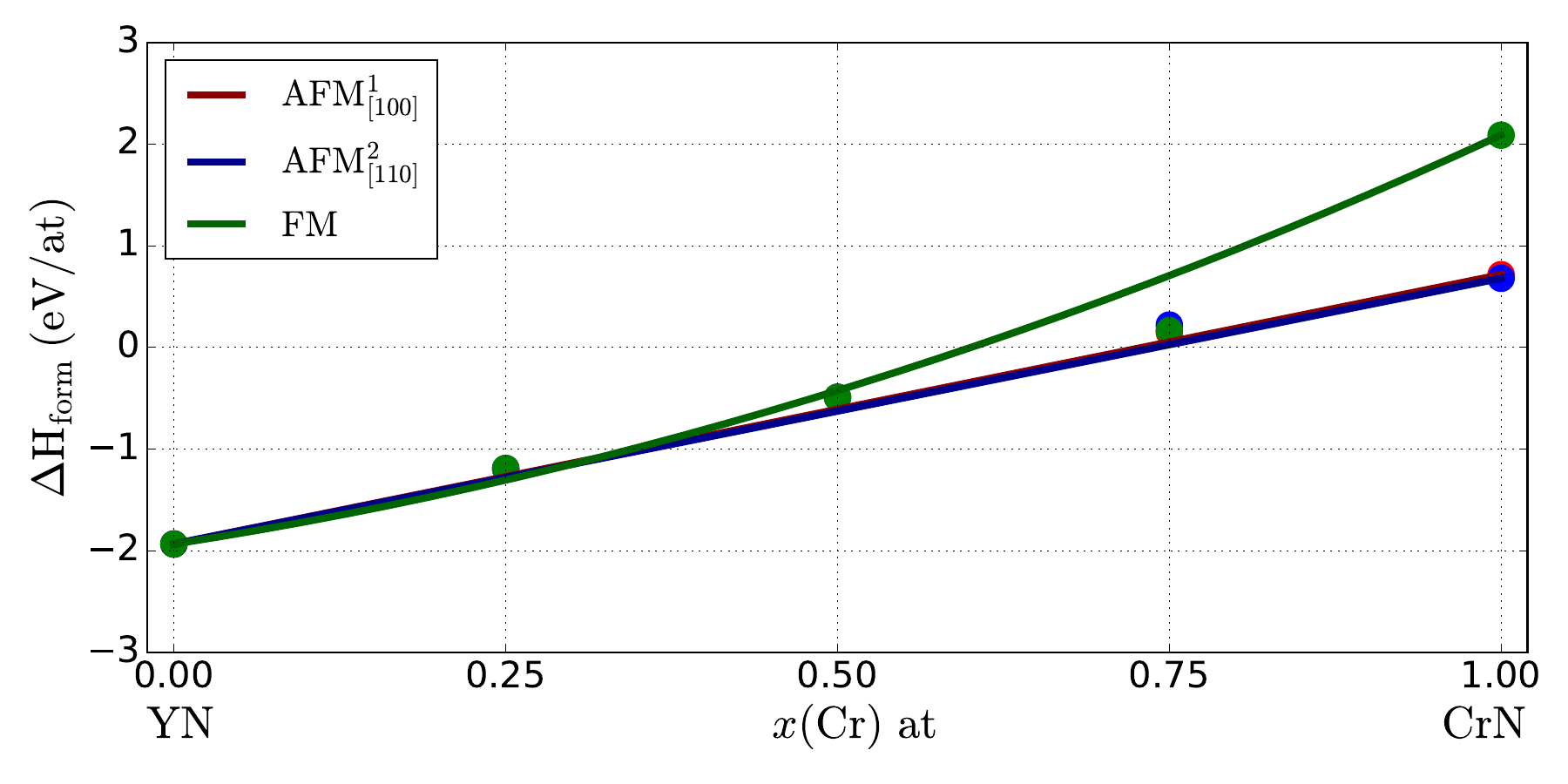}
  \caption{Formation enthalpy}\label{f:Hform}
\end{figure}

Then, the enthalpy of mixing of the solid substitutional solutions represented in the SQS must be calculated. The mixing enthalpy was calculated from equation \ref{ec:mix}. To find it, the energy values of the substitutional solid solution and of the individual structures of the nitrides that make up the alloy must be known. In this equation, the subscript indicates that the three energy values of the alloy and the nitrides must be obtained from the same phase, in this case, the cubic phase FCC.

\begin{align}
\centering
\mathrm{H}_{mix}&=\mathrm{E}_{\mathrm{Cr}_{1-x}\mathrm{Y}_x\mathrm{N\text -FCC}}-x\mathrm{E}_{\mathrm{YN\text -FCC}}-(1-x)\mathrm{E}_{\mathrm{CrN\text -FCC}}
\label{ec:mix}
\end{align}

\begin{figure}[htbp]
\centering
  \includegraphics[width=.5\textwidth]{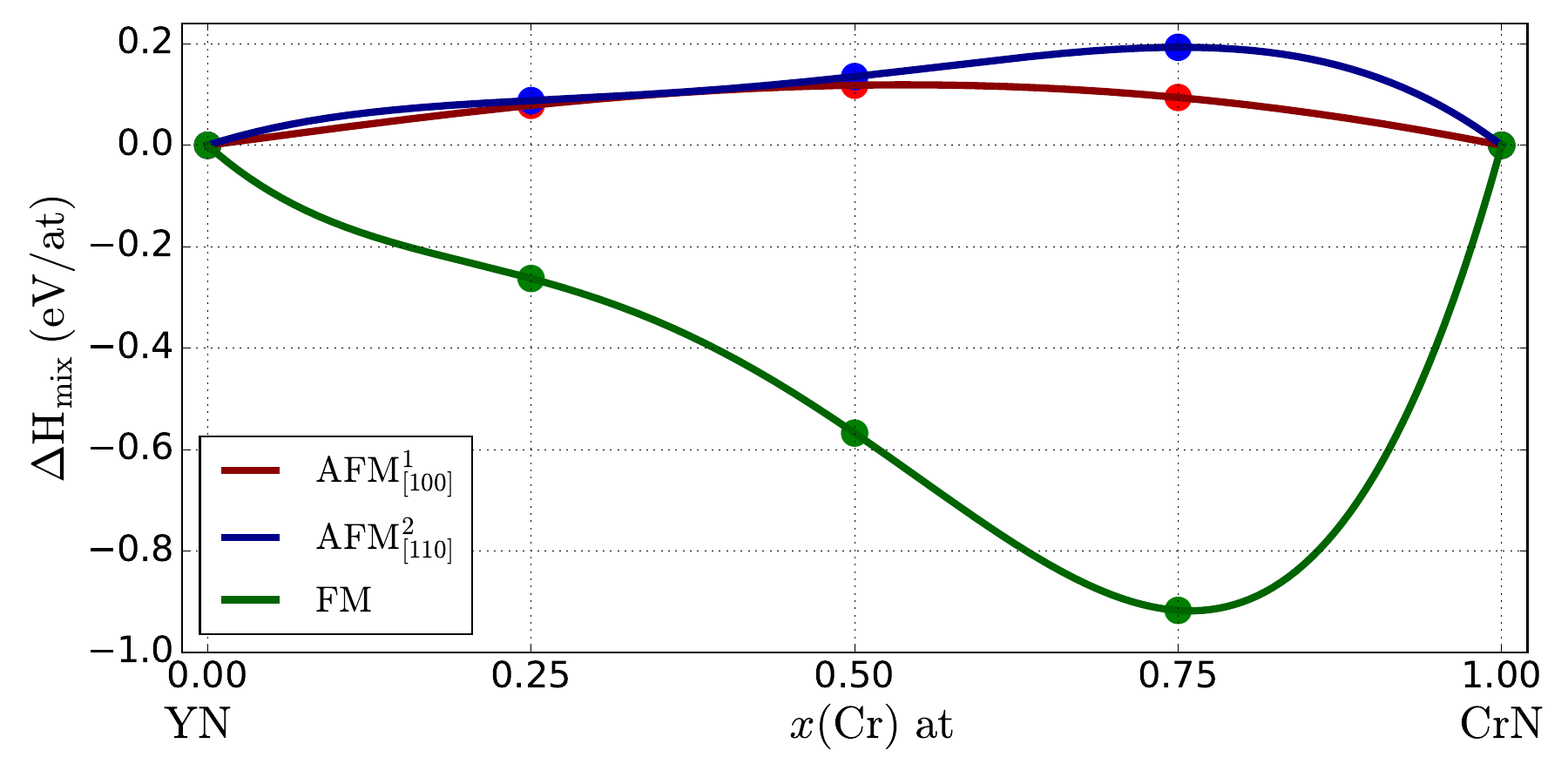}
  \caption{Mixing enthalpy}\label{f:Hmix}
\end{figure}

Figure \ref{f:Hmix} shows the mixing enthalpy as a function of composition. In this case, the free energy is equivalent to the mixing enthalpy because the calculations are made at 0 K, which allows fitting the values obtained by magnetic structure with the Redlich-Kister equation \cite{Redlich1948}, which is useful in this case in the absence of explicit experimental precise information and which has been shown in several studies to be a good fit with respect to experiments for substitutional solid solutions \cite{Ansara1998,Ramesh2014}. However, there are some theoretical studies on CrYN in PM state that behave similarly to the results obtained here for $\mathrm{AFM^{1}}$ and $\mathrm{AFM^{2}}$. As in \cite{Rovere2010} and \cite{Zhou2013}, the mixing enthalpy shows positive values for AFM configurations, indicating that there is a tendency for the alloys to separate into their binary compounds, which indicates that to form this mixture it is necessary to add energy, since under normal reaction conditions it does not seem to be possible for the solid solution to be generated. 

For the two AFM structures with different Cr compositions, it is evident that those that started with an initial $\mathrm{AFM_{[110]}^{2}}$ configuration have lower mixing energy, indicating that alloys starting in this way tend to be more stable. However, those starting with an FM configuration apparently have negative mixing energies, which means that the mixing of the components is energetically favorable. In this case, the mixing process would release energy, suggesting that the components would spontaneously combine and form a stable solid solution. But if the results of enthalpies of formation are integrated, it is clear that the $\mathrm{CrN}$-$\mathrm{AFM^{2}}$ structure has the lowest formation energy, which in real conditions means that it is the easiest to obtain in a laboratory. Therefore, YN and $\mathrm{CrN}$-$\mathrm{AFM^{2}}$ compounds seem to be easier to mix than YN and CrN-FM. In addition, the enthalpy of mixing values of $\mathrm{CrN}$-$\mathrm{AFM^{2}}$ are close to zero. This reflects an energy balance where the mixture is fairly neutral. It could imply that the components mix easily, but do not have a very pronounced stability or a strong tendency to separate. This suggests that this alloy could be stable under certain conditions, but is still less favorable overall compared to CrYN-FM.

\subsection{Analysis of octahedrons}

\begin{figure}[htbp]
\centering 
\begin{subfigure}[t]{0.38\linewidth}
    \includegraphics[width=\linewidth]{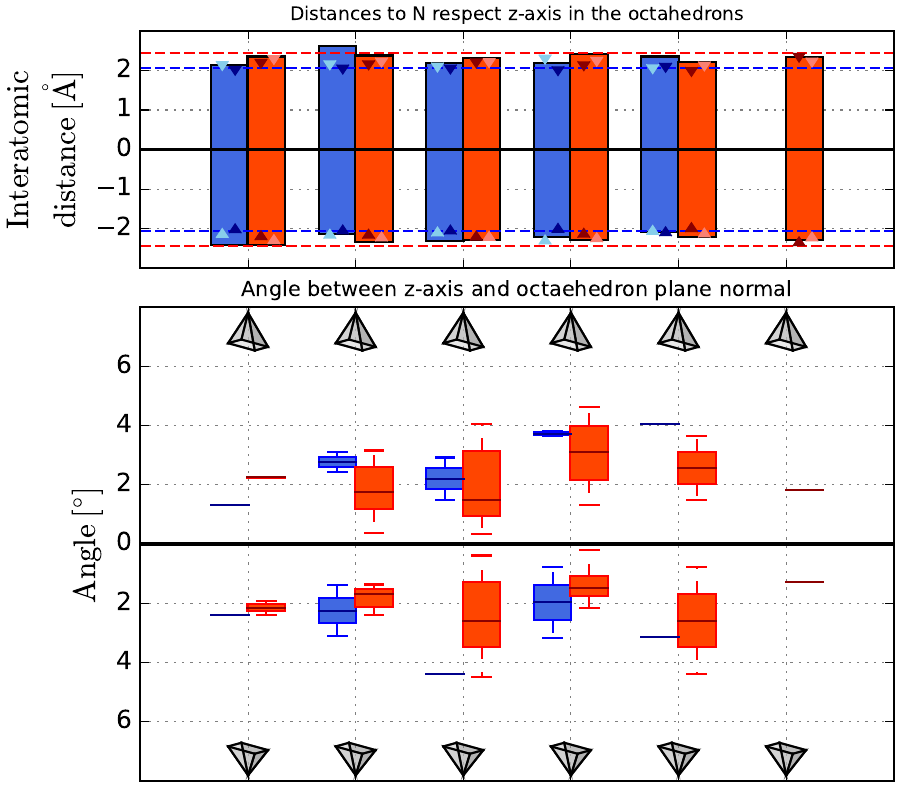}
    \caption{$\mathrm{Cr_{0.25}Y_{0.75}N}$-$\mathrm{AFM_{[100]}^{1}}$}
    \label{fig:sv25AFM1}
\end{subfigure}
\begin{subfigure}[t]{0.26\linewidth}
    \includegraphics[width=\linewidth]{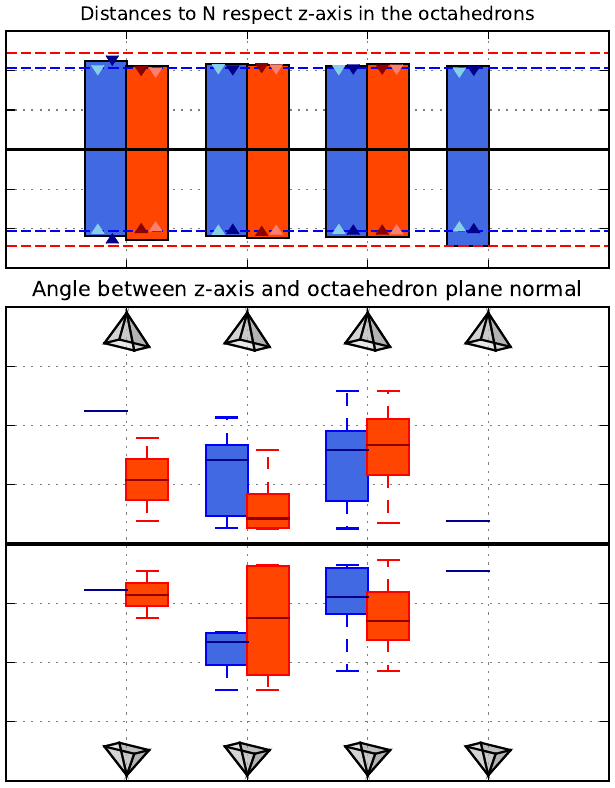}
    \caption{$\mathrm{Cr_{0.5}Y_{0.5}N}$-$\mathrm{AFM_{[100]}^{1}}$}
    \label{fig:sv50AFM1}
\end{subfigure}
\begin{subfigure}[t]{0.33\linewidth}
    \includegraphics[width=\linewidth]{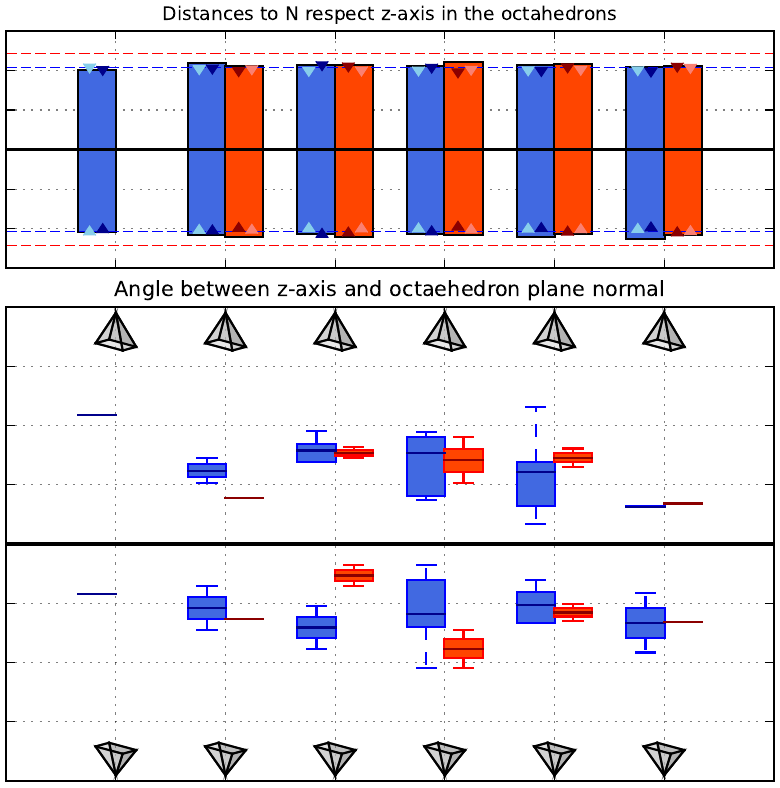}
    \caption{$\mathrm{Cr_{0.75}Y_{0.25}N}$-$\mathrm{AFM_{[100]}^{1}}$}
    \label{fig:sv75AFM1}
\end{subfigure}
\begin{subfigure}[t]{0.38\linewidth}
    \includegraphics[width=\linewidth]{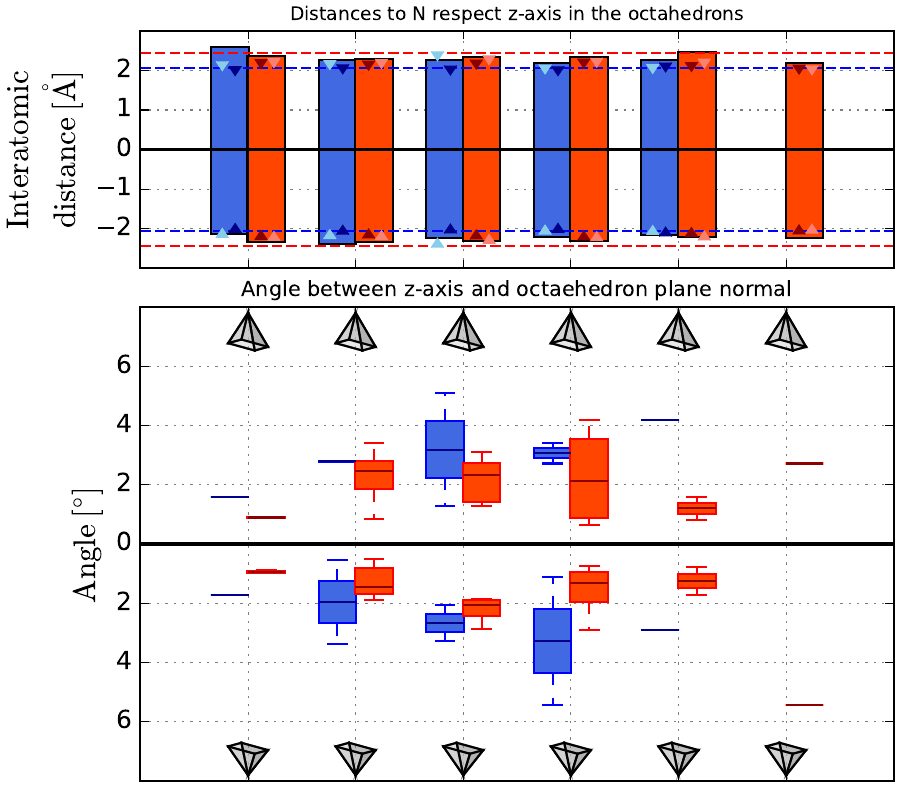}
    \caption{$\mathrm{Cr_{0.25}Y_{0.75}N}$-$\mathrm{AFM_{[110]}^{2}}$}
    \label{fig:sv25AFM2}
\end{subfigure}
\begin{subfigure}[t]{0.26\linewidth}
    \includegraphics[width=\linewidth]{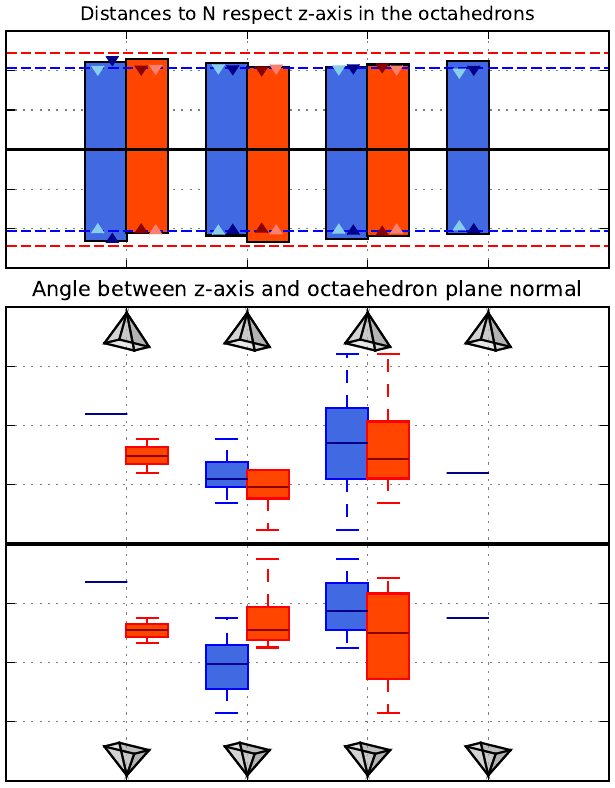}
    \caption{$\mathrm{Cr_{0.5}Y_{0.5}N}$-$\mathrm{AFM_{[110]}^{2}}$}
    \label{fig:sv50AFM2}
\end{subfigure}
\begin{subfigure}[t]{0.33\linewidth}
    \includegraphics[width=\linewidth]{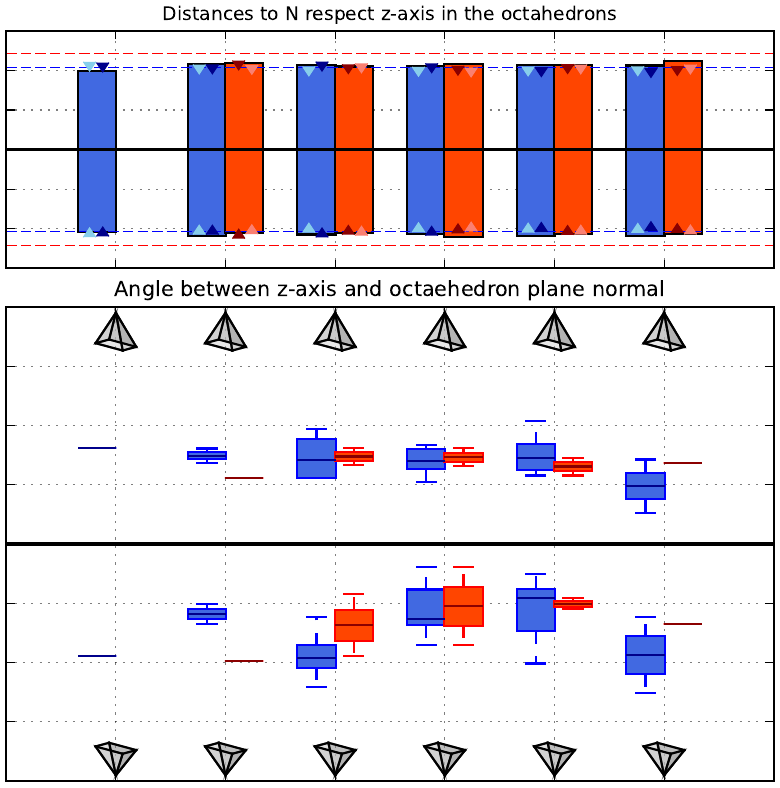}
    \caption{$\mathrm{Cr_{0.75}Y_{0.25}N}$-$\mathrm{AFM_{[110]}^{2}}$}
    \label{fig:sv75AFM2}
\end{subfigure} 
\begin{subfigure}[t]{0.38\linewidth}
    \includegraphics[width=\linewidth]{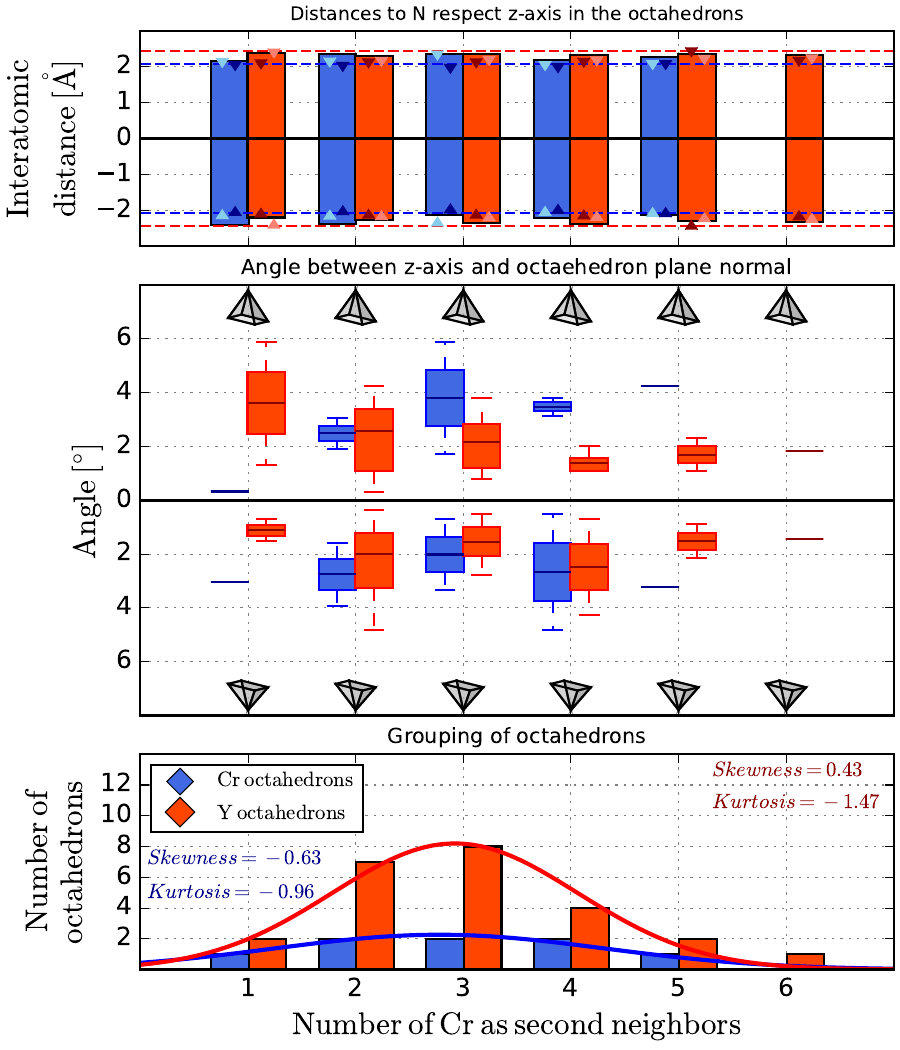}
    \caption{$\mathrm{Cr_{0.25}Y_{0.75}N}$-$\mathrm{FM}$}
    \label{fig:sv25FM}
\end{subfigure}
\begin{subfigure}[t]{0.26\linewidth}
    \includegraphics[width=\linewidth]{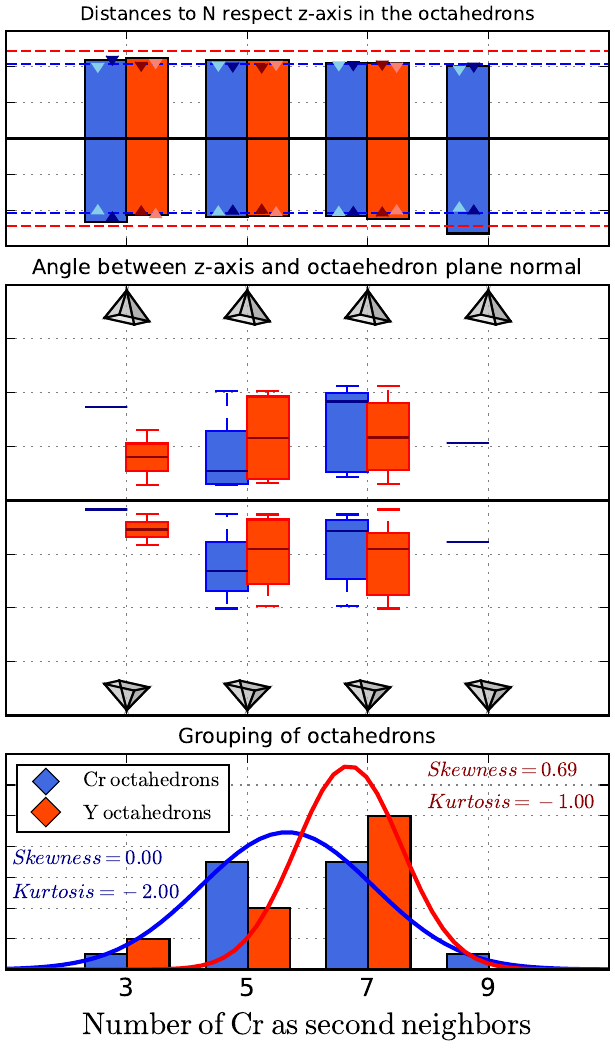}
    \caption{$\mathrm{Cr_{0.5}Y_{0.5}N}$-$\mathrm{FM}$}
    \label{fig:sv50FM}
\end{subfigure}
\begin{subfigure}[t]{0.33\linewidth}
    \includegraphics[width=\linewidth]{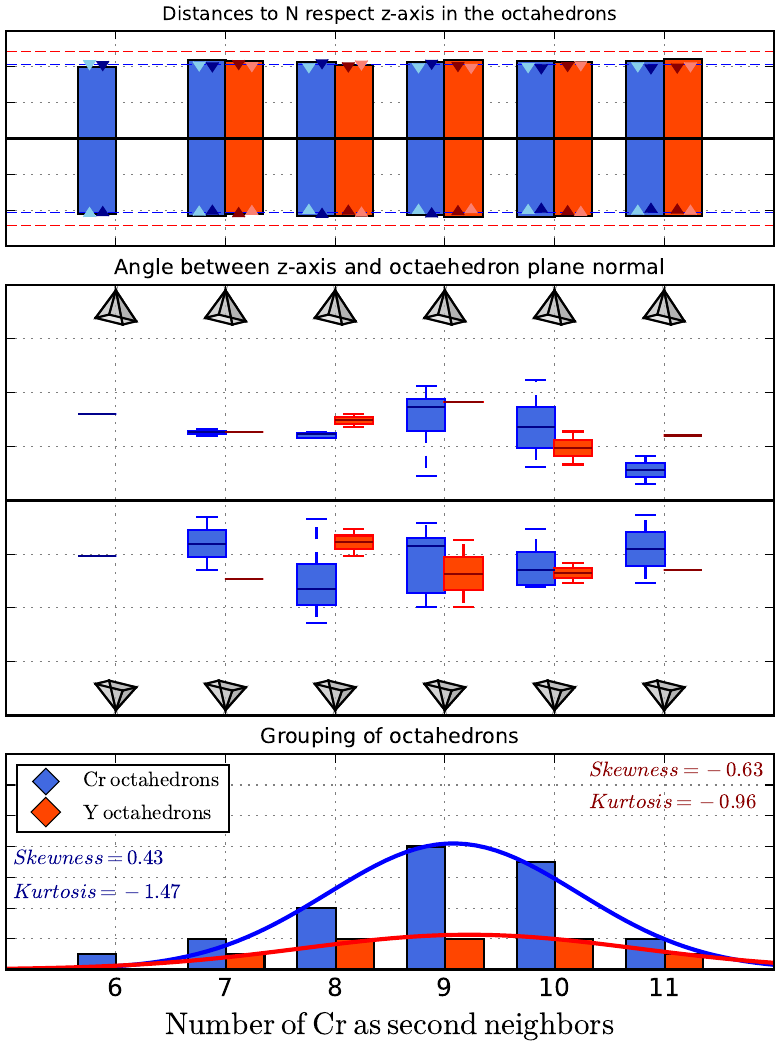}
    \caption{$\mathrm{Cr_{0.75}Y_{0.25}N}$-$\mathrm{FM}$}
    \label{fig:sv75FM}
\end{subfigure}        
\caption{Bond distances, distortion angle and number of octahedra with respect to second neighbors. The octahedra are classified according to the number of Cr atoms surrounding the octahedron as second neighbors. This quantity is represented on the horizontal axis. The results presented with blue color represent the octahedra surrounding Cr atoms and with red color the octahedra surrounding Y atoms.}
\label{fig:sv}
\end{figure}

In this section a detailed analysis of the chemical environment of the transition metals is made, with the objective of seeing if there are any influences from the second neighbors around the octahedra that can explain the magnetic behavior of the different alloys. To this end, three types of plots can be found in Figure \ref{fig:sv} for each of the magnetic structures discussed. In all of them, the octahedra are classified according to the number of Cr atoms surrounding the octahedron as second neighbors. This quantity is represented on the horizontal axis. The results presented with blue color represent the octahedra surrounding Cr atoms and with red color the octahedra surrounding Y atoms.

The first is a bar diagram representing the bond distance of a transition metal to its first N neighbors. The bars represent the bonds in the z-axis direction and the 2 triangles above the bars represent the bonds in the x-axis and y-axis directions respectively. It is necessary to clarify that these values are an average value of the octahedra that are surrounded by a certain number of Cr atoms as second neighbors (horizontal axis). The dashed lines represent the same bond distances in the CrN and YN. Since in pure compounds the bond distances are the same in all directions, it is only necessary to put one. These lines serve as a reference to observe how the octahedra in the alloys are deformed with respect to the pure compounds. 

The second type of graph is a box-and-whisker plot showing the deformation angles of the two links in the z-axis of the octahedron with respect to the ideal direction of the z-axis (one in the +z direction and the other in the -z direction). The pyramids at the upper and lower ends of the graph schematize the direction of the bond within the octahedron.  

The third type is another bar chart that classifies octahedra according to the number of Cr atoms they have as second neighbors. In addition, Gaussian fitting curves are presented to show the distribution of second neighbors and their respective equations indicating the mean (mu), standard deviation and maximum amplitude. These graphs only talk about the composition, so it is not necessary to make one for each magnetic structure. 

In the first instance, we will review the third type of graph, which is found at the bottom of figures \ref{fig:sv25FM}, \ref{fig:sv50FM} and \ref{fig:sv75FM}. From here we can see that the three compositions try to follow a normal distribution, represented by the Gaussian curve that passes over the bars of the graph. A Gaussian distribution is typical of random processes. Here, the probability of finding a particular number of Cr atoms as second neighbors should follow a symmetric distribution around a mean value. This would implies that there is no systematic or preferential pattern in the substitution of atoms within the crystal lattice, but that substitution occurs randomly. 

By observing each composition, various diagnoses can be made regarding the distribution of atoms within the alloys: It is clear from Figures \ref{fig:sv25FM} and \ref{fig:sv75FM} that the second neighbor distributions are inverted with respect to the Cr and Y octahedra. In Figure \ref{fig:sv25FM}, the second neighbor distribution for the 25\% Cr concentration shows a negative skewness with respect to the Cr octahedra, suggesting a lower probability of finding Cr atoms as second neighbors. The negative kurtosis indicates that the distribution has less heavy tails than a normal distribution, suggesting less variability in the extreme values. The high negative skewness and low amplitude in the Cr octahedra suggest a more localized distribution of Cr atoms. For Y octahedra, the distribution is skewed to the right, suggesting a higher probability of finding more Cr atoms as second neighbors. The negative kurtosis indicates that the distribution of Cr second neighbors is rather flat compared to a normal distribution. A higher concentration of values is seen around the mean and less at the extremes. The skewness with respect to the Y octahedra is less than 0.5, which can be considered as a homogeneous distribution. In Figure \ref{fig:sv75FM}, for a concentration of 75\% Cr, the observations are identical to those in Figure \ref{fig:sv25FM}, only now the number of Y and Cr octahedra are reversed, and thus there is a larger number of Cr atoms as second neighbors. In this case, the Y octahedra have a more localized distribution and that of the Cr octahedra can be considered homogeneous. For 50\% Cr, the distribution is symmetric around the Cr octahedra, indicating an equal distribution of Cr atoms as second neighbors. The very negative kurtosis suggests a very flat distribution with low peaks and light tails, reflecting a lower concentration of values at the extremes. For 50\% Cr, the distribution is symmetric around the Cr octahedra, indicating an equal distribution of Cr atoms as second neighbors. The very negative kurtosis suggests a very flat distribution with low peaks and light tails, reflecting a lower concentration of values at the extremes. The symmetrical distribution in the Cr octahedra suggests a more homogeneous alloy, while the positive skewness and kurtosis in the Y octahedra indicate greater localization.

On the other hand, reviewing the second type of graph where the angles between z-axis and octaehedron plane normal are plotted, it can be seen in all the compositions and in all the magnetic structures that all the octahedra broke with their ideal symmetry. In various octahedra, the angle in the +z-direction is different from the angle in the -z-direction. Moreover, they seem to be opposite, i.e. there are different signs, either in the 'x' or in the 'y' coordinate, if the metal atom of each octahedron is put as the origin. This distortion in the angles in different ways is typical of the Jahn-Teller deformation. This distortion allows the system to minimize the total electronic energy by lengthening and shortening certain bonds in an unequal manner. Furthermore, it is a common distortion in compounds containing transition ions (electrons in the 3d orbital), such as those found in the CrYN alloy \cite{H.A.JahnandE.Teller1937}. To confirm the above, it is necessary to observe the bond distances and check if some of these deformations are of a particular Jahn-Teller type where one of the octahedron axes is elongated or compressed, while the other two axes remain the same, resulting in a deformation that preserves tetragonal symmetry, called tetragonal distortion. To make an in-depth analysis of the electronic properties, as is done below, would improve the comprehension of this results.

Finally, the first type of graph is reviewed. It can be seen that the interatomic distances between the transition metals and the ligands (N atoms) change according to the composition, but according to the magnetic structure no significant differences are seen. Starting with the 25\% Cr alloys, it seems that the Y octahedra tend to maintain their bond distances in the three directions with respect to pure YN homogeneously, although with some exceptions where they seem to compact a little, approaching the bond distances of pure CrN, coinciding with the tails of the distribution curve; while in the CrN octahedra, most of these bonds seem to lengthen towards YN values, especially in the z direction, although many times it is more towards one direction of the z axis than the other. In this case, a Jahn-Teller type deformation seems to predominate, since some bonds in the Cr octahedra are elongated, while in the Y octahedra they are maintained or compacted. This deformation may also present some few tetragonal components, especially if the deformations are observed predominantly in the z-axis.

At 50\% Cr, it appears that around the center of the gaussian, the bond distances in the Cr octahedra remain close to those of pure CrN. At the extremes of the Gaussian it appears that, on the contrary, the Cr octahedron bonds are elongated towards YN values. These deformations in some cases only occur in the z-axis, while they remain with the same length in the x and y directions, while in a few other cases, these deformations in the three directions are different. The Y octahedra appear to compact also toward CrN values to the right of the Gaussian, but to the left, some bond distances resemble that of the YN, either in the +z or -z direction but not in both. A more complex mixture is observed here. There may be a combination of tetragonal and Jahn-Teller deformation. The compaction of the Y octahedra suggests tetragonal deformation, while the elongation of Cr bonds towards the YN values at the ends suggests Jahn-Teller type deformation components.

For 75\% Cr, a similar case occurs as at 25\% Cr, but in reverse, i.e. the bonds of the Y octahedra are compacted towards CrN values and with the particularity that the bonds in the x- and y-directions of the Y octahedra are more uniformly compacted than in z-directions. With the exception of a couple of bonds, there are no Cr octahedra that elongate to YN values in large proportions. Tetragonal deformation predominates, since most Cr octahedra keep their bond distances close to those of pure CrN. Exceptions with elongated bonds may exhibit Jahn-Teller type components, but in general, tetragonal deformation seems to be dominant.

\subsection{Electronic properties}

\begin{table}[htbp]
    \centering
    \adjustbox{max width=\textwidth}{
    \begin{tabular}{|c|c|c|c|} \hline 
         Structure & 
         \shortstack[c]{$\mathrm{Bg \uparrow}$ \\ $\mathrm{(meV)}$} &
         \shortstack[c]{$\mathrm{Bg \downarrow}$ \\ $\mathrm{(meV)}$} &
         \shortstack[c]{$\mathrm{Bg}$ \\ $\mathrm{(meV)}$} \\ \hline\hline 
         $\mathrm{YN}$&  -&  -&  -\\ \hline
         $\mathrm{CrN}$-$\mathrm{NM}$&  -&  -&  -\\ \hline
         $\mathrm{CrN}$-$\mathrm{PM}$&  -& -&-\\ \hline
         \multicolumn{4}{|l|}{$\mathrm{AFM^{1}_{[100]}}$} \\ \hline
         $\mathrm{Cr_{0.25}Y_{0.75}N}$&  2.9&  273.9&  2.9\\ \hline 
         $\mathrm{Cr_{0.50}Y_{0.50}N}$&  0.2*&  6.7& 0.2\\ \hline 
         $\mathrm{Cr_{0.75}Y_{0.25}N}$&  1.1&  1.4&  0.2\\ \hline 
         $\mathrm{CrN}$&  2.9&  2.9&  2.9\\ \hline 
         \multicolumn{4}{|l|}{$\mathrm{AFM^{2}_{[110]}}$} \\ \hline
         $\mathrm{Cr_{0.25}Y_{0.75}N }$&  1.9&  14.2& 1.9\\ \hline 
         $\mathrm{Cr_{0.50}Y_{0.50}N }$&  1.4& 0.1&  0.1\\ \hline 
         $\mathrm{Cr_{0.75}Y_{0.25}N }$& 0.8& 0.2& 0.1\\ \hline 
         $\mathrm{CrN}$&  2.7&  2.7&  2.7\\ \hline 
         \multicolumn{4}{|l|}{$\mathrm{FM}$} \\ \hline
         $\mathrm{Cr_{0.25}Y_{0.75}N}$&  0.3& 261.7&  0.3\\ \hline 
         $\mathrm{Cr_{0.50}Y_{0.50}N}$&  6.7& 0.8& 0.8\\ \hline 
         $\mathrm{Cr_{0.75}Y_{0.25}N}$&  1.5& 0.2&  0.2\\ \hline 
         $\mathrm{CrN }$&  1.7&  19.1&  1.7\\ \hline
         \multicolumn{4}{|l|}{$\mathrm{Theoric}$}\\ \hline
         $\mathrm{YN}$ \cite{Stampfl2001}&  -&  -&  850\\ \hline
         $\mathrm{YN}$ \cite{Cherchab2008}&  -&  -&  220\\ \hline
         $\mathrm{YN}$ \cite{Louhadj2009}&  -&  -&  1070\\ \hline
         $\mathrm{YN}$ \cite{Amrani2007}&  -&  -&  300\\ \hline
         $\mathrm{YN}$ \cite{Salguero2003}&  -&  -&  0\\ \hline
         \multicolumn{4}{|l|}{$\mathrm{Experimental}$}\\ \hline 
         $\mathrm{CrN}$-$\mathrm{PM}$ \cite{Gall2002}&  -&  -&  700\\ \hline
    \end{tabular}}
    \caption{Energy gap ($\mathrm{B_g}$) per spin channel of different magnetic structures}
    \label{t:brecha}
\end{table}

\begin{figure}[htbp]
\centering
  \includegraphics[width=.5\textwidth]{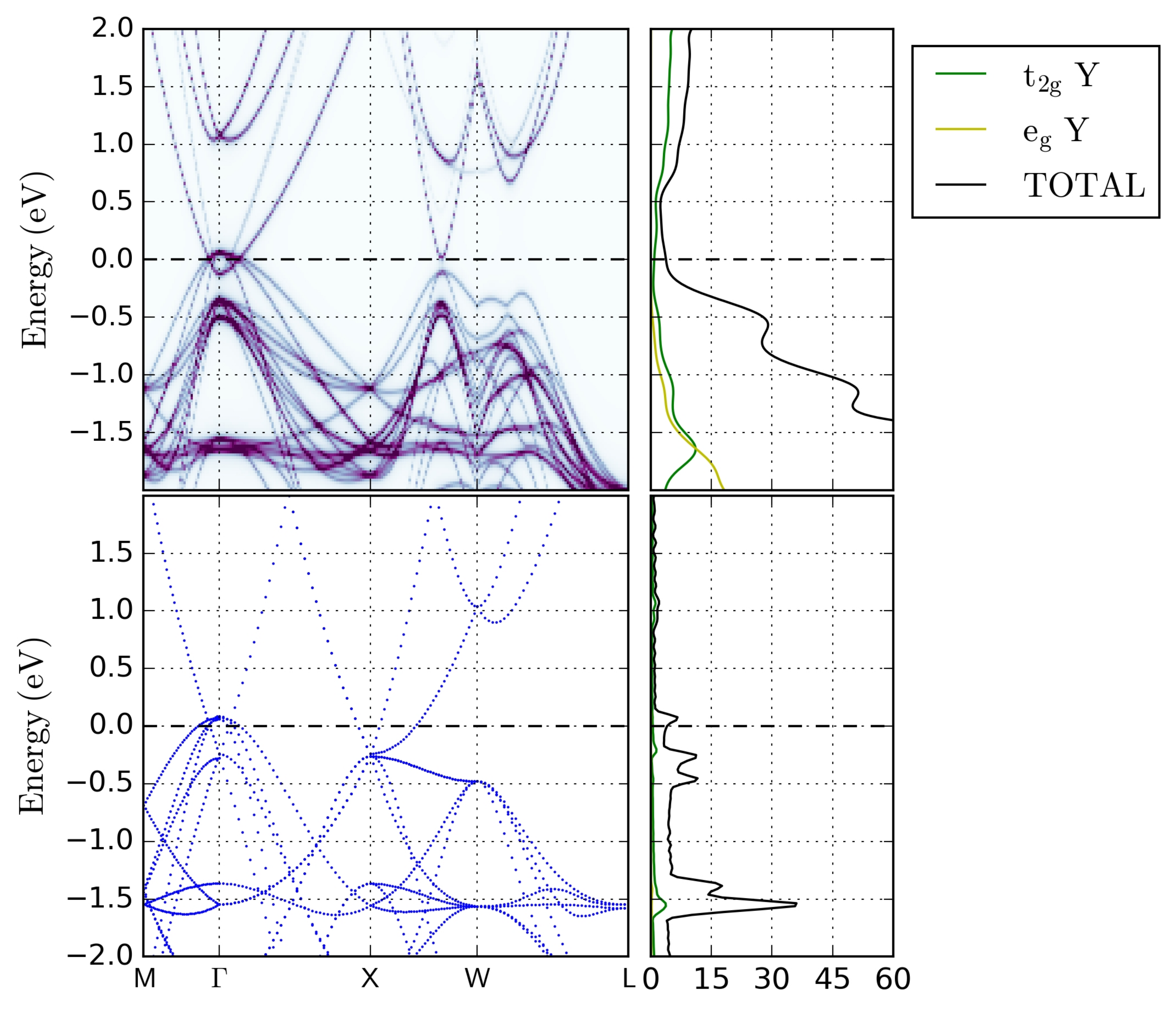}
  \caption{YN DOS and band structure}\label{f:BYN}
\end{figure}

\begin{figure}[htbp]
\centering
  \includegraphics[width=.5\textwidth]{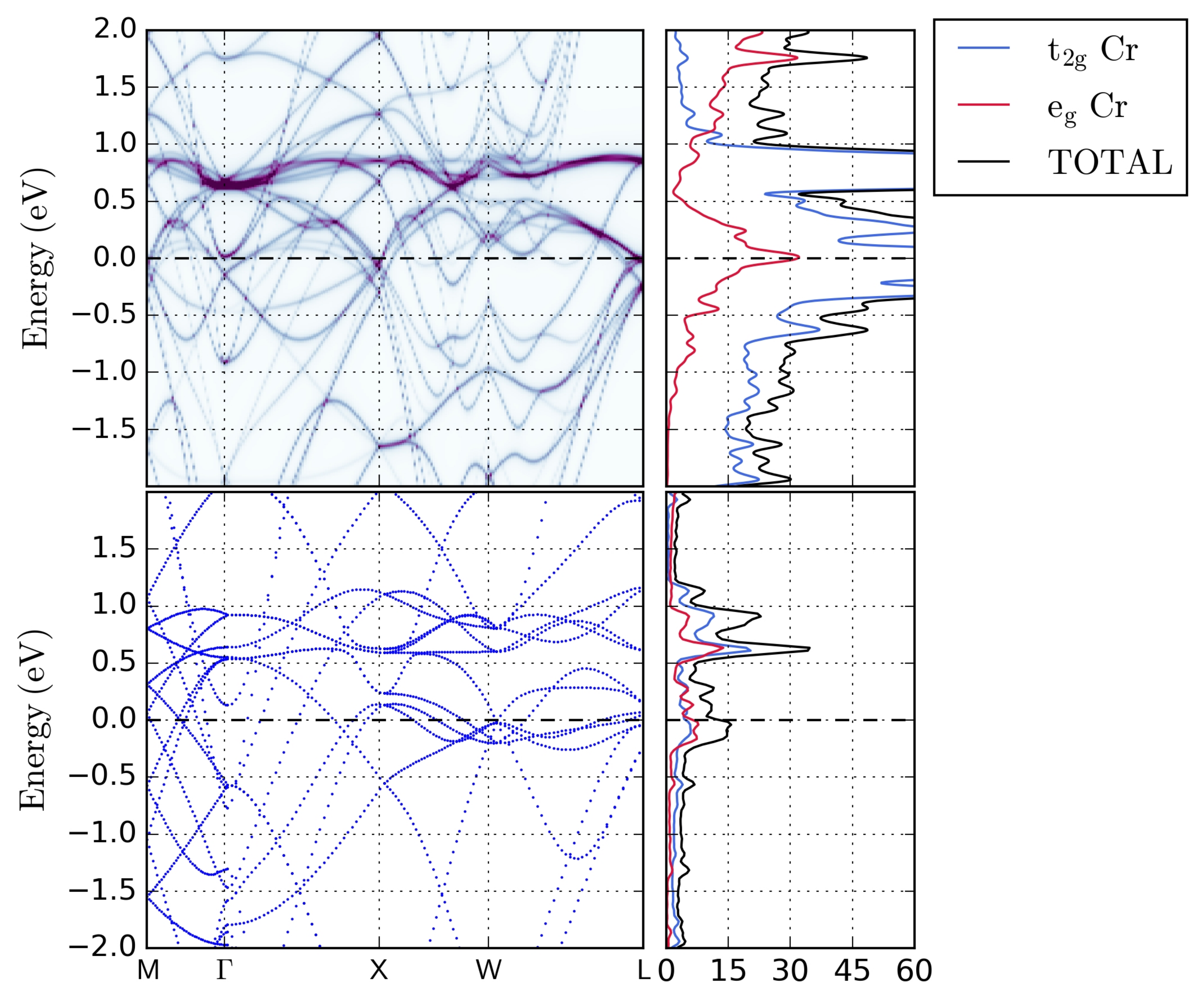}
  \caption{CrN $\mathrm{NM}$ DOS and band structure}\label{f:BNM}
\end{figure}

\begin{figure}[htbp]
\centering
   \includegraphics[width=.5\textwidth]{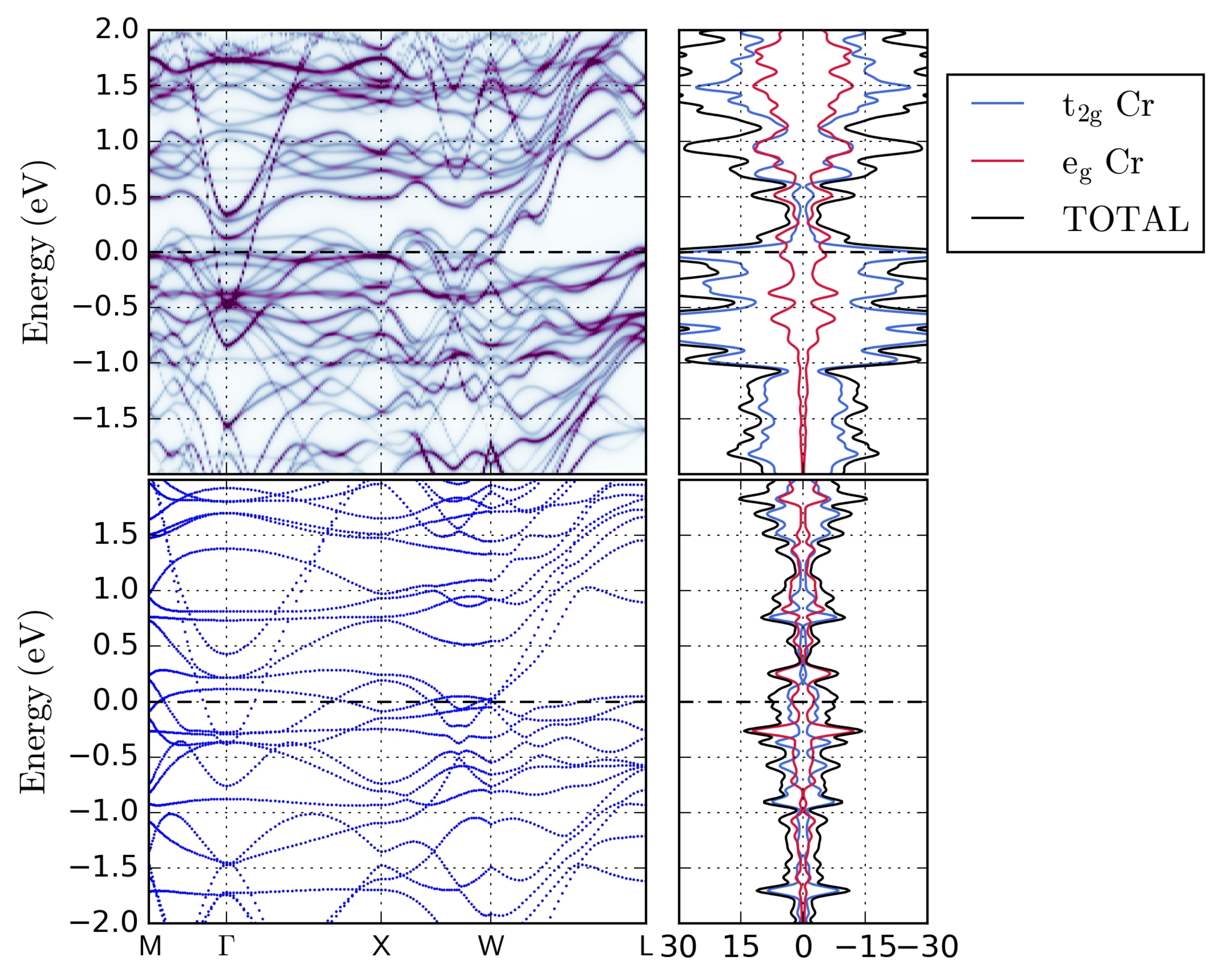}
  \caption{CrN $\mathrm{AFM_{[100]}^{1}}$ DOS and band structure}\label{f:BAFM1}
\end{figure}

\begin{figure}[htbp]
\centering
  \includegraphics[width=.5\textwidth]{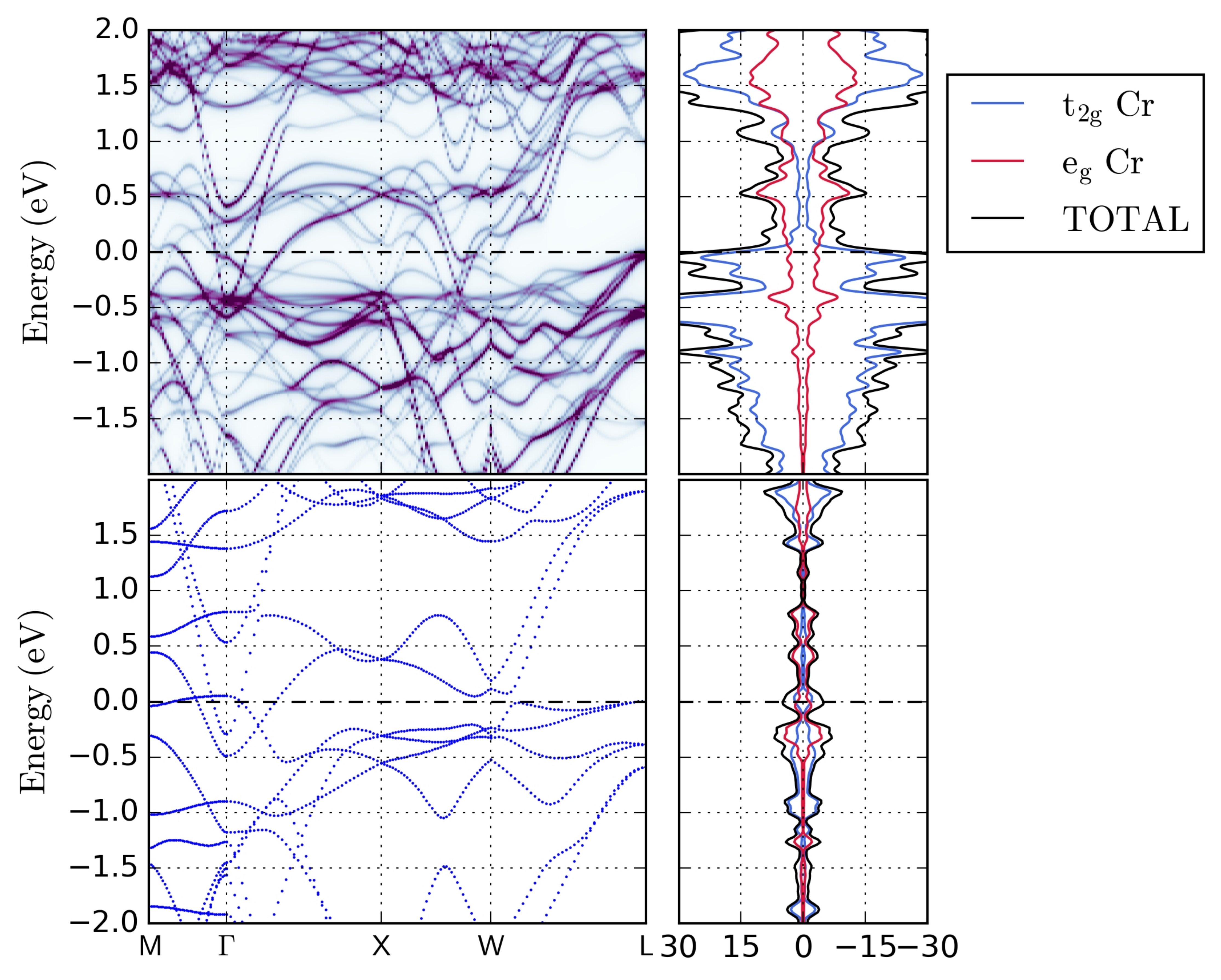}
  \caption{CrN $\mathrm{AFM_{[110]}^{2}}$ DOS and band structure}\label{f:BAFM2}
\end{figure}

\begin{figure}[htbp]
\centering
  \includegraphics[width=.73\textwidth]{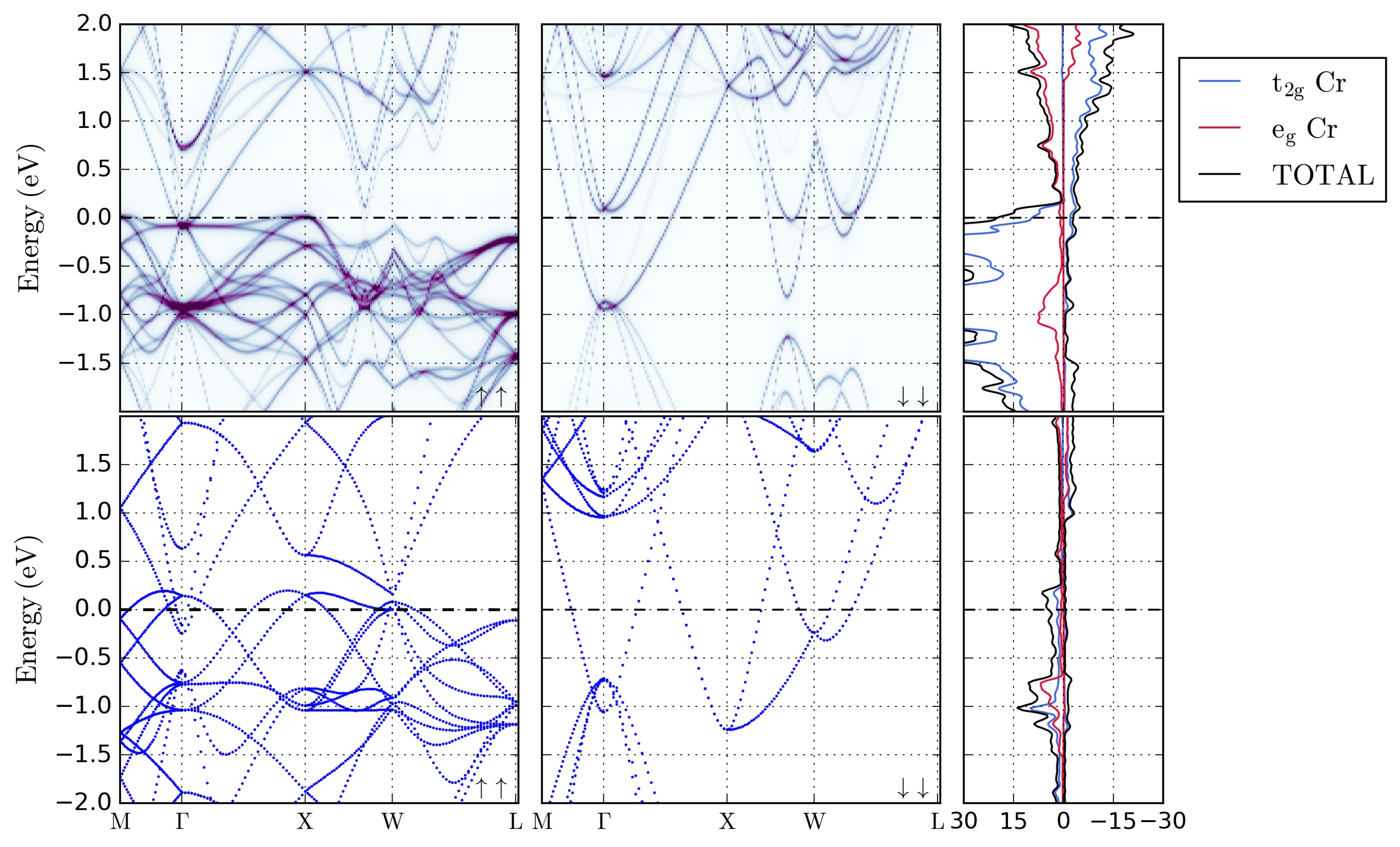}
  \caption{CrN $\mathrm{FM}$ DOS and band structure}\label{f:BFM}
\end{figure}

\begin{figure}[htbp]
\centering
  \includegraphics[width=0.9\textwidth]{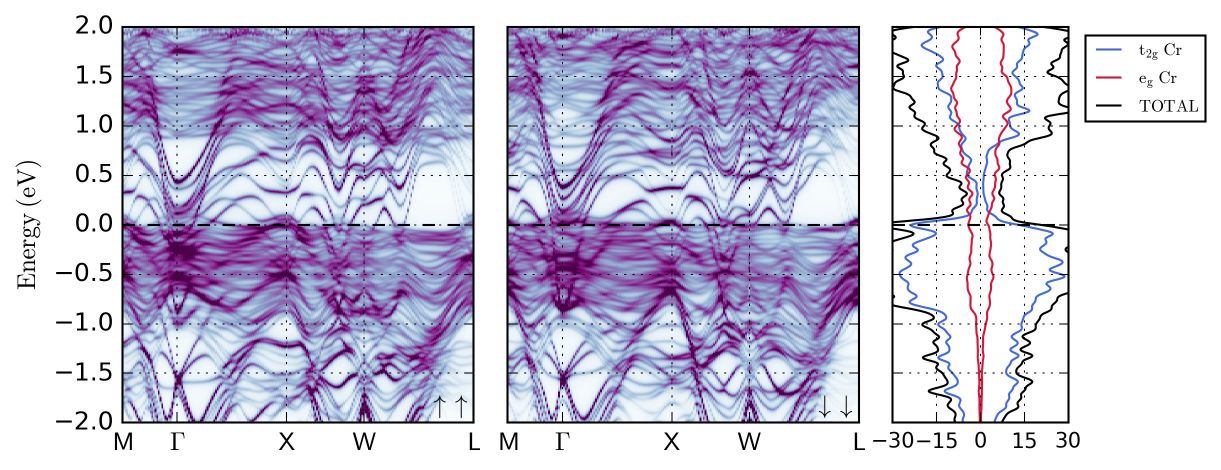}
  \caption{CrN $\mathrm{PM}$ DOS and band structure}\label{f:BPM}
\end{figure}

\begin{figure}[htbp]
\centering
\begin{subfigure}{0.9\linewidth}
    \includegraphics[width=\linewidth]{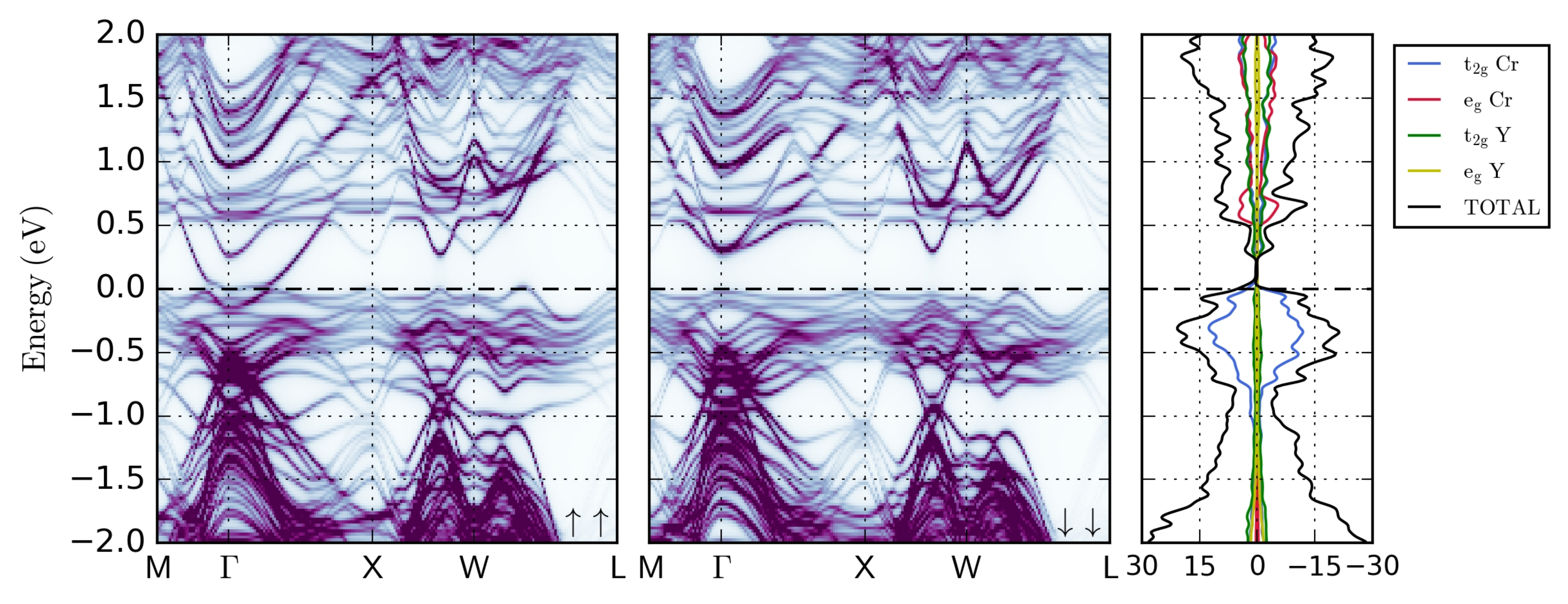}
    \caption{$\mathrm{AFM_{[100]}^{1}}$}
    \label{fig:B25AFM1}
\end{subfigure}
\begin{subfigure}{0.9\linewidth}
    \includegraphics[width=\linewidth]{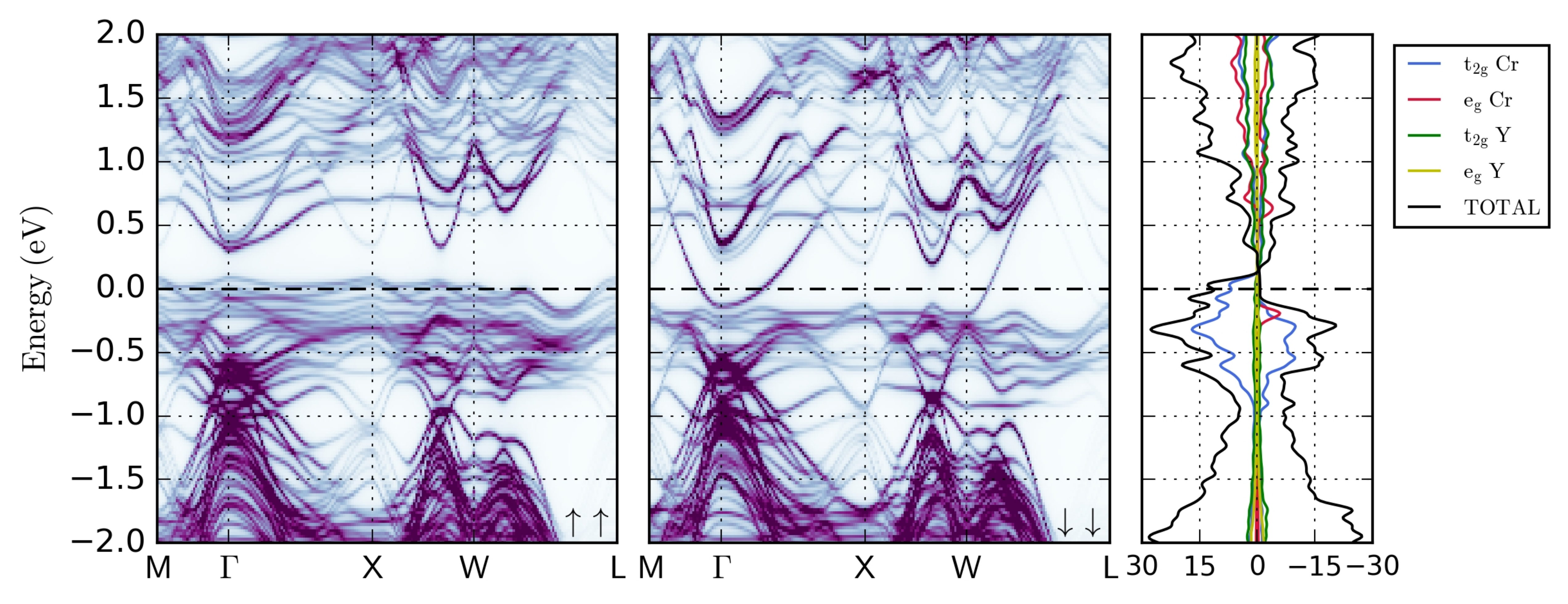}
    \caption{$\mathrm{AFM_{[110]}^{2}}$}
    \label{fig:B25AFM2}
\end{subfigure}
\begin{subfigure}{0.9\linewidth}
    \includegraphics[width=\linewidth]{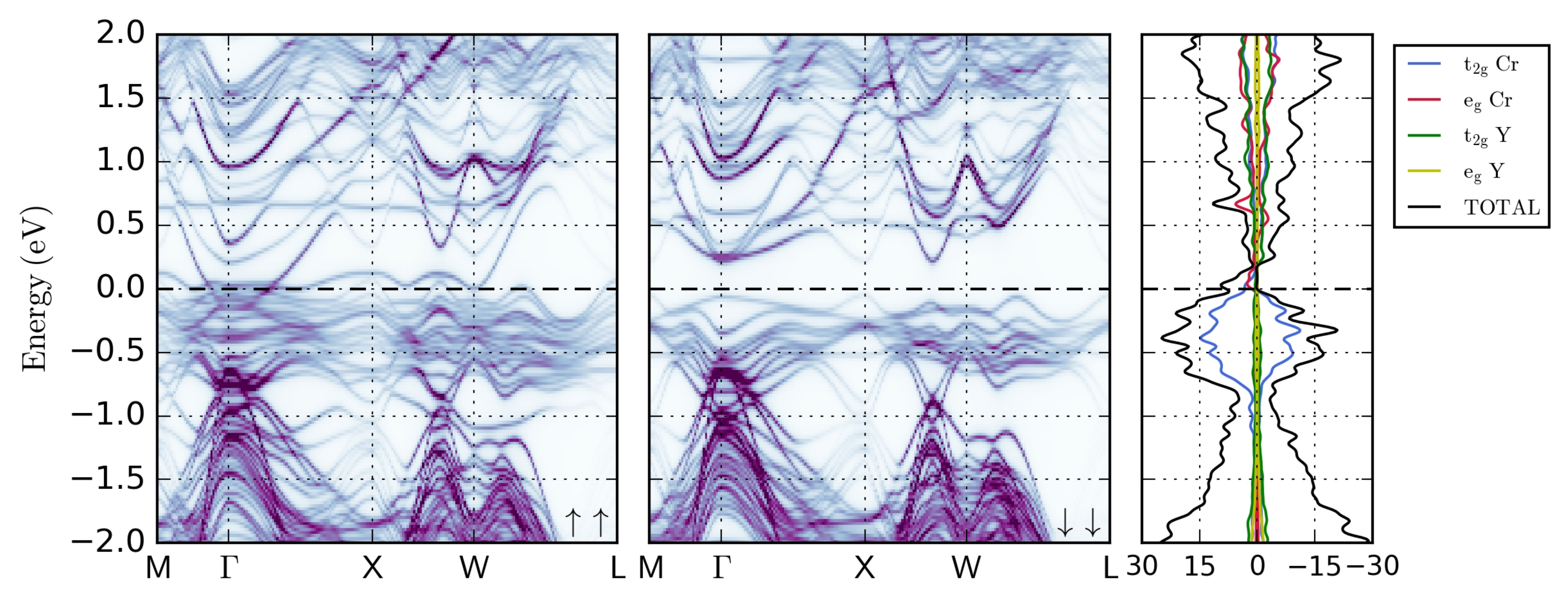}
    \caption{$\mathrm{FM}$}
    \label{fig:B25FM}
\end{subfigure} 
\caption{$\mathrm{Cr_{25}Y_{75}N}$ DOS and band structure}
\label{fig:B25}
\end{figure}

\begin{figure}[htbp]
\centering
\begin{subfigure}{0.9\linewidth}
    \includegraphics[width=\linewidth]{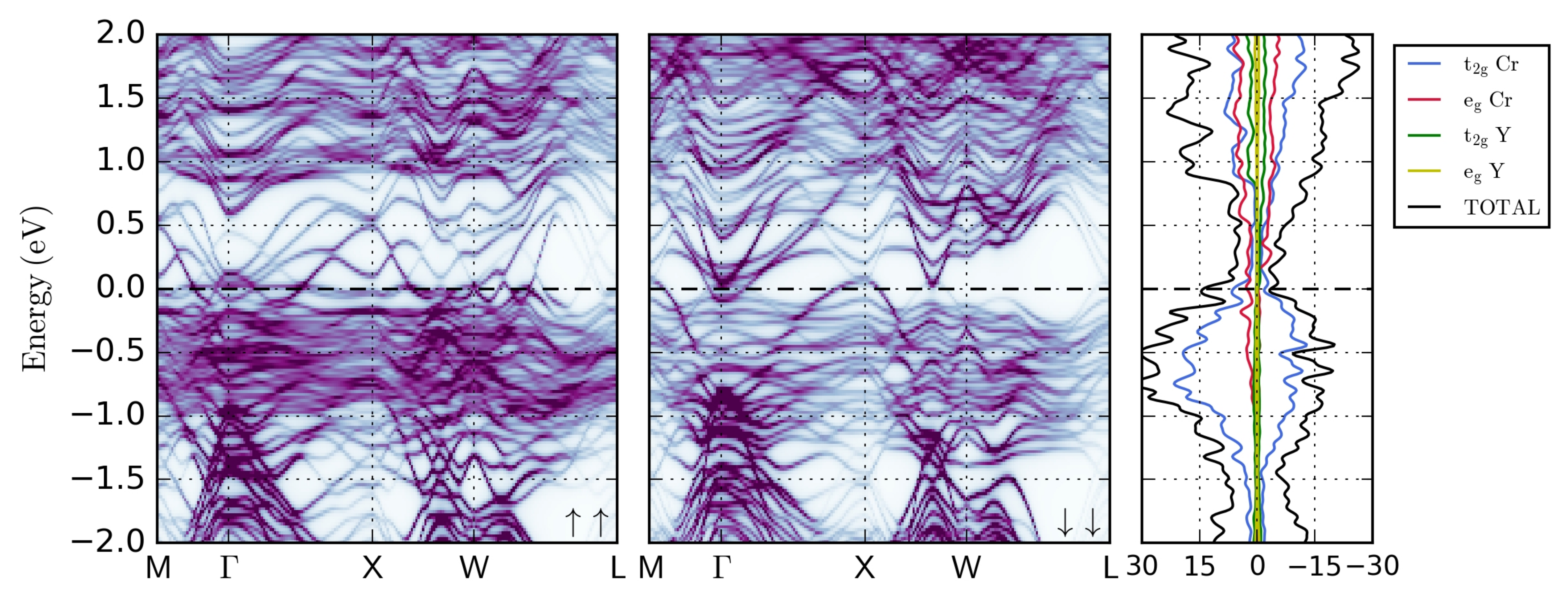}
    \caption{$\mathrm{AFM_{[100]}^{1}}$}
    \label{fig:B50AFM1}
\end{subfigure}
\begin{subfigure}{0.9\linewidth}
    \includegraphics[width=\linewidth]{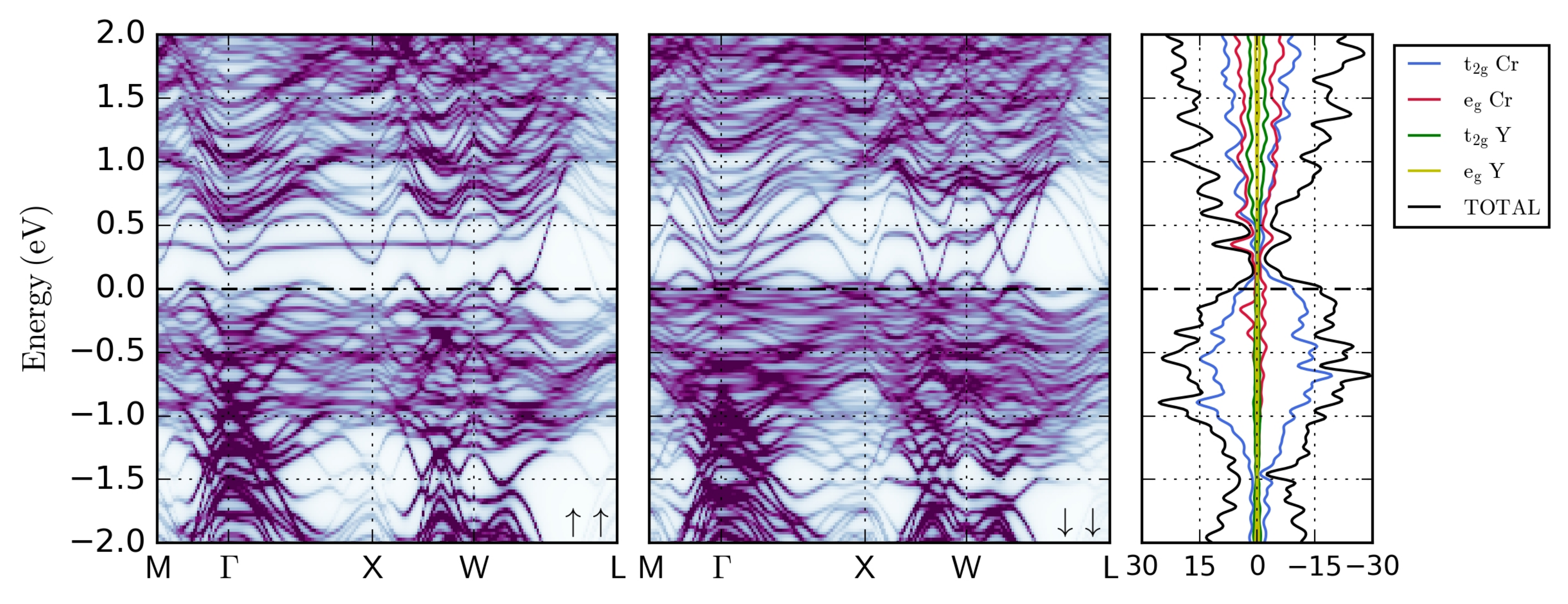}
    \caption{$\mathrm{AFM_{[110]}^{2}}$}
    \label{fig:B50AFM2}
\end{subfigure}
\begin{subfigure}{0.9\linewidth}
    \includegraphics[width=\linewidth]{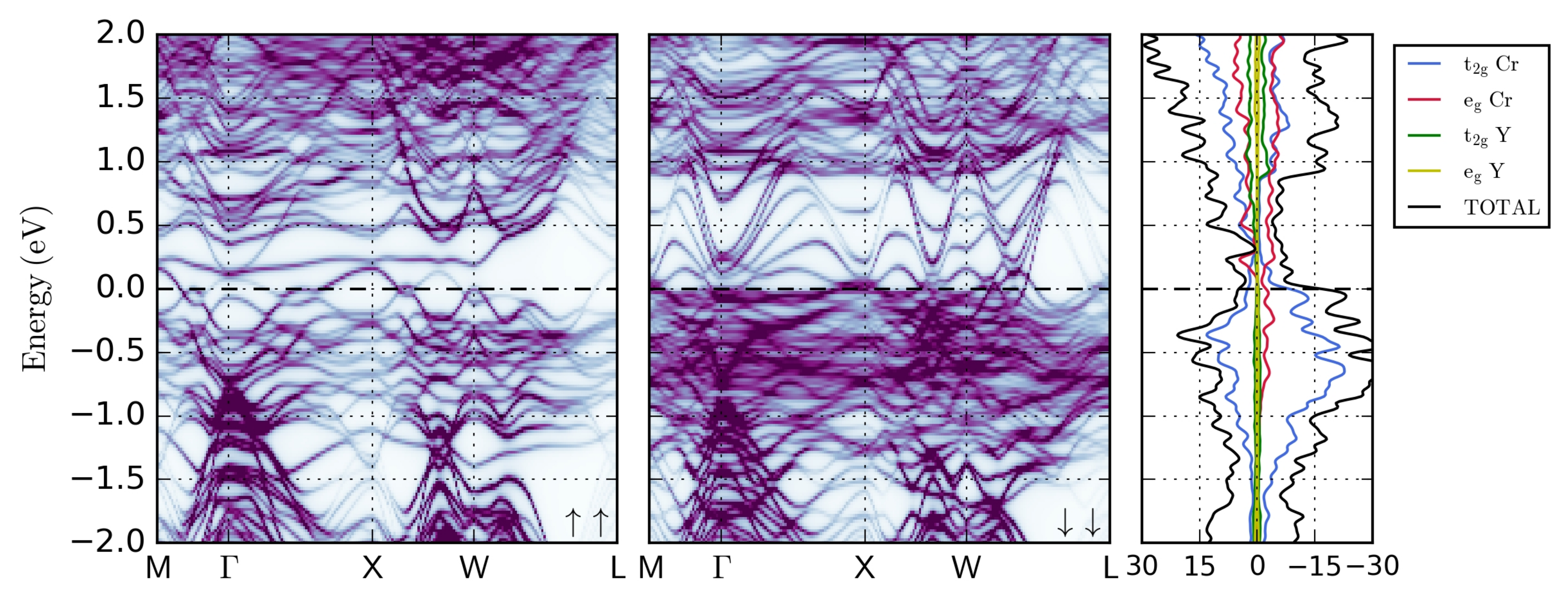}
    \caption{$\mathrm{FM}$}
    \label{fig:B50FM}
\end{subfigure}        
\caption{$\mathrm{Cr_{50}Y_{50}N}$ DOS and band structure}
\label{fig:B50}
\end{figure}

\begin{figure}[htbp]
\centering
\begin{subfigure}{0.9\linewidth}
    \includegraphics[width=\linewidth]{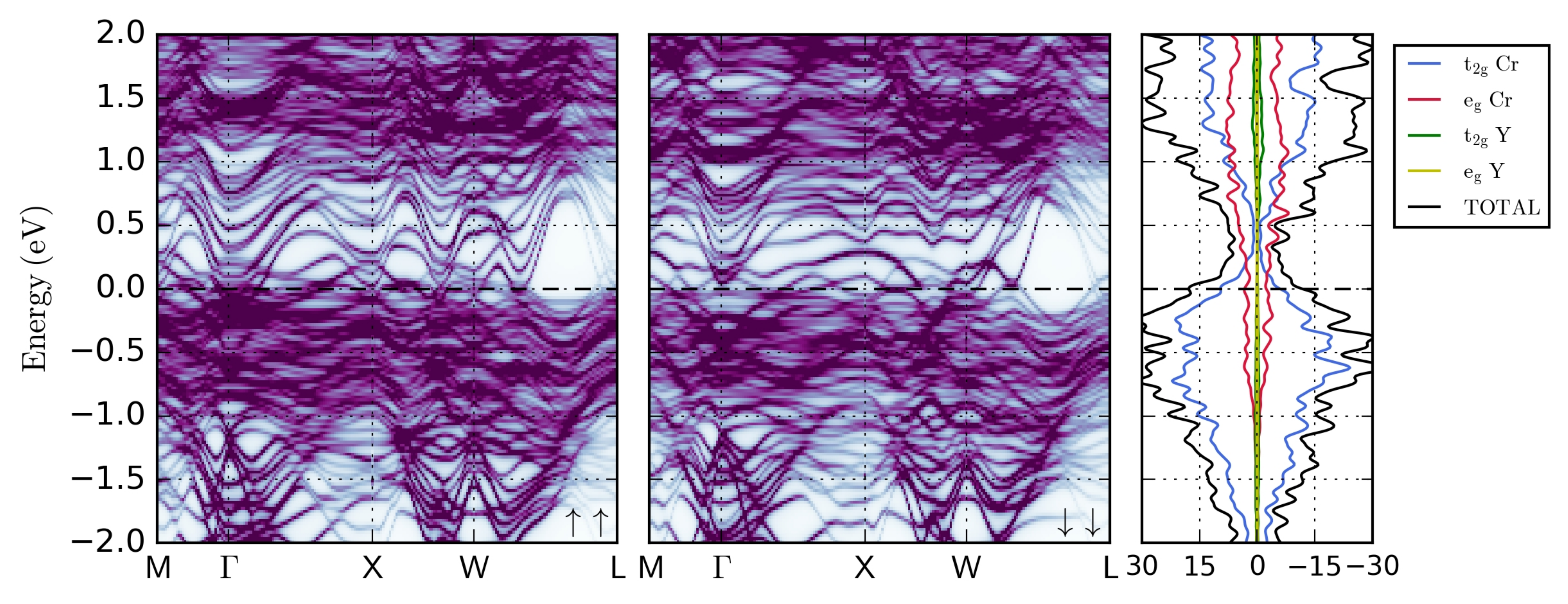}
    \caption{$\mathrm{AFM_{[100]}^{1}}$}
    \label{fig:B75AFM1}
\end{subfigure}
\begin{subfigure}{0.9\linewidth}
    \includegraphics[width=\linewidth]{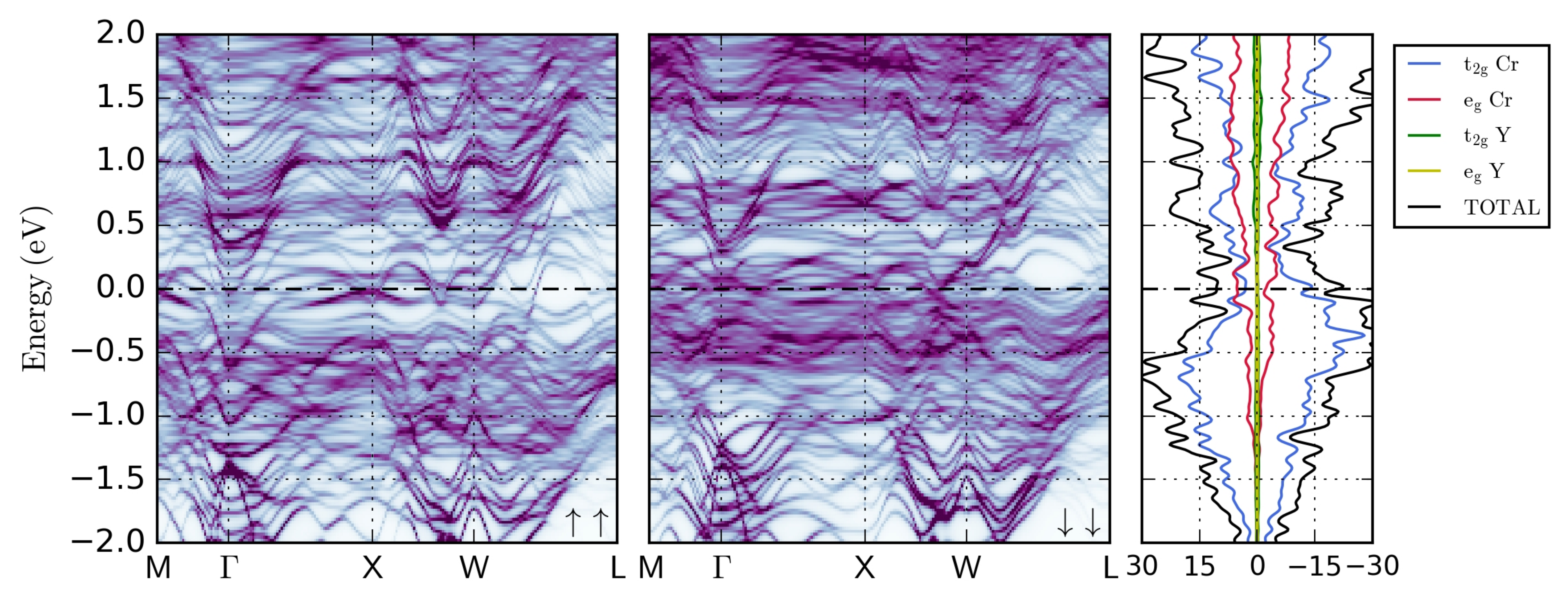}
    \caption{$\mathrm{AFM_{[110]}^{2}}$}
    \label{fig:B75AFM2}
\end{subfigure} 
\begin{subfigure}{0.9\linewidth}
    \includegraphics[width=\linewidth]{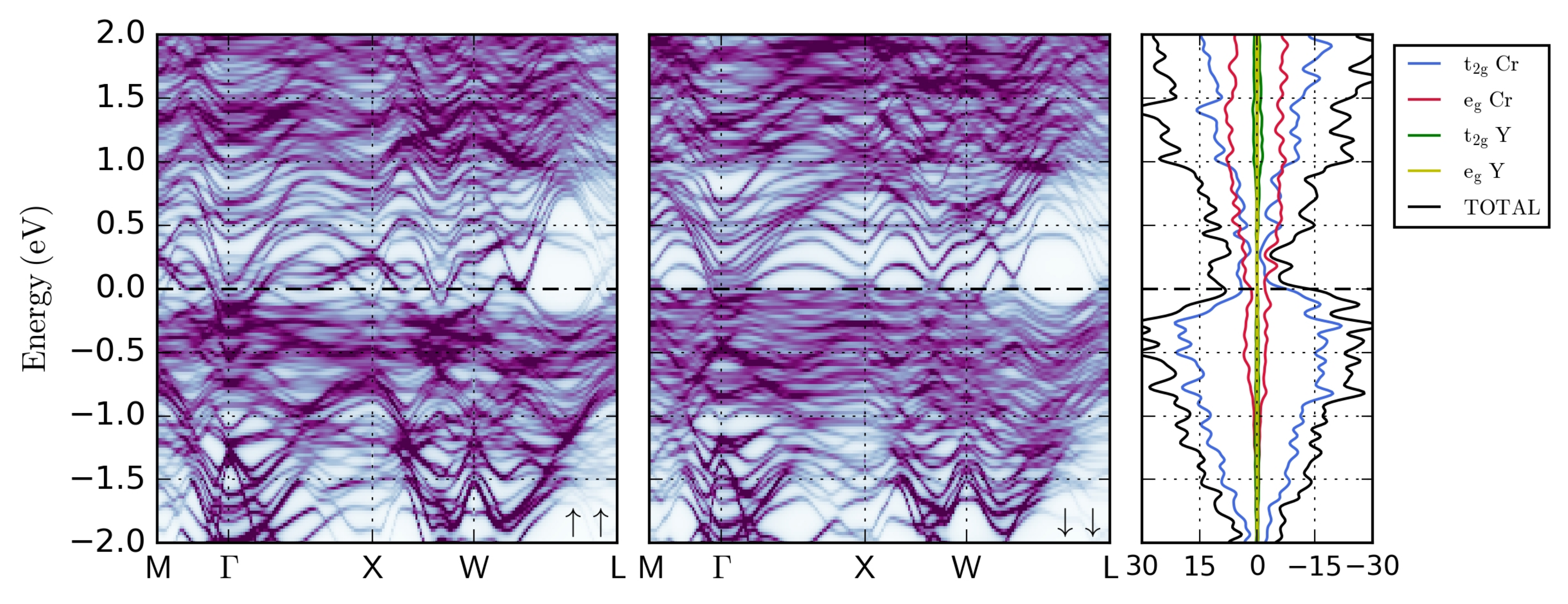}
    \caption{$\mathrm{FM}$}
    \label{fig:B75FM}
\end{subfigure}         
\caption{$\mathrm{Cr_{75}Y_{25}N}$ DOS and band structure}
\label{fig:B75}
\end{figure}

The electrons in the last valence shell in Cr and Y atoms are distributed as follows: $\mathrm{3d^54s^1}$ and $\mathrm{4d^15s^2}$ repectively within the pseudopotentials. In the first instance, the pure compounds are analyzed. The following figures show their band structures and densities of states (DOS) in the orthorhombic phase and the corresponding band unfolding and DOS of their respective cubic supercells are presented (See Figures from \ref{f:BAFM1} to \ref{f:BPM}). In each of the DOS the partial DOS (PDOS) of the `d' orbitals is shown separating their contributions by the $e_g$ and $t_{2g}$ orbital groups as indicated by crystal field theory (CFT) \cite{Bethe1929,Vanvleck1932}. Due to the octahedral coordination of the TMs, the splitting of the `d' orbitals into $e_g$ and $t_{2g}$ orbitals is expected in order to explain the magnetism present in the structures. For binary compounds, only the band structures of the spin-up bands are presented for the CrN-AFM structures, since in all cases they are the same for both types of spin. Secondly, CrYN alloys are analyzed. The band unfolding of each of the magnetic alloys is presented below. These unfolding are made from the orthorhombic cells. Each band structure is accompanied by its respective DOS as for the pure compounds. For the alloys, the band structures for spin-up and spin-down are presented separately (See Figures from \ref{fig:B25AFM1} to \ref{fig:B75FM}). The distribution of band weights is plotted according to the darkness of the band. The lighter the band, the lower the contribution to DOS, and the darker the band, the higher the contribution. Table \ref{t:brecha} shows the energy gap values for each of the supercells studied.

The complex structural and magnetic properties of CrN have already been demonstrated in several studies \cite{Corliss1960,Filippetti2000}. At high tempeartures it has already been proven that CrN crystallizes in cubic phase, with PM ordering and is semiconducting \cite{Kerdsongpanya2016,Quintela2010,Rivadulla2009},. However, at low temperatures, the discussion is quite extensive. In general it has been found that the excess \cite{Quintela2015} or deficiency \cite{MuhammedSabeer2021} of N influences the CrN structure in drastic ways, as well as the method of synthesis. On the one hand it has been possible to maintain the cubic, paramagnetic and semiconducting phase \cite{Zhang2011b}, while on the other hand the cubic phase undergoes a phase transition to orthorhombic and the magnetic moments are aligned in an AFM configuration. About this phase transition, there is no consensus regarding the conductivity, since in some studies it has been determined that it is metallic or a semi-metal \cite{Quintela2015,Constantin2004a,Bhobe2010} and that it can undergo a transition from metal to insulator or also called Mott-Hubbard transition, and in others that it is semiconductor like a mott insulator \cite{Zhang2010,Gall2002,Botana2012} or a charge transfer insulator \cite{Herwadkar2009,Ebad-Allah2016}. In all energy regions of the DOS in all the structures studied, there are pronounced peaks showing a high localization of the electrons, which coincides in the band structure with flat bands. This behaviour may suggest that CrN and the CrYN may be highly correlated materials, which may open the door to a Hubbard parameter treatment and spin-orbit coupling to make a detailed study of the band topology and observe more complex electronic states that can explain a possible Mott transition.

Starting with the analysis of the AFM binary compounds, it can be seen in the total DOS of CrN in both AFM structures (see Figures \ref{f:BAFM1} and \ref{f:BAFM2}), that before the Fermi level, there is an evident splitting of the $e_g$ and $t_{2g}$ orbitals, indicating a strong field and therefore, the structure is low spin, i.e. the spins prefer to occupy lower energy levels instead of occupying the anti-bonding orbitals. Furthermore, it is evident that the $t_{2g}$ orbitals contribute the most states before the Fermi level, while after the Fermi level the $e_g$ orbitals contribute the most states. In addition to this, a symmetry can be seen with respect to the spin-up and spin-down DOS, which, added to the above, indicates that AFM structures are present in both cases. In both AFM structures, the Fermi level is located at pronounced peaks in the density of states, indicating that these are highly locally, suggesting a high correlation. Just after the Fermi level there is a sharp decrease of states, but no energy gap occurs. In this region, the number of states is relatively small with respect to the whole energy spectrum, so they could also be considered semi-metallic, as indicated in \cite{Cheiwchanchamnangij2020}. In general, there is a narrowing of the DOS of about 1 eV, known as 'pseudogap', observed in AFM structures. This feature has been shown in different cases to be important in thermoelectric properties showing improvements in the Seebeck coefficient \cite{MOTT1995,Khandy2021,Hsieh2014,Nishino2004}, but also a signal of superconducting behavior . This discussion will be addressed in the final section of this paper. The band structure shows how the band density is lower along the high symmetry path just above Fermi, coinciding with the pseudogap shown in the DOS. 

In figure \ref{f:BFM}, for the CrN-FM structure, the DOS shows a metallic character in the spin majority channel, as there is a high DOS at the Fermi level. Here, too, there is a rapid decay after the Fermi level in the DOS. Regarding the $e_g$ and $t_{2g}$ orbitals in the spin majority channel it can be seen that, as in the AFM cases, there is a marked splitting. Moreover, before the Fermi level most of the states are contributed by the $t_{2g}$ orbitals, and after that it is the $e_g$ orbitals that contribute most of the states. Looking at the band structure, a decrease in the number of bands above the Fermi level can also be seen. 

On the CrN NM side (see Figure \ref{f:BNM}), a strong metallic character is observed in the DOS due to the presence of a strong peak at the Fermi level, without presenting abrupt decreases after the Fermi level as in the previous cases. This result serves to highlight the impact of including the magnetic moments in the CrN structure for a more accurate description of its electronic, magnetic and structural properties. Here we also observe a strong splitting of the $e_g$ and $t_{2g}$ orbitals, however, the $t_{2g}$ orbitals are the ones that contribute most of the states before and after the Fermi level. In this case, the band structure has a high number of bands above the Fermi level, coinciding with the high DOS peaks. 

In CrN-PM (see Figure \ref{f:BPM}), things are very similar to the AFM structures, with the difference that the symmetry in the DOS no longer exists, however, similar curves are observed on both sides of the DOS, which allows inferring that, although there is no definite magnetic symmetry, the amount of spins up and down are equal, which makes sense that the state distribution is quite similar in both directions of the DOS. In this case, unlike other studies \cite{Kerdsongpanya2016,Cheiwchanchamnangij2020}, the DOS shows a metallic character due to the presence of a high DOS just at the Fermi level, however, a rapid decrease in the DOS also occurs just after this level, presenting a narrowing of the DOS, or as mentioned before, a pseudogap is formed indicating interesting conduction properties that could influence the thermoelectric properties as indicated in \cite{MOTT1995}. In this case a decrease in the number of bands after the Fermi level also coincides with the DOS pseudogap.

Finishing with the pure compounds, YN shows characteristics in its DOS different from those of all CrN structures (see Figure \ref{f:BYN}). In this case, at the fermi level, although there are states available showing a possibly conducting character, this number of states is quite low, and it can also be considered a semi-metal. This last statement can be supported by observing that in the band structure very few bands cross the Fermi level. Another important feature is that the $e_g$ and $t_{2g}$ levels do not split as in the previous cases, evidencing a weak field and high spin system. The d orbitals do not contribute the majority of states, which can be inferred at a glance by the difference with the total DOS. This majority of state is then attributed to the p orbitals. This observation coincides with the studies of \cite{Salguero2003}.

With regard to the magnetic structures of the alloys, the following general observations can be made first. Depending on the composition, it is clearly seen that as the Cr content in the structure increases, the number of bands found after the Fermi level also increases. In all cases it can be seen that the DOS are not symmetric and that there are differences, although not very drastic, in the band structures per spin channel. 

Regarding $\mathrm{Cr_{0.25}Y_{0.75}N}$, as in the studies \cite{Sukkabot2019,Kerdsongpanya2016} , it can be seen how the energy gap opens just at the Fermi level in the spin-down channel, while in the spin-up channel states at the Fermi level are appreciated, giving a semi-metallic character to these alloys with this composition. The band structures of these alloys bear a strong resemblance to those of YN. Comparing between the two, it can be said that substituting few Cr atoms in the YN structure can induce the energy gap to open. This evident semi-metallic behavior of the band structure evidences that there are few bands crossing the Fermi level in the spin channel above, opening the doors to more complex features of this band structure that might require a deeper study of the topology of these bands, which is not in the scope of this study. 

Looking at this composition of the resulting magnetic configurations, appear some differences between the three structures, like the size of the energy gap and the band weights. In the case of the band weights, it can be seen that, although all structures seem to converge to similar results, the initial magnetic structure has an impact on the band weights. A higher weight is seen in the AFM configurations than in the FM configuration. This behavior may be due to the fact that since initially the moments are aligned in the same direction, it may be more difficult for the structure to form antiparallel hybridizations, which in the end may cost in stability and dispersion in the weights of the bands. These observations demonstrate that the way in which magnetic atoms are diluted in the nonmagnetic cell is important in the electronic properties. 

In the case of the AFM$^{1}$ structure, it can be seen that the Fermi level coincides in both spin channels with an abrupt decrease of the DOS, and just after the formation of a gap in the spin-down channel. The spin-up channel can be considered semi-metallic due to the presence of a few bands crossing from the valence band to the conduction band. In the band structure there is no evidence of a difference in band weights in both channels, confirming the absence of magnetization per cell presented in Figure \ref{fig:B25AFM1}. Similar features are present in the AFM$^{2}$ and FM structures. Before the Fermi level there is a clear majority contribution from the spin-up channels. The Fermi level in this channel is above a flat region of DOS showing a delocalization of electrons, giving a metallic character. On the side of the spin-down channel, the Fermi level is just above where an energy gap is formed. These alloys are then also considered to be semi-metallic. The difference is in the location of the Fermi level. It is expected that by Mott's approximation \cite{MOTT1995} (this discussion will be made in the next section), in structures having their Fermi level over energy regions with a pronounced decrease of states and a consequent appearance of a pseudogap, there will be a significant impact on the thermoelectric properties. These semi-metallic characteristics open the door to possible non-trivial topological properties in these alloys that could be addressed in future studies.

Now, talking about the $\mathrm{Cr_{0.5}Y_{0.5}N}$ alloys, the formation of energy gaps no longer occurs in any of the channels, however, the presence of pseudogaps mentioned above for the AFM structures of CrN is evidenced. These pseudgogaps can be clearly seen after the Fermi level in the band structure, where the band density just above is seen to decrease and then increase again at higher energy values. The size of these pseudogaps depends on the spin channel and the initial magnetic configuration. The presence of this pseudogap can ensure that there is a minimum number of states to guarantee a carrier mobility that allows high electronic conductivities depending on the deep. In all these cases, small indirect energy gaps can be observed in the band structure, depending on the spin channel and the region of the reciprocal space, which could explain possible optical properties at the experimental level. 

In the case of the AFM$^{1}$ structure, it can be seen that the pseudogap is higher in the spin-up channel, and also, below the Fermi level, this channel is the one that contributes the most number of states, an effect that is reflected in the band weights in this region. In this channel, the Fermi level is located above a steep decrease of available states. On the other hand, in the spin-down channel the Fermi level is located over a valley, a region of high electronic delocalization. The AFM$^{2}$ structure shows a different case. Here the contribution of states before the Fermi level is mostly from the spin-down channel, although without much difference as in the previous case. Here, the Fermi level is located in a region of a pronounced decrease of the DOS, and evidencing pseudogaps of similar sizes, although because of the band structure, the pseudogap from the spin-up channel is deeper. In the FM structure we see a behavior quite similar to that of the AFM$^{1}$ structure, with the difference that this time it is the spin-down channel that contributes the largest number of available states and where the Fermi level is located at a rapidly decreasing DOS and with a larger pseudogap as seen in the band structure. The Fermi level in the spin-up channel is located in a valley as well. 

To finish this discussion, looking at the $\mathrm{Cr_{0.75}Y_{0.25}N}$ magnetic alloys, no band structure shows possible indirect energy gaps. But, marked differences can be found. The AFM$^{2}$ structure shows a more uniform distribution of bands, with a higher weight in the spin-down channel bands, with no clear evidence of a pseudogap formation. Clearly it can be seen in the DOS that the spin-down channel is the largest contributor of states before the Fermi level, and furthermore no marked narrowing of the DOS is observed after the Fermi level. The Fermi level, in neither of the two spin channels, lies above a sharp decrease of the DOS. From these results it is expected that this structure does not perform well thermoelectrically as will be seen in the next section. This behaviour of the electronic structure could indicate that this structure has a higher degree of deformation than other structures of the same composition. This may also coincide with the suppression of magnetic moments of Cr, as shown in table \ref{t:param_magmom}. On the other hand, in the AFM$^{1}$ and FM structures, a pseudogap is visible, larger in the case of the spin-up channels in both cases. This pseudogap is clearly seen in the band structure of both structures after the Fermi level. In the case of the AFM$^{1}$ structure, the Fermi level in both spin channels is above steep DOS dips, and the pseudogap of the spin channel below is deeper. In the case of the FM structure, only in the spin-down channel the Fermi level is on a steep slope of DOS. The other channel has it over a valley. Here the pseudogap of the spin-down channel is also deeper. 

In all the DOS calculated for each alloy, a strong splitting of the $e_g$ and $t_{2g}$ orbitals was found, demonstrating that the presence of Cr has the greatest impact on the CrYN structure. In all cases it can be seen that before the Fermi level, the $t_{2g}$ orbitals are the ones that provide the highest number of available states, well above the other types of orbitals. With this result it can be said that all these structures are strong-field, low-spin magnetic alloys. 

\subsection{Thermoelectric properties}

\begin{figure}[htbp]
\centering
  \includegraphics[width=1\textwidth]{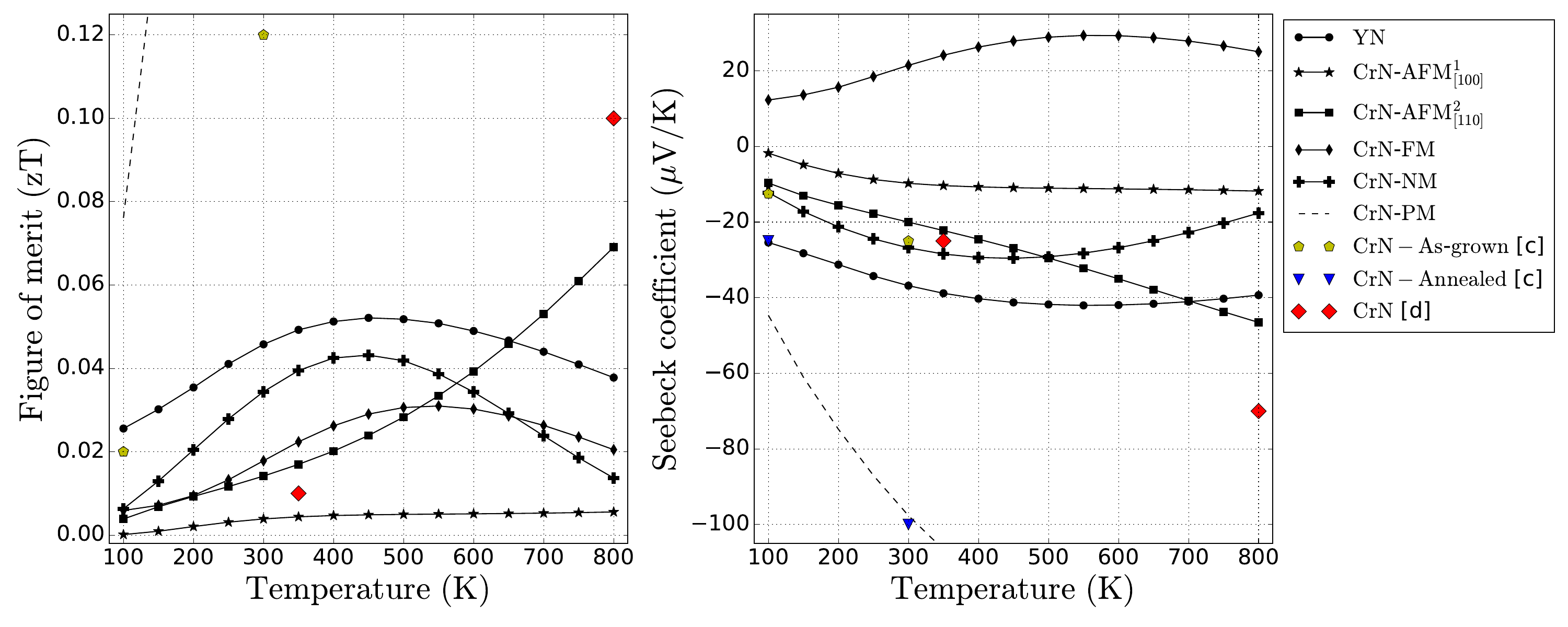}
  \caption{Figure of merit zT and Seebeck coefficient with respect to temperature (T) for each of the different magnetic structures. [c] correspond to reference \cite{Quintela2015}. [d] corresponds to reference \cite{Jiang2020}}\label{f:zTCr}
\end{figure}

\begin{figure}[htbp]
\centering
  \includegraphics[width=1\textwidth]{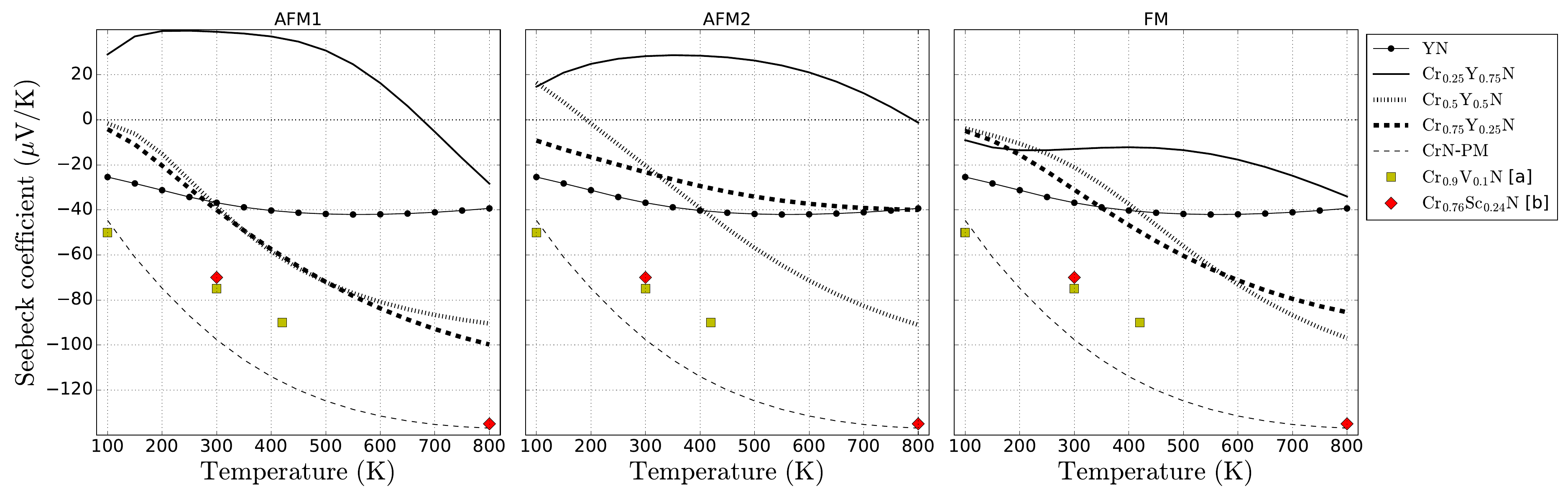}
  \caption{Seebeck coefficient (S) with respect to temperature (T) for each of the different magnetic structures. [a] corresponds to reference \cite{Quintela2009}. [b] corresponds to reference \cite{Kerdsongpanya2016}}\label{f:S}
\end{figure}

\begin{figure}[htbp]
\centering
  \includegraphics[width=1\textwidth]{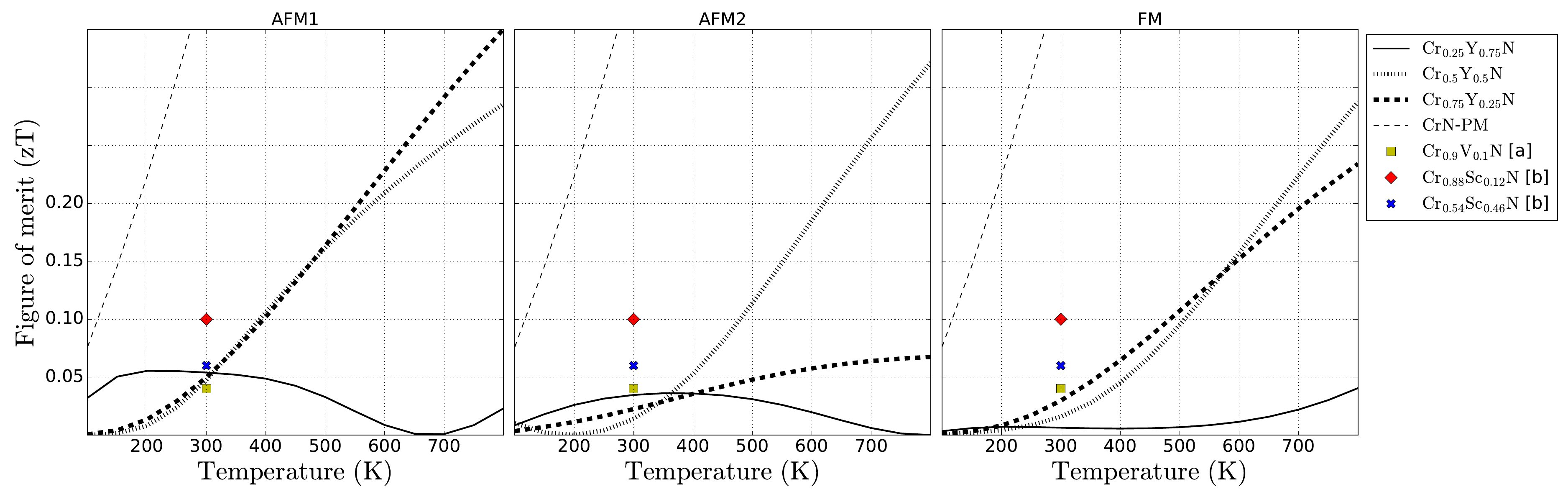}
  \caption{Figure of merit zT with respect to temperature (T) for each of the different magnetic structures. [a] corresponds to reference \cite{Quintela2009}. [b] corresponds to reference \cite{Kerdsongpanya2016}}\label{f:zT}
\end{figure}

The performance of a thermoelectric (TE) material can be efficiently measured from the dimensionless figure of merit (ZT), as explained in \cite{Disalvo1999}. This ZT is defined by:

\begin{align}
\centering
ZT=\frac{S^{2}\sigma T}{\kappa_{el}+\kappa_{latt}}
\label{ec:zt}
\end{align}
where $\sigma$ is the electrical conductivity [$\mathrm{S cm^{-1}}$], $\kappa_{el}$ is the electronic thermal conductivity [$\mathrm{W m^{-1} K^{-1}}$], $\kappa_{latt}$ is the phononic thermal conductivity [$\mathrm{W m^{-1} K^{-1}}$] and $S$ is the Seebeck coefficient [$\mathrm{\mu V K^{-1}}$] defined by $S=V/\Delta T$. For practical reasons, since we are working at 0K, the phonon thermal conductivity will be considered negligible, assuming that the level of deformation in the alloys is sufficient to be negligible. The values of $S$, $\sigma$ and $k_{el}$ are calculated with BoltzTrap2 \cite{BoltzTraP2}. Taking into account that S is the basis for the design of thermoelectric devices, Figures \ref{f:zTCr}-left and \ref{f:S} present the results of $S$ with respect to temperature for pure composites and for alloys respectively, in a range from 100 to 800K in order to observe the TE performance at low temperatures (100K), room temperature (300K) and high temperatures (800K). The zT is also presented in Figures \ref{f:zTCr}-right and \ref{f:zT} in the same temperature ranges. Some experimental results found for both CrN and other CrN alloys are plotted on these results.

According to Mott's approximation \cite{MOTT1995,Jonson1980}, the density of state can be related to the Seebeck coefficient as follows:

\begin{align}
\centering
S \propto \frac{\partial \mathrm{Ln} \sigma (E)}{\partial E}
\label{ec:see}
\end{align}
where $\sigma (E)$ is the electronic condictivity that depends on the energy and thus on the DOS. This equation suggests that at high DOS slopes, high values of $S$ are expected, and lower slopes correspond to low values of $S$.

Figure \ref{f:zTCr} shows the ZT and Seebeck coefficient plots of the pure compounds. Observing the $S$ plot, it can be seen that CrN-FM has positive values of $S$, which indicates that it is a p-type material, while all the others correspond to n-type. The absolute value of $S$ of CrN is the largest in the whole calculated temperature spectrum and at 300K it coincides with the values measured for CrN annealed experimentally in \cite{Quintela2015}. CrN-AFM$^2$ presents values similar to those measured by \cite{Quintela2015} at 100 and 300K for CrN As-grown and that measured by \cite{Jiang2020} near 300K. YN presents higher $|S|$ values in almost all the temperature range, surpassed only by AFM$^2$ above 700K. It should be noted that in this study the magnetic effects on the thermoelectric properties are being analyzed. The cases of AFMs and FM magnetic structures are hypothetical above room temperature.

Looking at the zT plot, it can be seen that these values are consistent with the calculated values of $S$. CrN-PM has the highest zT over the entire temperature range. It is followed by YN, CrN-NM and with very similar values are CrN-AFM$^2$ and CrN-FM up to a little after 500K. CrN-AFM$^1$ has the lowest $S$ values, hence the lowest zT. CrN-AFM$^2$ is the one that presents values more in line with those measured by \cite{Jiang2020}.

Up to this point, it can be said then that the structures with the best thermoelectric performance are CrN-PM by far, followed by YN. In the case of the alloys, $S$ and zT values are expected to be between these two reference ranges. Starting with $S$, Figure \ref{f:S} shows the values of $S$ versus temperature for each initial magnetic configuration. Together with these values, the values obtained for CrN-PM and YN are plotted again for reference and experimental values measured for some other CrN alloys doped with other TM at different concentrations are also included. 

In general, none of the proposed alloys obtained an $S$ value higher than those measured experimentally. However, the alloy in general terms increases $S$ values above YN in certain structures. Starting with the AFM$^{1}$ structures, it can be seen that at 25\% of Cr the $S$ values are positive until before 700K, evidencing a change in the type of charge carriers, being this structure of type p. This behavior was already seen in \cite{MuhammedSabeer2021}. This same behavior is seen in the AFM$^{2}$ structure of the same composition. At 300K, the $S$ of these structures is 40 and 29 $\mu$V/K. These characteristics, together with the opening of the energy gap mentioned in the previous section, these alloys may have potential applications as dilute magnetic semiconductors. For this purpose, it would be necessary to study other composition ranges in that area. In spite of this, the $|S|$ values remain between those calculated for CrN-PM and YN, showing an increase in $S$ with respect to the latter until before 600K. The FM structure corresponding to this composition shows the lowest $S$ values with respect to all the other structures. As these values are lower than those of YN, this is the least suitable structure for thermoelectric applications.

With respect to the alloys with a concentration of 50\% Cr, it can be seen that in all cases there is an increase with respect to the $S$ value of YN. In the case of AFM$^{1}$ this increase is above 300K, and in the case of AFM$^{2}$ and FM it is above 400K. At 800K these $S$ values become higher than 80 $\mu$V/K. In the particular case of the AFM$^{2}$ structure, it can be observed that at low temperatures, below 200K, the S values are positive, showing a small range of being p-type material. Not counting this, for this concentration these structures are n-type. These high $S$ values coincide with the fact that the Fermi levels in the DOS of these structures in either of the two spin channels are on steep DOS slopes, which according to Mott's approximation, should favor high $S$ values.

As far as the 75\% Cr alloys are concerned, something very similar occurs as in the 50\% Cr concentration, excluding the AFM$^{2}$ structure. The $S$ values between them are very similar, with a small difference in the FM, where the $S$ values start to be higher at lower temperature, near 300K. In the case of AFM$^{2}$ it is a very different case. Here in no $S$ value is achieved to be above the YN values, also does not show positive $S$ values which also makes these alloys type n. The case of the AFM$^{2}$ structure coincides with the fact that in its band structure is not clearly seen the presence of a pseudogap, and in addition, the Fermi level in neither of the two spin channels is on a steep slope, which does not favor the thermoelectric properties. 

Finally, as far as zT is concerned (see Figure \ref{f:zT}), excluding $\mathrm{Cr_{0.25}Y_{0.75}N}$-FM, all structures show better performance than YN. In the case of AFM$^{1}$ structures, all compositions show zT values higher than that measured by \cite{Quintela2009} for $\mathrm{Cr_{0.9}V_{0.1}N}$, and values very close to $\mathrm{Cr_{0.54}Sc_{0.46}N}$ at 300K. In AFM$^{2}$ structures, the 25\% Cr composition is the one showing a zT value at 300K higher and closer to that of $\mathrm{Cr_{0.9}V_{0.1}N}$. Regarding this same value, in FMs, the 75\% composition is the one showing the highest value. At 800K, $\mathrm{Cr_{0.75}Y_{0.25}N}$-AFM$^{1}$ exhibits the best zT with a value of 0.35. The $\mathrm{Cr_{0.5}Y_{0.5}N}$-AFM$^{1}$, $\mathrm{Cr_{0.5}Y_{0.5}N}$-AFM$^{2}$, $\mathrm{Cr_{0.5}Y_{0.5}N}$-FM$^{2}$ and $\mathrm{Cr_{0.75}Y_{0.25}N}$-FM structures also show high zT values above 0.25. These calculated structures then present zT values comparable with other alloys made in previous studies, and it is suggested that CrYN alloys may have potential applications in thermoelectric devices, both at room temperature and at high temperatures.

\section{Conclusions}

In this work, supercells were constructed for YN and CrN in AFM$^{1}_{[100]}$, AFM$^{2}_{[110]}$, FM and PM configurations. With the AFM$^{1}$, AFM$^{2}$ and FM configurations taken from \cite{Corliss1960}, 9 SQS were constructed for 25\%, 50\% and 75\% Cr compositions to simulate random substitutional solid solutions of (Cr,Y)N. From first principles calculations with SCAN functional was determined that all alloys comply with Vegaard's law and that the resulting magnetic configurations are fM with the exception of composition at 25\% Cr which was considered PM. The magnetic structure that is possibly the easiest to mix with YN was determined to be the AFM$^{2}$configuration despite the hypothetical CrN-FM being the lowest formation energy. It was determined that the distribution of the second neighbors of the central atoms of the octahedra follow a Gaussian distribution. It was found that the chemical environment influences the deformation of the bonds causing most Jhan-Teller type distortions at 25\% Cr, tetragonal distortions at 75\% and a combination of both at 50\% depending precisely on their second neighbors. Despite the high number of Cr atoms, it can be seen that the structures tend to maintain their cubic structure. The initial magnetic configuration can influence the electronic properties of the alloys. Thanks to CFT, it was possible to verify by means of DOS and band structure calculations that there is a strong splitting in the $e_g$ and $t_{2g}$ orbitals, causing these structures to be of strong field and low spin. The $t_{2g}$ orbitals have the most contribution to the DOS before Fermi level. It was also possible to observe the presence of pseudogaps that coincide with the decrease in the number of bands after the Fermi level, and that this Fermi level is found in several cases on steep DOS slopes. Thanks to Mott's approximation it was possible to determine that these electronic features have a strong impact on the thermoelectric properties, increasing S values with respect to YN and showing zT values comparable to other CrN alloys with other TMs at room temperature. At room temperature it was determined that in all compositions can reach significant zT values depending on the initial magnetic configuration. At high temperatures only compositions higher than 50\% Cr show significant zT values. At compositions of 50\% Cr the presence of indirect energy gaps can be seen which could be signs of optical effects at the experimental level. It is hoped that this material can be synthesized and characterized in thermoelectric properties to contrast with the calculations made in this work. Although this material does not show the best thermoelectric performance, it is expected that this study can contribute to the understanding of the thermoelectric properties of semiconductors and semimetals in magnetic alloys. The CrN-PM structure shows the best thermoelectric performance, agreeing with other studies that SQS are a good approach to study disordered and random magnetic alloys.

Finally, due to the observations made on the band structures, Hubbard parameter studies may be suggested to analyze the impact of high correlation that Cr atoms may have. It is also suggested to mix with band topology studies with Wannier functions to search for non-trivial topological properties.

\section{Acknowledgments}

We are grateful for the help of the University of Guadalajara, which made its computing capacity available to CIMAV in order to carry out this research.

\bibliography{REE_cleaned6.bib}

\begin{thebibliography}{75}%
\makeatletter
\providecommand \@ifxundefined [1]{%
 \@ifx{#1\undefined}
}%
\providecommand \@ifnum [1]{%
 \ifnum #1\expandafter \@firstoftwo
 \else \expandafter \@secondoftwo
 \fi
}%
\providecommand \@ifx [1]{%
 \ifx #1\expandafter \@firstoftwo
 \else \expandafter \@secondoftwo
 \fi
}%
\providecommand \natexlab [1]{#1}%
\providecommand \enquote  [1]{``#1''}%
\providecommand \bibnamefont  [1]{#1}%
\providecommand \bibfnamefont [1]{#1}%
\providecommand \citenamefont [1]{#1}%
\providecommand \href@noop [0]{\@secondoftwo}%
\providecommand \href [0]{\begingroup \@sanitize@url \@href}%
\providecommand \@href[1]{\@@startlink{#1}\@@href}%
\providecommand \@@href[1]{\endgroup#1\@@endlink}%
\providecommand \@sanitize@url [0]{\catcode `\\12\catcode `\$12\catcode
  `\&12\catcode `\#12\catcode `\^12\catcode `\_12\catcode `\%12\relax}%
\providecommand \@@startlink[1]{}%
\providecommand \@@endlink[0]{}%
\providecommand \url  [0]{\begingroup\@sanitize@url \@url }%
\providecommand \@url [1]{\endgroup\@href {#1}{\urlprefix }}%
\providecommand \urlprefix  [0]{URL }%
\providecommand \Eprint [0]{\href }%
\providecommand \doibase [0]{https://doi.org/}%
\providecommand \selectlanguage [0]{\@gobble}%
\providecommand \bibinfo  [0]{\@secondoftwo}%
\providecommand \bibfield  [0]{\@secondoftwo}%
\providecommand \translation [1]{[#1]}%
\providecommand \BibitemOpen [0]{}%
\providecommand \bibitemStop [0]{}%
\providecommand \bibitemNoStop [0]{.\EOS\space}%
\providecommand \EOS [0]{\spacefactor3000\relax}%
\providecommand \BibitemShut  [1]{\csname bibitem#1\endcsname}%
\let\auto@bib@innerbib\@empty
\bibitem [{\citenamefont {Nita}\ \emph {et~al.}(2016)\citenamefont {Nita},
  \citenamefont {Mastail},\ and\ \citenamefont {Abadias}}]{Nita2016}%
  \BibitemOpen
  \bibfield  {author} {\bibinfo {author} {\bibfnamefont {F.}~\bibnamefont
  {Nita}}, \bibinfo {author} {\bibfnamefont {C.}~\bibnamefont {Mastail}},\ and\
  \bibinfo {author} {\bibfnamefont {G.}~\bibnamefont {Abadias}},\ }\href
  {https://doi.org/10.1103/PhysRevB.93.064107} {\bibfield  {journal} {\bibinfo
  {journal} {Physical Review B}\ }\textbf {\bibinfo {volume} {93}},\ \bibinfo
  {pages} {1} (\bibinfo {year} {2016})}\BibitemShut {NoStop}%
\bibitem [{\citenamefont {Nagao}\ \emph {et~al.}(2006)\citenamefont {Nagao},
  \citenamefont {Nordlund},\ and\ \citenamefont {Nowak}}]{Nagao2006}%
  \BibitemOpen
  \bibfield  {author} {\bibinfo {author} {\bibfnamefont {S.}~\bibnamefont
  {Nagao}}, \bibinfo {author} {\bibfnamefont {K.}~\bibnamefont {Nordlund}},\
  and\ \bibinfo {author} {\bibfnamefont {R.}~\bibnamefont {Nowak}},\ }\href
  {https://doi.org/10.1103/PhysRevB.73.144113} {\bibfield  {journal} {\bibinfo
  {journal} {Physical Review B - Condensed Matter and Materials Physics}\
  }\textbf {\bibinfo {volume} {73}},\ \bibinfo {pages} {1} (\bibinfo {year}
  {2006})}\BibitemShut {NoStop}%
\bibitem [{\citenamefont {Gupta}\ \emph {et~al.}(2014)\citenamefont {Gupta},
  \citenamefont {Gupta}, \citenamefont {Soni}, \citenamefont {Mankad},\ and\
  \citenamefont {Jha}}]{Gupta2014}%
  \BibitemOpen
  \bibfield  {author} {\bibinfo {author} {\bibfnamefont {S.~K.}\ \bibnamefont
  {Gupta}}, \bibinfo {author} {\bibfnamefont {S.~D.}\ \bibnamefont {Gupta}},
  \bibinfo {author} {\bibfnamefont {H.~R.}\ \bibnamefont {Soni}}, \bibinfo
  {author} {\bibfnamefont {V.}~\bibnamefont {Mankad}},\ and\ \bibinfo {author}
  {\bibfnamefont {P.~K.}\ \bibnamefont {Jha}},\ }\href
  {https://doi.org/10.1016/j.matchemphys.2013.08.046} {\bibfield  {journal}
  {\bibinfo  {journal} {Materials Chemistry and Physics}\ }\textbf {\bibinfo
  {volume} {143}},\ \bibinfo {pages} {503} (\bibinfo {year}
  {2014})}\BibitemShut {NoStop}%
\bibitem [{\citenamefont {Fernondez~Guillermet}\ \emph
  {et~al.}(1992)\citenamefont {Fernondez~Guillermet}, \citenamefont {Haglund},\
  and\ \citenamefont {Grimvall}}]{FernondezGuillermet1992}%
  \BibitemOpen
  \bibfield  {author} {\bibinfo {author} {\bibfnamefont {A.}~\bibnamefont
  {Fernondez~Guillermet}}, \bibinfo {author} {\bibfnamefont {J.}~\bibnamefont
  {Haglund}},\ and\ \bibinfo {author} {\bibfnamefont {G.}~\bibnamefont
  {Grimvall}},\ }\href {https://doi.org/10.1103/PhysRevB.45.11557} {\bibfield
  {journal} {\bibinfo  {journal} {Physical Review B}\ }\textbf {\bibinfo
  {volume} {45}},\ \bibinfo {pages} {11557} (\bibinfo {year}
  {1992})}\BibitemShut {NoStop}%
\bibitem [{\citenamefont {Papaconstantopoulos}\ \emph
  {et~al.}(1985)\citenamefont {Papaconstantopoulos}, \citenamefont {Pickett},
  \citenamefont {Klein},\ and\ \citenamefont
  {Boyer}}]{papaconstantopoulos1985}%
  \BibitemOpen
  \bibfield  {author} {\bibinfo {author} {\bibfnamefont {D.~A.}\ \bibnamefont
  {Papaconstantopoulos}}, \bibinfo {author} {\bibfnamefont {W.~E.}\
  \bibnamefont {Pickett}}, \bibinfo {author} {\bibfnamefont {B.}~\bibnamefont
  {Klein}},\ and\ \bibinfo {author} {\bibfnamefont {L.~L.}\ \bibnamefont
  {Boyer}},\ }\href@noop {} {\bibfield  {journal} {\bibinfo  {journal}
  {PHYSICAL REVIEW B}\ }\textbf {\bibinfo {volume} {31}},\ \bibinfo {pages}
  {752} (\bibinfo {year} {1985})}\BibitemShut {NoStop}%
\bibitem [{\citenamefont {Shimizu}\ \emph {et~al.}(1997)\citenamefont
  {Shimizu}, \citenamefont {Shirai},\ and\ \citenamefont
  {Suzuki}}]{Shimizu1997}%
  \BibitemOpen
  \bibfield  {author} {\bibinfo {author} {\bibfnamefont {H.}~\bibnamefont
  {Shimizu}}, \bibinfo {author} {\bibfnamefont {M.}~\bibnamefont {Shirai}},\
  and\ \bibinfo {author} {\bibfnamefont {N.}~\bibnamefont {Suzuki}},\ }\href
  {https://doi.org/10.1016/S0921-4526(97)00223-8} {\bibinfo {title}
  {{Electronic structure and magnetism of transition-metal mononitrides}}}
  (\bibinfo {year} {1997})\BibitemShut {NoStop}%
\bibitem [{\citenamefont {Filippetti}\ \emph {et~al.}(1999)\citenamefont
  {Filippetti}, \citenamefont {Pickett},\ and\ \citenamefont
  {Klein}}]{Filippetti1999}%
  \BibitemOpen
  \bibfield  {author} {\bibinfo {author} {\bibfnamefont {A.}~\bibnamefont
  {Filippetti}}, \bibinfo {author} {\bibfnamefont {W.~E.}\ \bibnamefont
  {Pickett}},\ and\ \bibinfo {author} {\bibfnamefont {B.~M.}\ \bibnamefont
  {Klein}},\ }\href {https://doi.org/10.1103/PhysRevB.59.7043} {\bibfield
  {journal} {\bibinfo  {journal} {Physical Review B - Condensed Matter and
  Materials Physics}\ }\textbf {\bibinfo {volume} {59}},\ \bibinfo {pages}
  {7043} (\bibinfo {year} {1999})},\ \Eprint {https://arxiv.org/abs/9808226}
  {arXiv:9808226 [cond-mat]} \BibitemShut {NoStop}%
\bibitem [{\citenamefont {Stampfl}\ \emph {et~al.}(2001)\citenamefont
  {Stampfl}, \citenamefont {Mannstadt}, \citenamefont {Asahi},\ and\
  \citenamefont {Freeman}}]{Stampfl2001}%
  \BibitemOpen
  \bibfield  {author} {\bibinfo {author} {\bibfnamefont {C.}~\bibnamefont
  {Stampfl}}, \bibinfo {author} {\bibfnamefont {W.}~\bibnamefont {Mannstadt}},
  \bibinfo {author} {\bibfnamefont {R.}~\bibnamefont {Asahi}},\ and\ \bibinfo
  {author} {\bibfnamefont {A.~J.}\ \bibnamefont {Freeman}},\ }\href
  {https://doi.org/10.1103/PhysRevB.63.155106} {\bibfield  {journal} {\bibinfo
  {journal} {Physical Review B - Condensed Matter and Materials Physics}\
  }\textbf {\bibinfo {volume} {63}},\ \bibinfo {pages} {1} (\bibinfo {year}
  {2001})}\BibitemShut {NoStop}%
\bibitem [{\citenamefont {Isaev}\ \emph {et~al.}(2005)\citenamefont {Isaev},
  \citenamefont {Ahuja}, \citenamefont {Simak}, \citenamefont {Lichtenstein},
  \citenamefont {Vekilov}, \citenamefont {Johansson},\ and\ \citenamefont
  {Abrikosov}}]{Isaev2005}%
  \BibitemOpen
  \bibfield  {author} {\bibinfo {author} {\bibfnamefont {E.~I.}\ \bibnamefont
  {Isaev}}, \bibinfo {author} {\bibfnamefont {R.}~\bibnamefont {Ahuja}},
  \bibinfo {author} {\bibfnamefont {S.~I.}\ \bibnamefont {Simak}}, \bibinfo
  {author} {\bibfnamefont {A.~I.}\ \bibnamefont {Lichtenstein}}, \bibinfo
  {author} {\bibfnamefont {Y.~K.}\ \bibnamefont {Vekilov}}, \bibinfo {author}
  {\bibfnamefont {B.}~\bibnamefont {Johansson}},\ and\ \bibinfo {author}
  {\bibfnamefont {I.~A.}\ \bibnamefont {Abrikosov}},\ }\href
  {https://doi.org/10.1103/PhysRevB.72.064515} {\bibfield  {journal} {\bibinfo
  {journal} {Physical Review B - Condensed Matter and Materials Physics}\
  }\textbf {\bibinfo {volume} {72}},\ \bibinfo {pages} {1} (\bibinfo {year}
  {2005})}\BibitemShut {NoStop}%
\bibitem [{\citenamefont {Holec}\ \emph {et~al.}(2012)\citenamefont {Holec},
  \citenamefont {Friak}, \citenamefont {Neugebauer},\ and\ \citenamefont
  {Mayrhofer}}]{Holec2012}%
  \BibitemOpen
  \bibfield  {author} {\bibinfo {author} {\bibfnamefont {D.}~\bibnamefont
  {Holec}}, \bibinfo {author} {\bibfnamefont {M.}~\bibnamefont {Friak}},
  \bibinfo {author} {\bibfnamefont {J.}~\bibnamefont {Neugebauer}},\ and\
  \bibinfo {author} {\bibfnamefont {P.~H.}\ \bibnamefont {Mayrhofer}},\
  }\bibfield  {journal} {\bibinfo  {journal} {Physical Review B - Condensed
  Matter and Materials Physics}\ }\textbf {\bibinfo {volume} {85}},\ \href
  {https://doi.org/10.1103/PhysRevB.85.064101} {10.1103/PhysRevB.85.064101}
  (\bibinfo {year} {2012}),\ \Eprint {https://arxiv.org/abs/1111.2737}
  {arXiv:1111.2737} \BibitemShut {NoStop}%
\bibitem [{\citenamefont {Pierson}(1996)}]{Pierson1996}%
  \BibitemOpen
  \bibfield  {author} {\bibinfo {author} {\bibfnamefont {H.~O.}\ \bibnamefont
  {Pierson}},\ }\href {https://doi.org/10.1016/b978-081551392-6.50007-6}
  {\bibfield  {journal} {\bibinfo  {journal} {Handbook of Refractory Carbides
  and Nitrides}\ ,\ \bibinfo {pages} {100}} (\bibinfo {year}
  {1996})}\BibitemShut {NoStop}%
\bibitem [{\citenamefont {Alam}\ \emph {et~al.}(2017)\citenamefont {Alam},
  \citenamefont {Disseler}, \citenamefont {Ratcliff}, \citenamefont {Borchers},
  \citenamefont {Ponce-P{{e}}rez}, \citenamefont {Cocoletzi}, \citenamefont
  {Takeuchi}, \citenamefont {Foley}, \citenamefont {Richard}, \citenamefont
  {Ingram},\ and\ \citenamefont {Smith}}]{Alam2017}%
  \BibitemOpen
  \bibfield  {author} {\bibinfo {author} {\bibfnamefont {K.}~\bibnamefont
  {Alam}}, \bibinfo {author} {\bibfnamefont {S.~M.}\ \bibnamefont {Disseler}},
  \bibinfo {author} {\bibfnamefont {W.~D.}\ \bibnamefont {Ratcliff}}, \bibinfo
  {author} {\bibfnamefont {J.~A.}\ \bibnamefont {Borchers}}, \bibinfo {author}
  {\bibfnamefont {R.}~\bibnamefont {Ponce-P{{e}}rez}}, \bibinfo {author}
  {\bibfnamefont {G.~H.}\ \bibnamefont {Cocoletzi}}, \bibinfo {author}
  {\bibfnamefont {N.}~\bibnamefont {Takeuchi}}, \bibinfo {author}
  {\bibfnamefont {A.}~\bibnamefont {Foley}}, \bibinfo {author} {\bibfnamefont
  {A.}~\bibnamefont {Richard}}, \bibinfo {author} {\bibfnamefont {D.~C.}\
  \bibnamefont {Ingram}},\ and\ \bibinfo {author} {\bibfnamefont {A.~R.}\
  \bibnamefont {Smith}},\ }\href {https://doi.org/10.1103/PhysRevB.96.104433}
  {\bibfield  {journal} {\bibinfo  {journal} {Physical Review B}\ }\textbf
  {\bibinfo {volume} {96}},\ \bibinfo {pages} {1} (\bibinfo {year} {2017})},\
  \Eprint {https://arxiv.org/abs/1703.03829} {arXiv:1703.03829} \BibitemShut
  {NoStop}%
\bibitem [{\citenamefont {Zhou}\ \emph {et~al.}(2014)\citenamefont {Zhou},
  \citenamefont {K{{o}}rmann}, \citenamefont {Holec}, \citenamefont {Bartosik},
  \citenamefont {Grabowski}, \citenamefont {Neugebauer},\ and\ \citenamefont
  {Mayrhofer}}]{Zhou2014}%
  \BibitemOpen
  \bibfield  {author} {\bibinfo {author} {\bibfnamefont {L.}~\bibnamefont
  {Zhou}}, \bibinfo {author} {\bibfnamefont {F.}~\bibnamefont {K{{o}}rmann}},
  \bibinfo {author} {\bibfnamefont {D.}~\bibnamefont {Holec}}, \bibinfo
  {author} {\bibfnamefont {M.}~\bibnamefont {Bartosik}}, \bibinfo {author}
  {\bibfnamefont {B.}~\bibnamefont {Grabowski}}, \bibinfo {author}
  {\bibfnamefont {J.}~\bibnamefont {Neugebauer}},\ and\ \bibinfo {author}
  {\bibfnamefont {P.~H.}\ \bibnamefont {Mayrhofer}},\ }\href
  {https://doi.org/10.1103/PhysRevB.90.184102} {\bibfield  {journal} {\bibinfo
  {journal} {Physical Review B - Condensed Matter and Materials Physics}\
  }\textbf {\bibinfo {volume} {90}},\ \bibinfo {pages} {1} (\bibinfo {year}
  {2014})}\BibitemShut {NoStop}%
\bibitem [{\citenamefont {Ebad-Allah}\ \emph {et~al.}(2016)\citenamefont
  {Ebad-Allah}, \citenamefont {Kugelmann}, \citenamefont {Rivadulla},\ and\
  \citenamefont {Kuntscher}}]{Ebad-Allah2016}%
  \BibitemOpen
  \bibfield  {author} {\bibinfo {author} {\bibfnamefont {J.}~\bibnamefont
  {Ebad-Allah}}, \bibinfo {author} {\bibfnamefont {B.}~\bibnamefont
  {Kugelmann}}, \bibinfo {author} {\bibfnamefont {F.}~\bibnamefont
  {Rivadulla}},\ and\ \bibinfo {author} {\bibfnamefont {C.~A.}\ \bibnamefont
  {Kuntscher}},\ }\href {https://doi.org/10.1103/PhysRevB.94.195118} {\bibfield
   {journal} {\bibinfo  {journal} {Physical Review B}\ }\textbf {\bibinfo
  {volume} {94}},\ \bibinfo {pages} {18} (\bibinfo {year} {2016})}\BibitemShut
  {NoStop}%
\bibitem [{\citenamefont {Mozafari}\ \emph {et~al.}(2015)\citenamefont
  {Mozafari}, \citenamefont {Alling}, \citenamefont {Steneteg},\ and\
  \citenamefont {Abrikosov}}]{Mozafari2015}%
  \BibitemOpen
  \bibfield  {author} {\bibinfo {author} {\bibfnamefont {E.}~\bibnamefont
  {Mozafari}}, \bibinfo {author} {\bibfnamefont {B.}~\bibnamefont {Alling}},
  \bibinfo {author} {\bibfnamefont {P.}~\bibnamefont {Steneteg}},\ and\
  \bibinfo {author} {\bibfnamefont {I.~A.}\ \bibnamefont {Abrikosov}},\ }\href
  {https://doi.org/10.1103/PhysRevB.91.094101} {\bibfield  {journal} {\bibinfo
  {journal} {Physical Review B - Condensed Matter and Materials Physics}\
  }\textbf {\bibinfo {volume} {91}},\ \bibinfo {pages} {1} (\bibinfo {year}
  {2015})},\ \Eprint {https://arxiv.org/abs/1410.5346} {arXiv:1410.5346}
  \BibitemShut {NoStop}%
\bibitem [{\citenamefont {Rojas}\ and\ \citenamefont
  {Ulloa}(2017)}]{Rojas2017}%
  \BibitemOpen
  \bibfield  {author} {\bibinfo {author} {\bibfnamefont {T.}~\bibnamefont
  {Rojas}}\ and\ \bibinfo {author} {\bibfnamefont {S.~E.}\ \bibnamefont
  {Ulloa}},\ }\href {https://doi.org/10.1103/PhysRevB.96.125203} {\bibfield
  {journal} {\bibinfo  {journal} {Physical Review B}\ }\textbf {\bibinfo
  {volume} {96}},\ \bibinfo {pages} {1} (\bibinfo {year} {2017})}\BibitemShut
  {NoStop}%
\bibitem [{\citenamefont {Rojas}\ and\ \citenamefont
  {Ulloa}(2018)}]{Rojas2018}%
  \BibitemOpen
  \bibfield  {author} {\bibinfo {author} {\bibfnamefont {T.}~\bibnamefont
  {Rojas}}\ and\ \bibinfo {author} {\bibfnamefont {S.~E.}\ \bibnamefont
  {Ulloa}},\ }\href {https://doi.org/10.1103/PhysRevB.98.214111} {\bibfield
  {journal} {\bibinfo  {journal} {Physical Review B}\ }\textbf {\bibinfo
  {volume} {98}},\ \bibinfo {pages} {1} (\bibinfo {year} {2018})}\BibitemShut
  {NoStop}%
\bibitem [{\citenamefont {Quintela}\ \emph {et~al.}(2015)\citenamefont
  {Quintela}, \citenamefont {Podkaminer}, \citenamefont {Luckyanova},
  \citenamefont {Paudel}, \citenamefont {Thies}, \citenamefont {Hillsberry},
  \citenamefont {Tenne}, \citenamefont {Tsymbal}, \citenamefont {Chen},
  \citenamefont {Eom},\ and\ \citenamefont {Rivadulla}}]{Quintela2015}%
  \BibitemOpen
  \bibfield  {author} {\bibinfo {author} {\bibfnamefont {C.~X.}\ \bibnamefont
  {Quintela}}, \bibinfo {author} {\bibfnamefont {J.~P.}\ \bibnamefont
  {Podkaminer}}, \bibinfo {author} {\bibfnamefont {M.~N.}\ \bibnamefont
  {Luckyanova}}, \bibinfo {author} {\bibfnamefont {T.~R.}\ \bibnamefont
  {Paudel}}, \bibinfo {author} {\bibfnamefont {E.~L.}\ \bibnamefont {Thies}},
  \bibinfo {author} {\bibfnamefont {D.~A.}\ \bibnamefont {Hillsberry}},
  \bibinfo {author} {\bibfnamefont {D.~A.}\ \bibnamefont {Tenne}}, \bibinfo
  {author} {\bibfnamefont {E.~Y.}\ \bibnamefont {Tsymbal}}, \bibinfo {author}
  {\bibfnamefont {G.}~\bibnamefont {Chen}}, \bibinfo {author} {\bibfnamefont
  {C.~B.}\ \bibnamefont {Eom}},\ and\ \bibinfo {author} {\bibfnamefont
  {F.}~\bibnamefont {Rivadulla}},\ }\href
  {https://doi.org/10.1002/adma.201500110} {\bibfield  {journal} {\bibinfo
  {journal} {Advanced Materials}\ }\textbf {\bibinfo {volume} {27}},\ \bibinfo
  {pages} {3032} (\bibinfo {year} {2015})}\BibitemShut {NoStop}%
\bibitem [{\citenamefont {Lee}\ \emph {et~al.}(2014)\citenamefont {Lee},
  \citenamefont {Esfarjani}, \citenamefont {Luo}, \citenamefont {Zhou},
  \citenamefont {Tian},\ and\ \citenamefont {Chen}}]{Lee2014}%
  \BibitemOpen
  \bibfield  {author} {\bibinfo {author} {\bibfnamefont {S.}~\bibnamefont
  {Lee}}, \bibinfo {author} {\bibfnamefont {K.}~\bibnamefont {Esfarjani}},
  \bibinfo {author} {\bibfnamefont {T.}~\bibnamefont {Luo}}, \bibinfo {author}
  {\bibfnamefont {J.}~\bibnamefont {Zhou}}, \bibinfo {author} {\bibfnamefont
  {Z.}~\bibnamefont {Tian}},\ and\ \bibinfo {author} {\bibfnamefont
  {G.}~\bibnamefont {Chen}},\ }\href {https://doi.org/10.1038/ncomms4525}
  {\bibfield  {journal} {\bibinfo  {journal} {Nature Communications}\ }\textbf
  {\bibinfo {volume} {5}},\ \bibinfo {pages} {1} (\bibinfo {year}
  {2014})}\BibitemShut {NoStop}%
\bibitem [{\citenamefont {Cherchab}\ \emph {et~al.}(2008)\citenamefont
  {Cherchab}, \citenamefont {Amrani}, \citenamefont {Sekkal}, \citenamefont
  {Ghezali},\ and\ \citenamefont {Talbi}}]{Cherchab2008}%
  \BibitemOpen
  \bibfield  {author} {\bibinfo {author} {\bibfnamefont {Y.}~\bibnamefont
  {Cherchab}}, \bibinfo {author} {\bibfnamefont {B.}~\bibnamefont {Amrani}},
  \bibinfo {author} {\bibfnamefont {N.}~\bibnamefont {Sekkal}}, \bibinfo
  {author} {\bibfnamefont {M.}~\bibnamefont {Ghezali}},\ and\ \bibinfo {author}
  {\bibfnamefont {K.}~\bibnamefont {Talbi}},\ }\href
  {https://doi.org/10.1016/j.physe.2007.08.122} {\bibfield  {journal} {\bibinfo
   {journal} {Physica E: Low-Dimensional Systems and Nanostructures}\ }\textbf
  {\bibinfo {volume} {40}},\ \bibinfo {pages} {606} (\bibinfo {year}
  {2008})}\BibitemShut {NoStop}%
\bibitem [{\citenamefont {P.~Villars}(1985)}]{Villars1985}%
  \BibitemOpen
  \bibfield  {author} {\bibinfo {author} {\bibfnamefont {L.~D.~C.}\
  \bibnamefont {P.~Villars}},\ }\href@noop {} {\emph {\bibinfo {title}
  {Pearsons Handbook of Crystallographic Data for Intermetallic Phases}}},\
  \bibinfo {edition} {1st}\ ed.,\ Vol.\ \bibinfo {volume} {1; 2; 3}\ (\bibinfo
  {publisher} {Asm Intl},\ \bibinfo {year} {1985})\BibitemShut {NoStop}%
\bibitem [{\citenamefont {Morris}\ \emph {et~al.}(1985)\citenamefont {Morris},
  \citenamefont {McMurdie}, \citenamefont {Evans}, \citenamefont {Paretzkin},
  \citenamefont {Parker}, \citenamefont {Wong-Ng}, \citenamefont {Gladhill},\
  and\ \citenamefont {Hubbard}}]{Morris1985}%
  \BibitemOpen
  \bibfield  {author} {\bibinfo {author} {\bibfnamefont {M.~C.}\ \bibnamefont
  {Morris}}, \bibinfo {author} {\bibfnamefont {H.~F.}\ \bibnamefont
  {McMurdie}}, \bibinfo {author} {\bibfnamefont {E.~H.}\ \bibnamefont {Evans}},
  \bibinfo {author} {\bibfnamefont {B.}~\bibnamefont {Paretzkin}}, \bibinfo
  {author} {\bibfnamefont {H.~S.}\ \bibnamefont {Parker}}, \bibinfo {author}
  {\bibfnamefont {W.}~\bibnamefont {Wong-Ng}}, \bibinfo {author} {\bibfnamefont
  {D.~M.}\ \bibnamefont {Gladhill}},\ and\ \bibinfo {author} {\bibfnamefont
  {C.~R.}\ \bibnamefont {Hubbard}},\ }\href
  {https://doi.org/https://doi.org/10.6028/NBS.MONO.25-21} {\bibinfo {title}
  {Standard x-ray diffraction powder patterns ::section 21- data for 92
  substances}} (\bibinfo {year} {1985})\BibitemShut {NoStop}%
\bibitem [{\citenamefont {Rovere}\ \emph {et~al.}(2010)\citenamefont {Rovere},
  \citenamefont {Music}, \citenamefont {Schneider},\ and\ \citenamefont
  {Mayrhofer}}]{Rovere2010}%
  \BibitemOpen
  \bibfield  {author} {\bibinfo {author} {\bibfnamefont {F.}~\bibnamefont
  {Rovere}}, \bibinfo {author} {\bibfnamefont {D.}~\bibnamefont {Music}},
  \bibinfo {author} {\bibfnamefont {J.~M.}\ \bibnamefont {Schneider}},\ and\
  \bibinfo {author} {\bibfnamefont {P.~H.}\ \bibnamefont {Mayrhofer}},\ }\href
  {https://doi.org/10.1016/j.actamat.2010.01.005} {\bibfield  {journal}
  {\bibinfo  {journal} {Acta Materialia}\ }\textbf {\bibinfo {volume} {58}},\
  \bibinfo {pages} {2708} (\bibinfo {year} {2010})}\BibitemShut {NoStop}%
\bibitem [{\citenamefont {Zhou}\ \emph {et~al.}(2013)\citenamefont {Zhou},
  \citenamefont {Holec},\ and\ \citenamefont {Mayrhofer}}]{Zhou2013}%
  \BibitemOpen
  \bibfield  {author} {\bibinfo {author} {\bibfnamefont {L.}~\bibnamefont
  {Zhou}}, \bibinfo {author} {\bibfnamefont {D.}~\bibnamefont {Holec}},\ and\
  \bibinfo {author} {\bibfnamefont {P.~H.}\ \bibnamefont {Mayrhofer}},\
  }\bibfield  {journal} {\bibinfo  {journal} {Journal of Physics D: Applied
  Physics}\ }\textbf {\bibinfo {volume} {46}},\ \href
  {https://doi.org/10.1088/0022-3727/46/36/365301}
  {10.1088/0022-3727/46/36/365301} (\bibinfo {year} {2013})\BibitemShut
  {NoStop}%
\bibitem [{\citenamefont {Salguero}\ \emph {et~al.}(2003)\citenamefont
  {Salguero}, \citenamefont {Mancera}, \citenamefont {Rodr{{i}}guez},\ and\
  \citenamefont {Takeuchi}}]{Salguero2003}%
  \BibitemOpen
  \bibfield  {author} {\bibinfo {author} {\bibfnamefont {L.~A.}\ \bibnamefont
  {Salguero}}, \bibinfo {author} {\bibfnamefont {L.}~\bibnamefont {Mancera}},
  \bibinfo {author} {\bibfnamefont {J.~A.}\ \bibnamefont {Rodr{{i}}guez}},\
  and\ \bibinfo {author} {\bibfnamefont {N.}~\bibnamefont {Takeuchi}},\ }\href
  {https://doi.org/10.1002/pssb.200541013} {\bibfield  {journal} {\bibinfo
  {journal} {Physica Status Solidi (B) Basic Research}\ }\textbf {\bibinfo
  {volume} {243}},\ \bibinfo {pages} {1808} (\bibinfo {year}
  {2003})}\BibitemShut {NoStop}%
\bibitem [{\citenamefont {Louhadj}\ \emph {et~al.}(2009)\citenamefont
  {Louhadj}, \citenamefont {Ghezali}, \citenamefont {Badi}, \citenamefont
  {Mehnane}, \citenamefont {Cherchab}, \citenamefont {Amrani}, \citenamefont
  {Abid},\ and\ \citenamefont {Sekkal}}]{Louhadj2009}%
  \BibitemOpen
  \bibfield  {author} {\bibinfo {author} {\bibfnamefont {A.}~\bibnamefont
  {Louhadj}}, \bibinfo {author} {\bibfnamefont {M.}~\bibnamefont {Ghezali}},
  \bibinfo {author} {\bibfnamefont {F.}~\bibnamefont {Badi}}, \bibinfo {author}
  {\bibfnamefont {N.}~\bibnamefont {Mehnane}}, \bibinfo {author} {\bibfnamefont
  {Y.}~\bibnamefont {Cherchab}}, \bibinfo {author} {\bibfnamefont
  {B.}~\bibnamefont {Amrani}}, \bibinfo {author} {\bibfnamefont
  {H.}~\bibnamefont {Abid}},\ and\ \bibinfo {author} {\bibfnamefont
  {N.}~\bibnamefont {Sekkal}},\ }\href
  {https://doi.org/10.1016/j.spmi.2009.05.001} {\bibfield  {journal} {\bibinfo
  {journal} {Superlattices and Microstructures}\ }\textbf {\bibinfo {volume}
  {46}},\ \bibinfo {pages} {435} (\bibinfo {year} {2009})}\BibitemShut
  {NoStop}%
\bibitem [{\citenamefont {Kerdsongpanya}\ \emph {et~al.}(2016)\citenamefont
  {Kerdsongpanya}, \citenamefont {Sun}, \citenamefont {Eriksson}, \citenamefont
  {Jensen}, \citenamefont {Lu}, \citenamefont {Koh}, \citenamefont {Nong},
  \citenamefont {Balke}, \citenamefont {Alling},\ and\ \citenamefont
  {Eklund}}]{Kerdsongpanya2016}%
  \BibitemOpen
  \bibfield  {author} {\bibinfo {author} {\bibfnamefont {S.}~\bibnamefont
  {Kerdsongpanya}}, \bibinfo {author} {\bibfnamefont {B.}~\bibnamefont {Sun}},
  \bibinfo {author} {\bibfnamefont {F.}~\bibnamefont {Eriksson}}, \bibinfo
  {author} {\bibfnamefont {J.}~\bibnamefont {Jensen}}, \bibinfo {author}
  {\bibfnamefont {J.}~\bibnamefont {Lu}}, \bibinfo {author} {\bibfnamefont
  {Y.~K.}\ \bibnamefont {Koh}}, \bibinfo {author} {\bibfnamefont {N.~V.}\
  \bibnamefont {Nong}}, \bibinfo {author} {\bibfnamefont {B.}~\bibnamefont
  {Balke}}, \bibinfo {author} {\bibfnamefont {B.}~\bibnamefont {Alling}},\ and\
  \bibinfo {author} {\bibfnamefont {P.}~\bibnamefont {Eklund}},\ }\bibfield
  {journal} {\bibinfo  {journal} {Journal of Applied Physics}\ }\textbf
  {\bibinfo {volume} {120}},\ \href {https://doi.org/10.1063/1.4968570}
  {10.1063/1.4968570} (\bibinfo {year} {2016})\BibitemShut {NoStop}%
\bibitem [{\citenamefont {Sukkabot}(2019)}]{Sukkabot2019}%
  \BibitemOpen
  \bibfield  {author} {\bibinfo {author} {\bibfnamefont {W.}~\bibnamefont
  {Sukkabot}},\ }\href {https://doi.org/10.1016/j.physb.2019.06.043} {\bibfield
   {journal} {\bibinfo  {journal} {Physica B: Condensed Matter}\ }\textbf
  {\bibinfo {volume} {570}},\ \bibinfo {pages} {236} (\bibinfo {year}
  {2019})}\BibitemShut {NoStop}%
\bibitem [{\citenamefont {Felipe}(2021)}]{unal_85410}%
  \BibitemOpen
  \bibfield  {author} {\bibinfo {author} {\bibfnamefont {L.~P.~L.}\
  \bibnamefont {Felipe}},\ }\href
  {https://repositorio.unal.edu.co/handle/unal/85410} {\bibinfo {title}
  {Estudio de las propiedades electr{\\'{o}}nicas y magn{\\'{e}}ticas del crn
  dopado con y por medio de la teor{\\'{i}}a del funcional de la densidad}}
  (\bibinfo {year} {2021})\BibitemShut {NoStop}%
\bibitem [{\citenamefont {Kresse}\ \emph {et~al.}(2021)\citenamefont {Kresse},
  \citenamefont {Vogtenhuber}, \citenamefont {Marsman}, \citenamefont {Kaltak},
  \citenamefont {Karsai},\ and\ \citenamefont {Schlipf}}]{VASP}%
  \BibitemOpen
  \bibfield  {author} {\bibinfo {author} {\bibfnamefont {G.}~\bibnamefont
  {Kresse}}, \bibinfo {author} {\bibfnamefont {D.}~\bibnamefont {Vogtenhuber}},
  \bibinfo {author} {\bibfnamefont {M.}~\bibnamefont {Marsman}}, \bibinfo
  {author} {\bibfnamefont {M.}~\bibnamefont {Kaltak}}, \bibinfo {author}
  {\bibfnamefont {F.}~\bibnamefont {Karsai}},\ and\ \bibinfo {author}
  {\bibfnamefont {M.}~\bibnamefont {Schlipf}},\ }\href {https://www.vasp.at/}
  {\bibinfo {title} {{Vienna Ab initio Simulation Package} (vasp)}} (\bibinfo
  {year} {2021})\BibitemShut {NoStop}%
\bibitem [{\citenamefont {Bl{{o}}chl}(1994)}]{Blochl1994}%
  \BibitemOpen
  \bibfield  {author} {\bibinfo {author} {\bibfnamefont {P.~E.}\ \bibnamefont
  {Bl{{o}}chl}},\ }\href {https://doi.org/10.1103/PhysRevB.50.17953} {\bibfield
   {journal} {\bibinfo  {journal} {Physical Review B}\ }\textbf {\bibinfo
  {volume} {50}},\ \bibinfo {pages} {17953} (\bibinfo {year}
  {1994})}\BibitemShut {NoStop}%
\bibitem [{\citenamefont {Sun}\ \emph {et~al.}(2015)\citenamefont {Sun},
  \citenamefont {Ruzsinszky},\ and\ \citenamefont {Perdew}}]{Sun2015}%
  \BibitemOpen
  \bibfield  {author} {\bibinfo {author} {\bibfnamefont {J.}~\bibnamefont
  {Sun}}, \bibinfo {author} {\bibfnamefont {A.}~\bibnamefont {Ruzsinszky}},\
  and\ \bibinfo {author} {\bibfnamefont {J.~P.}\ \bibnamefont {Perdew}},\
  }\href {https://doi.org/10.1103/PhysRevLett.115.036402} {\bibfield  {journal}
  {\bibinfo  {journal} {Phys. Rev. Lett.}\ }\textbf {\bibinfo {volume} {115}},\
  \bibinfo {pages} {036402} (\bibinfo {year} {2015})}\BibitemShut {NoStop}%
\bibitem [{\citenamefont {Sun}\ \emph {et~al.}(2016)\citenamefont {Sun},
  \citenamefont {Remsing}, \citenamefont {Zhang}, \citenamefont {Sun},
  \citenamefont {Ruzsinszky}, \citenamefont {Peng}, \citenamefont {Yang},
  \citenamefont {Paul}, \citenamefont {Waghmare}, \citenamefont {Wu},
  \citenamefont {Klein},\ and\ \citenamefont {Perdew}}]{Sun2016}%
  \BibitemOpen
  \bibfield  {author} {\bibinfo {author} {\bibfnamefont {J.}~\bibnamefont
  {Sun}}, \bibinfo {author} {\bibfnamefont {R.~C.}\ \bibnamefont {Remsing}},
  \bibinfo {author} {\bibfnamefont {Y.}~\bibnamefont {Zhang}}, \bibinfo
  {author} {\bibfnamefont {Z.}~\bibnamefont {Sun}}, \bibinfo {author}
  {\bibfnamefont {A.}~\bibnamefont {Ruzsinszky}}, \bibinfo {author}
  {\bibfnamefont {H.}~\bibnamefont {Peng}}, \bibinfo {author} {\bibfnamefont
  {Z.}~\bibnamefont {Yang}}, \bibinfo {author} {\bibfnamefont {A.}~\bibnamefont
  {Paul}}, \bibinfo {author} {\bibfnamefont {U.}~\bibnamefont {Waghmare}},
  \bibinfo {author} {\bibfnamefont {X.}~\bibnamefont {Wu}}, \bibinfo {author}
  {\bibfnamefont {M.~L.}\ \bibnamefont {Klein}},\ and\ \bibinfo {author}
  {\bibfnamefont {J.~P.}\ \bibnamefont {Perdew}},\ }\href
  {https://doi.org/10.1038/nchem.2535} {\bibfield  {journal} {\bibinfo
  {journal} {Nature Chemistry}\ }\textbf {\bibinfo {volume} {8}},\ \bibinfo
  {pages} {831} (\bibinfo {year} {2016})}\BibitemShut {NoStop}%
\bibitem [{\citenamefont {Cheiwchanchamnangij}\ and\ \citenamefont
  {Lambrecht}(2020)}]{Cheiwchanchamnangij2020}%
  \BibitemOpen
  \bibfield  {author} {\bibinfo {author} {\bibfnamefont {T.}~\bibnamefont
  {Cheiwchanchamnangij}}\ and\ \bibinfo {author} {\bibfnamefont {W.~R.}\
  \bibnamefont {Lambrecht}},\ }\href
  {https://doi.org/10.1103/PhysRevB.101.085103} {\bibfield  {journal} {\bibinfo
   {journal} {Physical Review B}\ }\textbf {\bibinfo {volume} {101}},\ \bibinfo
  {pages} {1} (\bibinfo {year} {2020})}\BibitemShut {NoStop}%
\bibitem [{\citenamefont {Herwadkar}\ and\ \citenamefont
  {Lambrecht}(2009)}]{Herwadkar2009}%
  \BibitemOpen
  \bibfield  {author} {\bibinfo {author} {\bibfnamefont {A.}~\bibnamefont
  {Herwadkar}}\ and\ \bibinfo {author} {\bibfnamefont {W.~R.}\ \bibnamefont
  {Lambrecht}},\ }\href {https://doi.org/10.1103/PhysRevB.79.035125} {\bibfield
   {journal} {\bibinfo  {journal} {Physical Review B - Condensed Matter and
  Materials Physics}\ }\textbf {\bibinfo {volume} {79}},\ \bibinfo {pages} {1}
  (\bibinfo {year} {2009})}\BibitemShut {NoStop}%
\bibitem [{\citenamefont {Corliss}\ \emph {et~al.}(1960)\citenamefont
  {Corliss}, \citenamefont {Elliott},\ and\ \citenamefont
  {Hastings}}]{Corliss1960}%
  \BibitemOpen
  \bibfield  {author} {\bibinfo {author} {\bibfnamefont {L.~M.}\ \bibnamefont
  {Corliss}}, \bibinfo {author} {\bibfnamefont {N.}~\bibnamefont {Elliott}},\
  and\ \bibinfo {author} {\bibfnamefont {J.~M.}\ \bibnamefont {Hastings}},\
  }\href {https://doi.org/10.1103/PhysRev.117.929} {\bibfield  {journal}
  {\bibinfo  {journal} {Physical Review}\ }\textbf {\bibinfo {volume} {117}},\
  \bibinfo {pages} {929} (\bibinfo {year} {1960})}\BibitemShut {NoStop}%
\bibitem [{\citenamefont {Zunger}\ \emph {et~al.}(1990)\citenamefont {Zunger},
  \citenamefont {Wei}, \citenamefont {Ferreira},\ and\ \citenamefont
  {Bernard}}]{Zunger1990}%
  \BibitemOpen
  \bibfield  {author} {\bibinfo {author} {\bibfnamefont {A.}~\bibnamefont
  {Zunger}}, \bibinfo {author} {\bibfnamefont {S.~H.}\ \bibnamefont {Wei}},
  \bibinfo {author} {\bibfnamefont {L.~G.}\ \bibnamefont {Ferreira}},\ and\
  \bibinfo {author} {\bibfnamefont {J.~E.}\ \bibnamefont {Bernard}},\ }\href
  {https://doi.org/10.1103/PhysRevLett.65.353} {\bibfield  {journal} {\bibinfo
  {journal} {Physical Review Letters}\ }\textbf {\bibinfo {volume} {65}},\
  \bibinfo {pages} {353} (\bibinfo {year} {1990})}\BibitemShut {NoStop}%
\bibitem [{\citenamefont {van~de Walle}\ \emph {et~al.}(2002)\citenamefont
  {van~de Walle}, \citenamefont {Asta},\ and\ \citenamefont {Ceder}}]{atat}%
  \BibitemOpen
  \bibfield  {author} {\bibinfo {author} {\bibfnamefont {A.}~\bibnamefont
  {van~de Walle}}, \bibinfo {author} {\bibfnamefont {M.~D.}\ \bibnamefont
  {Asta}},\ and\ \bibinfo {author} {\bibfnamefont {G.}~\bibnamefont {Ceder}},\
  }\href {https://doi.org/10.1016/S0364-5916(02)80006-2} {\bibfield  {journal}
  {\bibinfo  {journal} {Calphad}\ }\textbf {\bibinfo {volume} {26}},\ \bibinfo
  {pages} {539} (\bibinfo {year} {2002})}\BibitemShut {NoStop}%
\bibitem [{\citenamefont {Zhu}\ \emph {et~al.}(2024)\citenamefont {Zhu},
  \citenamefont {Kavanagh},\ and\ \citenamefont {Scanlon}}]{Zhu2024}%
  \BibitemOpen
  \bibfield  {author} {\bibinfo {author} {\bibfnamefont {B.}~\bibnamefont
  {Zhu}}, \bibinfo {author} {\bibfnamefont {S.~R.}\ \bibnamefont {Kavanagh}},\
  and\ \bibinfo {author} {\bibfnamefont {D.}~\bibnamefont {Scanlon}},\ }\href
  {https://doi.org/10.21105/joss.05974} {\bibfield  {journal} {\bibinfo
  {journal} {Journal of Open Source Software}\ }\textbf {\bibinfo {volume}
  {9}},\ \bibinfo {pages} {5974} (\bibinfo {year} {2024})}\BibitemShut
  {NoStop}%
\bibitem [{\citenamefont {Madsen}\ \emph {et~al.}(2018)\citenamefont {Madsen},
  \citenamefont {Carrete},\ and\ \citenamefont {Verstraete}}]{BoltzTraP2}%
  \BibitemOpen
  \bibfield  {author} {\bibinfo {author} {\bibfnamefont {G.~K.~H.}\
  \bibnamefont {Madsen}}, \bibinfo {author} {\bibfnamefont {J.}~\bibnamefont
  {Carrete}},\ and\ \bibinfo {author} {\bibfnamefont {M.~J.}\ \bibnamefont
  {Verstraete}},\ }\href {https://doi.org/10.1016/j.cpc.2018.05.010} {\bibfield
   {journal} {\bibinfo  {journal} {Comput. Phys. Commun.}\ }\textbf {\bibinfo
  {volume} {231}},\ \bibinfo {pages} {140 } (\bibinfo {year}
  {2018})}\BibitemShut {NoStop}%
\bibitem [{\citenamefont {Matas}\ \emph {et~al.}(2025)\citenamefont {Matas},
  \citenamefont {Mayrhofer},\ and\ \citenamefont {Holec}}]{Matas2025}%
  \BibitemOpen
  \bibfield  {author} {\bibinfo {author} {\bibfnamefont {M.}~\bibnamefont
  {Matas}}, \bibinfo {author} {\bibfnamefont {P.~H.}\ \bibnamefont
  {Mayrhofer}},\ and\ \bibinfo {author} {\bibfnamefont {D.}~\bibnamefont
  {Holec}},\ }\href {https://doi.org/10.1016/j.surfcoat.2024.131540} {\bibfield
   {journal} {\bibinfo  {journal} {Surface and Coatings Technology}\ }\textbf
  {\bibinfo {volume} {496}},\ \bibinfo {pages} {131540} (\bibinfo {year}
  {2025})}\BibitemShut {NoStop}%
\bibitem [{\citenamefont {Shulumba}\ \emph {et~al.}(2014)\citenamefont
  {Shulumba}, \citenamefont {Alling}, \citenamefont {Hellman}, \citenamefont
  {Mozafari}, \citenamefont {Steneteg}, \citenamefont {Od{{e}}n},\ and\
  \citenamefont {Abrikosov}}]{Shulumba2014}%
  \BibitemOpen
  \bibfield  {author} {\bibinfo {author} {\bibfnamefont {N.}~\bibnamefont
  {Shulumba}}, \bibinfo {author} {\bibfnamefont {B.}~\bibnamefont {Alling}},
  \bibinfo {author} {\bibfnamefont {O.}~\bibnamefont {Hellman}}, \bibinfo
  {author} {\bibfnamefont {E.}~\bibnamefont {Mozafari}}, \bibinfo {author}
  {\bibfnamefont {P.}~\bibnamefont {Steneteg}}, \bibinfo {author}
  {\bibfnamefont {M.}~\bibnamefont {Od{{e}}n}},\ and\ \bibinfo {author}
  {\bibfnamefont {I.~A.}\ \bibnamefont {Abrikosov}},\ }\href
  {https://doi.org/10.1103/PhysRevB.89.174108} {\bibfield  {journal} {\bibinfo
  {journal} {Physical Review B - Condensed Matter and Materials Physics}\
  }\textbf {\bibinfo {volume} {89}},\ \bibinfo {pages} {1} (\bibinfo {year}
  {2014})},\ \Eprint {https://arxiv.org/abs/1403.4766} {arXiv:1403.4766}
  \BibitemShut {NoStop}%
\bibitem [{\citenamefont {Siegel}\ \emph {et~al.}(2003)\citenamefont {Siegel},
  \citenamefont {Hector},\ and\ \citenamefont {Adams}}]{Siegel2003}%
  \BibitemOpen
  \bibfield  {author} {\bibinfo {author} {\bibfnamefont {D.~J.}\ \bibnamefont
  {Siegel}}, \bibinfo {author} {\bibfnamefont {L.~G.}\ \bibnamefont {Hector}},\
  and\ \bibinfo {author} {\bibfnamefont {J.~B.}\ \bibnamefont {Adams}},\ }\href
  {https://doi.org/10.1103/PhysRevB.67.092105} {\bibfield  {journal} {\bibinfo
  {journal} {Physical Review B - Condensed Matter and Materials Physics}\
  }\textbf {\bibinfo {volume} {67}},\ \bibinfo {pages} {1} (\bibinfo {year}
  {2003})}\BibitemShut {NoStop}%
\bibitem [{\citenamefont {Zhang}\ \emph {et~al.}(2013)\citenamefont {Zhang},
  \citenamefont {Li}, \citenamefont {Daniel}, \citenamefont {Mitterer},\ and\
  \citenamefont {Dehm}}]{Zhang2013}%
  \BibitemOpen
  \bibfield  {author} {\bibinfo {author} {\bibfnamefont {Z.}~\bibnamefont
  {Zhang}}, \bibinfo {author} {\bibfnamefont {H.}~\bibnamefont {Li}}, \bibinfo
  {author} {\bibfnamefont {R.}~\bibnamefont {Daniel}}, \bibinfo {author}
  {\bibfnamefont {C.}~\bibnamefont {Mitterer}},\ and\ \bibinfo {author}
  {\bibfnamefont {G.}~\bibnamefont {Dehm}},\ }\href
  {https://doi.org/10.1103/PhysRevB.87.014104} {\bibfield  {journal} {\bibinfo
  {journal} {Physical Review B - Condensed Matter and Materials Physics}\
  }\textbf {\bibinfo {volume} {87}},\ \bibinfo {pages} {1} (\bibinfo {year}
  {2013})}\BibitemShut {NoStop}%
\bibitem [{\citenamefont {Alling}(2010)}]{Alling2010a}%
  \BibitemOpen
  \bibfield  {author} {\bibinfo {author} {\bibfnamefont {B.}~\bibnamefont
  {Alling}},\ }\href {https://doi.org/10.1103/PhysRevB.82.054408} {\bibfield
  {journal} {\bibinfo  {journal} {Physical Review B - Condensed Matter and
  Materials Physics}\ }\textbf {\bibinfo {volume} {82}},\ \bibinfo {pages} {1}
  (\bibinfo {year} {2010})}\BibitemShut {NoStop}%
\bibitem [{\citenamefont {Alling}\ \emph {et~al.}(2007)\citenamefont {Alling},
  \citenamefont {Marten}, \citenamefont {Abrikosov},\ and\ \citenamefont
  {Karimi}}]{Alling2007}%
  \BibitemOpen
  \bibfield  {author} {\bibinfo {author} {\bibfnamefont {B.}~\bibnamefont
  {Alling}}, \bibinfo {author} {\bibfnamefont {T.}~\bibnamefont {Marten}},
  \bibinfo {author} {\bibfnamefont {I.~A.}\ \bibnamefont {Abrikosov}},\ and\
  \bibinfo {author} {\bibfnamefont {A.}~\bibnamefont {Karimi}},\ }\bibfield
  {journal} {\bibinfo  {journal} {Journal of Applied Physics}\ }\textbf
  {\bibinfo {volume} {102}},\ \href {https://doi.org/10.1063/1.2773625}
  {10.1063/1.2773625} (\bibinfo {year} {2007})\BibitemShut {NoStop}%
\bibitem [{\citenamefont {Amrani}\ and\ \citenamefont {{El Haj
  Hassan}}(2007)}]{Amrani2007}%
  \BibitemOpen
  \bibfield  {author} {\bibinfo {author} {\bibfnamefont {B.}~\bibnamefont
  {Amrani}}\ and\ \bibinfo {author} {\bibfnamefont {F.}~\bibnamefont {{El Haj
  Hassan}}},\ }\href {https://doi.org/10.1016/j.commatsci.2006.08.009}
  {\bibfield  {journal} {\bibinfo  {journal} {Computational Materials Science}\
  }\textbf {\bibinfo {volume} {39}},\ \bibinfo {pages} {563} (\bibinfo {year}
  {2007})}\BibitemShut {NoStop}%
\bibitem [{\citenamefont {Alling}\ \emph {et~al.}(2010)\citenamefont {Alling},
  \citenamefont {Marten},\ and\ \citenamefont {Abrikosov}}]{Alling2010}%
  \BibitemOpen
  \bibfield  {author} {\bibinfo {author} {\bibfnamefont {B.}~\bibnamefont
  {Alling}}, \bibinfo {author} {\bibfnamefont {T.}~\bibnamefont {Marten}},\
  and\ \bibinfo {author} {\bibfnamefont {I.~A.}\ \bibnamefont {Abrikosov}},\
  }\href {https://doi.org/10.1103/PhysRevB.82.184430} {\bibfield  {journal}
  {\bibinfo  {journal} {Physical Review B - Condensed Matter and Materials
  Physics}\ }\textbf {\bibinfo {volume} {82}},\ \bibinfo {pages} {1} (\bibinfo
  {year} {2010})},\ \Eprint {https://arxiv.org/abs/1006.3460} {arXiv:1006.3460}
  \BibitemShut {NoStop}%
\bibitem [{\citenamefont {Gall}\ \emph {et~al.}(2002)\citenamefont {Gall},
  \citenamefont {Shin}, \citenamefont {Haasch}, \citenamefont {Petrov},\ and\
  \citenamefont {Greene}}]{Gall2002}%
  \BibitemOpen
  \bibfield  {author} {\bibinfo {author} {\bibfnamefont {D.}~\bibnamefont
  {Gall}}, \bibinfo {author} {\bibfnamefont {C.~S.}\ \bibnamefont {Shin}},
  \bibinfo {author} {\bibfnamefont {R.~T.}\ \bibnamefont {Haasch}}, \bibinfo
  {author} {\bibfnamefont {I.}~\bibnamefont {Petrov}},\ and\ \bibinfo {author}
  {\bibfnamefont {J.~E.}\ \bibnamefont {Greene}},\ }\href
  {https://doi.org/10.1063/1.1466528} {\bibfield  {journal} {\bibinfo
  {journal} {Journal of Applied Physics}\ }\textbf {\bibinfo {volume} {91}},\
  \bibinfo {pages} {5882} (\bibinfo {year} {2002})}\BibitemShut {NoStop}%
\bibitem [{\citenamefont {Rivadulla}\ \emph {et~al.}(2009)\citenamefont
  {Rivadulla}, \citenamefont {B{{a}}obre-L{{o}}pez}, \citenamefont {Quintela},
  \citenamefont {Peiro}, \citenamefont {Pardo}, \citenamefont {Baldomir},
  \citenamefont {L{{o}}pez-Quintela}, \citenamefont {Rivas}, \citenamefont
  {Ramos}, \citenamefont {Salva}, \citenamefont {Zhou},\ and\ \citenamefont
  {Goodenough}}]{Rivadulla2009}%
  \BibitemOpen
  \bibfield  {author} {\bibinfo {author} {\bibfnamefont {F.}~\bibnamefont
  {Rivadulla}}, \bibinfo {author} {\bibfnamefont {M.}~\bibnamefont
  {B{{a}}obre-L{{o}}pez}}, \bibinfo {author} {\bibfnamefont {C.~X.}\
  \bibnamefont {Quintela}}, \bibinfo {author} {\bibfnamefont {A.}~\bibnamefont
  {Peiro}}, \bibinfo {author} {\bibfnamefont {V.}~\bibnamefont {Pardo}},
  \bibinfo {author} {\bibfnamefont {D.}~\bibnamefont {Baldomir}}, \bibinfo
  {author} {\bibfnamefont {M.~A.}\ \bibnamefont {L{{o}}pez-Quintela}}, \bibinfo
  {author} {\bibfnamefont {J.}~\bibnamefont {Rivas}}, \bibinfo {author}
  {\bibfnamefont {C.~A.}\ \bibnamefont {Ramos}}, \bibinfo {author}
  {\bibfnamefont {H.}~\bibnamefont {Salva}}, \bibinfo {author} {\bibfnamefont
  {J.~S.}\ \bibnamefont {Zhou}},\ and\ \bibinfo {author} {\bibfnamefont
  {J.~B.}\ \bibnamefont {Goodenough}},\ }\href
  {https://doi.org/10.1038/nmat2549} {\bibfield  {journal} {\bibinfo  {journal}
  {Nature Materials}\ }\textbf {\bibinfo {volume} {8}},\ \bibinfo {pages} {947}
  (\bibinfo {year} {2009})}\BibitemShut {NoStop}%
\bibitem [{\citenamefont {Vegard}(1921)}]{Vegard1921}%
  \BibitemOpen
  \bibfield  {author} {\bibinfo {author} {\bibfnamefont {L.}~\bibnamefont
  {Vegard}},\ }\href {https://doi.org/10.1007/BF01349680} {\bibfield  {journal}
  {\bibinfo  {journal} {Zeitschrift fr Physik}\ }\textbf {\bibinfo {volume}
  {5}},\ \bibinfo {pages} {17} (\bibinfo {year} {1921})}\BibitemShut {NoStop}%
\bibitem [{\citenamefont {{H.A. Jahn and E.
  Teller}}(1937)}]{H.A.JahnandE.Teller1937}%
  \BibitemOpen
  \bibfield  {author} {\bibinfo {author} {\bibnamefont {{H.A. Jahn and E.
  Teller}}},\ }\href {https://doi.org/10.1098/rspa.1937.0142} {\bibfield
  {journal} {\bibinfo  {journal} {Proceedings of the Royal Society of London.
  Series A - Mathematical and Physical Sciences}\ }\textbf {\bibinfo {volume}
  {161}},\ \bibinfo {pages} {220} (\bibinfo {year} {1937})}\BibitemShut
  {NoStop}%
\bibitem [{\citenamefont {Filippetti}\ and\ \citenamefont
  {Hill}(2000)}]{Filippetti2000}%
  \BibitemOpen
  \bibfield  {author} {\bibinfo {author} {\bibfnamefont {A.}~\bibnamefont
  {Filippetti}}\ and\ \bibinfo {author} {\bibfnamefont {N.~A.}\ \bibnamefont
  {Hill}},\ }\href {https://doi.org/10.1103/PhysRevLett.85.5166} {\bibfield
  {journal} {\bibinfo  {journal} {Physical Review Letters}\ }\textbf {\bibinfo
  {volume} {85}},\ \bibinfo {pages} {5166} (\bibinfo {year} {2000})},\ \Eprint
  {https://arxiv.org/abs/0004252} {arXiv:0004252 [cond-mat]} \BibitemShut
  {NoStop}%
\bibitem [{\citenamefont {Tran}\ \emph {et~al.}(2020)\citenamefont {Tran},
  \citenamefont {Baudesson}, \citenamefont {Carrete}, \citenamefont {Madsen},
  \citenamefont {Blaha}, \citenamefont {Schwarz},\ and\ \citenamefont
  {Singh}}]{Tran2020}%
  \BibitemOpen
  \bibfield  {author} {\bibinfo {author} {\bibfnamefont {F.}~\bibnamefont
  {Tran}}, \bibinfo {author} {\bibfnamefont {G.}~\bibnamefont {Baudesson}},
  \bibinfo {author} {\bibfnamefont {J.}~\bibnamefont {Carrete}}, \bibinfo
  {author} {\bibfnamefont {G.~K.}\ \bibnamefont {Madsen}}, \bibinfo {author}
  {\bibfnamefont {P.}~\bibnamefont {Blaha}}, \bibinfo {author} {\bibfnamefont
  {K.}~\bibnamefont {Schwarz}},\ and\ \bibinfo {author} {\bibfnamefont {D.~J.}\
  \bibnamefont {Singh}},\ }\href {https://doi.org/10.1103/PhysRevB.102.024407}
  {\bibfield  {journal} {\bibinfo  {journal} {Physical Review B}\ }\textbf
  {\bibinfo {volume} {102}},\ \bibinfo {pages} {1} (\bibinfo {year} {2020})},\
  \Eprint {https://arxiv.org/abs/2004.04543} {arXiv:2004.04543} \BibitemShut
  {NoStop}%
\bibitem [{\citenamefont {ukauskait}\ \emph {et~al.}(2012)\citenamefont
  {ukauskait}, \citenamefont {Tholander}, \citenamefont {Palisaitis},
  \citenamefont {Persson}, \citenamefont {Darakchieva}, \citenamefont
  {Sedrine}, \citenamefont {Tasn{\\'{a}}di}, \citenamefont {Alling},
  \citenamefont {Birch},\ and\ \citenamefont {Hultman}}]{Zaukauskaite2021}%
  \BibitemOpen
  \bibfield  {author} {\bibinfo {author} {\bibfnamefont {A.}~\bibnamefont
  {ukauskait}}, \bibinfo {author} {\bibfnamefont {C.}~\bibnamefont
  {Tholander}}, \bibinfo {author} {\bibfnamefont {J.}~\bibnamefont
  {Palisaitis}}, \bibinfo {author} {\bibfnamefont {P.~O.}\ \bibnamefont
  {Persson}}, \bibinfo {author} {\bibfnamefont {V.}~\bibnamefont
  {Darakchieva}}, \bibinfo {author} {\bibfnamefont {N.~B.}\ \bibnamefont
  {Sedrine}}, \bibinfo {author} {\bibfnamefont {F.}~\bibnamefont
  {Tasn{\\'{a}}di}}, \bibinfo {author} {\bibfnamefont {B.}~\bibnamefont
  {Alling}}, \bibinfo {author} {\bibfnamefont {J.}~\bibnamefont {Birch}},\ and\
  \bibinfo {author} {\bibfnamefont {L.}~\bibnamefont {Hultman}},\ }\href
  {https://doi.org/10.1088/0022-3727/45/42/422001} {\bibfield  {journal}
  {\bibinfo  {journal} {Journal of Physics D: Applied Physics}\ }\textbf
  {\bibinfo {volume} {45}},\ \bibinfo {pages} {422001} (\bibinfo {year}
  {2012})}\BibitemShut {NoStop}%
\bibitem [{\citenamefont {Redlich}\ and\ \citenamefont
  {Kister}(1948)}]{Redlich1948}%
  \BibitemOpen
  \bibfield  {author} {\bibinfo {author} {\bibfnamefont {O.}~\bibnamefont
  {Redlich}}\ and\ \bibinfo {author} {\bibfnamefont {A.~T.}\ \bibnamefont
  {Kister}},\ }\href {https://doi.org/10.1021/ie50458a036} {\bibfield
  {journal} {\bibinfo  {journal} {Industrial Engineering Chemistry}\ }\textbf
  {\bibinfo {volume} {40}},\ \bibinfo {pages} {345} (\bibinfo {year}
  {1948})}\BibitemShut {NoStop}%
\bibitem [{\citenamefont {Ansara}(1998)}]{Ansara1998}%
  \BibitemOpen
  \bibfield  {author} {\bibinfo {author} {\bibfnamefont {I.}~\bibnamefont
  {Ansara}},\ }\href {https://doi.org/10.1351/pac199870020449} {\bibfield
  {journal} {\bibinfo  {journal} {Pure and Applied Chemistry}\ }\textbf
  {\bibinfo {volume} {70}},\ \bibinfo {pages} {449} (\bibinfo {year}
  {1998})}\BibitemShut {NoStop}%
\bibitem [{\citenamefont {Ramesh}\ \emph {et~al.}(2014)\citenamefont {Ramesh},
  \citenamefont {Hisyam}, \citenamefont {Sulaiman},\ and\ \citenamefont
  {Ramesh}}]{Ramesh2014}%
  \BibitemOpen
  \bibfield  {author} {\bibinfo {author} {\bibfnamefont {R.}~\bibnamefont
  {Ramesh}}, \bibinfo {author} {\bibfnamefont {A.}~\bibnamefont {Hisyam}},
  \bibinfo {author} {\bibfnamefont {A.}~\bibnamefont {Sulaiman}},\ and\
  \bibinfo {author} {\bibfnamefont {K.}~\bibnamefont {Ramesh}},\ }\href
  {https://doi.org/10.12691/ces-2-2-2} {\bibfield  {journal} {\bibinfo
  {journal} {Chemical Engineering and Science}\ }\textbf {\bibinfo {volume}
  {2}},\ \bibinfo {pages} {18} (\bibinfo {year} {2014})}\BibitemShut {NoStop}%
\bibitem [{\citenamefont {Bethe}(1929)}]{Bethe1929}%
  \BibitemOpen
  \bibfield  {author} {\bibinfo {author} {\bibfnamefont {H.}~\bibnamefont
  {Bethe}},\ }\href {https://doi.org/10.1002/andp.19293950202} {\bibfield
  {journal} {\bibinfo  {journal} {Annalen der Physik}\ }\textbf {\bibinfo
  {volume} {395}},\ \bibinfo {pages} {133} (\bibinfo {year}
  {1929})}\BibitemShut {NoStop}%
\bibitem [{\citenamefont {Van~Vleck}(1932)}]{Vanvleck1932}%
  \BibitemOpen
  \bibfield  {author} {\bibinfo {author} {\bibfnamefont {J.~H.}\ \bibnamefont
  {Van~Vleck}},\ }\href {https://doi.org/10.1103/PhysRev.41.208} {\bibfield
  {journal} {\bibinfo  {journal} {Phys. Rev.}\ }\textbf {\bibinfo {volume}
  {41}},\ \bibinfo {pages} {208} (\bibinfo {year} {1932})}\BibitemShut
  {NoStop}%
\bibitem [{\citenamefont {Quintela}\ \emph {et~al.}(2010)\citenamefont
  {Quintela}, \citenamefont {Rivadulla},\ and\ \citenamefont
  {Rivas}}]{Quintela2010}%
  \BibitemOpen
  \bibfield  {author} {\bibinfo {author} {\bibfnamefont {C.~X.}\ \bibnamefont
  {Quintela}}, \bibinfo {author} {\bibfnamefont {F.}~\bibnamefont
  {Rivadulla}},\ and\ \bibinfo {author} {\bibfnamefont {J.}~\bibnamefont
  {Rivas}},\ }\href {https://doi.org/10.1103/PhysRevB.82.245201} {\bibfield
  {journal} {\bibinfo  {journal} {Physical Review B - Condensed Matter and
  Materials Physics}\ }\textbf {\bibinfo {volume} {82}},\ \bibinfo {pages} {3}
  (\bibinfo {year} {2010})}\BibitemShut {NoStop}%
\bibitem [{\citenamefont {{Muhammed Sabeer}}\ and\ \citenamefont
  {Pradyumnan}(2021)}]{MuhammedSabeer2021}%
  \BibitemOpen
  \bibfield  {author} {\bibinfo {author} {\bibfnamefont {N.}~\bibnamefont
  {{Muhammed Sabeer}}}\ and\ \bibinfo {author} {\bibfnamefont {P.}~\bibnamefont
  {Pradyumnan}},\ }\href {https://doi.org/10.1016/j.mseb.2021.115428}
  {\bibfield  {journal} {\bibinfo  {journal} {Materials Science and
  Engineering: B}\ }\textbf {\bibinfo {volume} {273}},\ \bibinfo {pages}
  {115428} (\bibinfo {year} {2021})}\BibitemShut {NoStop}%
\bibitem [{\citenamefont {Zhang}\ \emph {et~al.}(2011)\citenamefont {Zhang},
  \citenamefont {Chawla}, \citenamefont {Howe},\ and\ \citenamefont
  {Gall}}]{Zhang2011b}%
  \BibitemOpen
  \bibfield  {author} {\bibinfo {author} {\bibfnamefont {X.~Y.}\ \bibnamefont
  {Zhang}}, \bibinfo {author} {\bibfnamefont {J.~S.}\ \bibnamefont {Chawla}},
  \bibinfo {author} {\bibfnamefont {B.~M.}\ \bibnamefont {Howe}},\ and\
  \bibinfo {author} {\bibfnamefont {D.}~\bibnamefont {Gall}},\ }\href
  {https://doi.org/10.1103/PhysRevB.83.165205} {\bibfield  {journal} {\bibinfo
  {journal} {Physical Review B - Condensed Matter and Materials Physics}\
  }\textbf {\bibinfo {volume} {83}},\ \bibinfo {pages} {1} (\bibinfo {year}
  {2011})}\BibitemShut {NoStop}%
\bibitem [{\citenamefont {Constantin}\ \emph {et~al.}(2004)\citenamefont
  {Constantin}, \citenamefont {Haider}, \citenamefont {Ingram},\ and\
  \citenamefont {Smith}}]{Constantin2004a}%
  \BibitemOpen
  \bibfield  {author} {\bibinfo {author} {\bibfnamefont {C.}~\bibnamefont
  {Constantin}}, \bibinfo {author} {\bibfnamefont {M.~B.}\ \bibnamefont
  {Haider}}, \bibinfo {author} {\bibfnamefont {D.}~\bibnamefont {Ingram}},\
  and\ \bibinfo {author} {\bibfnamefont {A.~R.}\ \bibnamefont {Smith}},\ }\href
  {https://doi.org/10.1063/1.1836878} {\bibfield  {journal} {\bibinfo
  {journal} {Applied Physics Letters}\ }\textbf {\bibinfo {volume} {85}},\
  \bibinfo {pages} {6371} (\bibinfo {year} {2004})}\BibitemShut {NoStop}%
\bibitem [{\citenamefont {Bhobe}\ \emph {et~al.}(2010)\citenamefont {Bhobe},
  \citenamefont {Chainani}, \citenamefont {Taguchi}, \citenamefont {Takeuchi},
  \citenamefont {Eguchi}, \citenamefont {Matsunami}, \citenamefont {Ishizaka},
  \citenamefont {Takata}, \citenamefont {Oura}, \citenamefont {Senba},
  \citenamefont {Ohashi}, \citenamefont {Nishino}, \citenamefont {Yabashi},
  \citenamefont {Tamasaku}, \citenamefont {Ishikawa}, \citenamefont {Takenaka},
  \citenamefont {Takagi},\ and\ \citenamefont {Shin}}]{Bhobe2010}%
  \BibitemOpen
  \bibfield  {author} {\bibinfo {author} {\bibfnamefont {P.~A.}\ \bibnamefont
  {Bhobe}}, \bibinfo {author} {\bibfnamefont {A.}~\bibnamefont {Chainani}},
  \bibinfo {author} {\bibfnamefont {M.}~\bibnamefont {Taguchi}}, \bibinfo
  {author} {\bibfnamefont {T.}~\bibnamefont {Takeuchi}}, \bibinfo {author}
  {\bibfnamefont {R.}~\bibnamefont {Eguchi}}, \bibinfo {author} {\bibfnamefont
  {M.}~\bibnamefont {Matsunami}}, \bibinfo {author} {\bibfnamefont
  {K.}~\bibnamefont {Ishizaka}}, \bibinfo {author} {\bibfnamefont
  {Y.}~\bibnamefont {Takata}}, \bibinfo {author} {\bibfnamefont
  {M.}~\bibnamefont {Oura}}, \bibinfo {author} {\bibfnamefont {Y.}~\bibnamefont
  {Senba}}, \bibinfo {author} {\bibfnamefont {H.}~\bibnamefont {Ohashi}},
  \bibinfo {author} {\bibfnamefont {Y.}~\bibnamefont {Nishino}}, \bibinfo
  {author} {\bibfnamefont {M.}~\bibnamefont {Yabashi}}, \bibinfo {author}
  {\bibfnamefont {K.}~\bibnamefont {Tamasaku}}, \bibinfo {author}
  {\bibfnamefont {T.}~\bibnamefont {Ishikawa}}, \bibinfo {author}
  {\bibfnamefont {K.}~\bibnamefont {Takenaka}}, \bibinfo {author}
  {\bibfnamefont {H.}~\bibnamefont {Takagi}},\ and\ \bibinfo {author}
  {\bibfnamefont {S.}~\bibnamefont {Shin}},\ }\href
  {https://doi.org/10.1103/PhysRevLett.104.236404} {\bibfield  {journal}
  {\bibinfo  {journal} {Physical Review Letters}\ }\textbf {\bibinfo {volume}
  {104}},\ \bibinfo {pages} {1} (\bibinfo {year} {2010})},\ \Eprint
  {https://arxiv.org/abs/1004.0042} {arXiv:1004.0042} \BibitemShut {NoStop}%
\bibitem [{\citenamefont {Zhang}\ \emph {et~al.}(2010)\citenamefont {Zhang},
  \citenamefont {Punkkinen}, \citenamefont {Johansson}, \citenamefont
  {Hertzman},\ and\ \citenamefont {Vitos}}]{Zhang2010}%
  \BibitemOpen
  \bibfield  {author} {\bibinfo {author} {\bibfnamefont {H.}~\bibnamefont
  {Zhang}}, \bibinfo {author} {\bibfnamefont {M.~P.}\ \bibnamefont
  {Punkkinen}}, \bibinfo {author} {\bibfnamefont {B.}~\bibnamefont
  {Johansson}}, \bibinfo {author} {\bibfnamefont {S.}~\bibnamefont
  {Hertzman}},\ and\ \bibinfo {author} {\bibfnamefont {L.}~\bibnamefont
  {Vitos}},\ }\href {https://doi.org/10.1103/PhysRevB.81.184105} {\bibfield
  {journal} {\bibinfo  {journal} {Physical Review B - Condensed Matter and
  Materials Physics}\ }\textbf {\bibinfo {volume} {81}},\ \bibinfo {pages} {1}
  (\bibinfo {year} {2010})}\BibitemShut {NoStop}%
\bibitem [{\citenamefont {Botana}\ \emph {et~al.}(2012)\citenamefont {Botana},
  \citenamefont {Tran}, \citenamefont {Pardo}, \citenamefont {Baldomir},\ and\
  \citenamefont {Blaha}}]{Botana2012}%
  \BibitemOpen
  \bibfield  {author} {\bibinfo {author} {\bibfnamefont {A.~S.}\ \bibnamefont
  {Botana}}, \bibinfo {author} {\bibfnamefont {F.}~\bibnamefont {Tran}},
  \bibinfo {author} {\bibfnamefont {V.}~\bibnamefont {Pardo}}, \bibinfo
  {author} {\bibfnamefont {D.}~\bibnamefont {Baldomir}},\ and\ \bibinfo
  {author} {\bibfnamefont {P.}~\bibnamefont {Blaha}},\ }\href
  {https://doi.org/10.1103/PhysRevB.85.235118} {\bibfield  {journal} {\bibinfo
  {journal} {Physical Review B - Condensed Matter and Materials Physics}\
  }\textbf {\bibinfo {volume} {85}},\ \bibinfo {pages} {1} (\bibinfo {year}
  {2012})},\ \Eprint {https://arxiv.org/abs/1205.2672} {arXiv:1205.2672}
  \BibitemShut {NoStop}%
\bibitem [{\citenamefont {Mott}(1995)}]{MOTT1995}%
  \BibitemOpen
  \bibfield  {author} {\bibinfo {author} {\bibfnamefont {N.~F.}\ \bibnamefont
  {Mott}},\ }in\ \href {https://doi.org/10.1142/9789812794086_0027} {\emph
  {\bibinfo {booktitle} {Conduction in Non-Crystalline Materials}}},\ \bibinfo
  {editor} {edited by\ \bibinfo {editor} {\bibfnamefont {R.~E.}\ \bibnamefont
  {Davies}}\ and\ \bibinfo {editor} {\bibfnamefont {H.}~\bibnamefont {Jones}}}\
  (\bibinfo  {publisher} {World Scientific},\ \bibinfo {year} {1995})\ pp.\
  \bibinfo {pages} {465--482}\BibitemShut {NoStop}%
\bibitem [{\citenamefont {Khandy}\ and\ \citenamefont
  {Chai}(2021)}]{Khandy2021}%
  \BibitemOpen
  \bibfield  {author} {\bibinfo {author} {\bibfnamefont {S.~A.}\ \bibnamefont
  {Khandy}}\ and\ \bibinfo {author} {\bibfnamefont {J.~D.}\ \bibnamefont
  {Chai}},\ }\href {https://doi.org/10.1016/j.jpcs.2021.110098} {\bibfield
  {journal} {\bibinfo  {journal} {Journal of Physics and Chemistry of Solids}\
  }\textbf {\bibinfo {volume} {154}},\ \bibinfo {pages} {110098} (\bibinfo
  {year} {2021})}\BibitemShut {NoStop}%
\bibitem [{\citenamefont {Hsieh}\ \emph {et~al.}(2014)\citenamefont {Hsieh},
  \citenamefont {Okazaki}, \citenamefont {Taniguchi},\ and\ \citenamefont
  {Terasaki}}]{Hsieh2014}%
  \BibitemOpen
  \bibfield  {author} {\bibinfo {author} {\bibfnamefont {Y.~C.}\ \bibnamefont
  {Hsieh}}, \bibinfo {author} {\bibfnamefont {R.}~\bibnamefont {Okazaki}},
  \bibinfo {author} {\bibfnamefont {H.}~\bibnamefont {Taniguchi}},\ and\
  \bibinfo {author} {\bibfnamefont {I.}~\bibnamefont {Terasaki}},\ }\href
  {https://doi.org/10.7566/JPSJ.83.054710} {\bibfield  {journal} {\bibinfo
  {journal} {Journal of the Physical Society of Japan}\ }\textbf {\bibinfo
  {volume} {83}},\ \bibinfo {pages} {2} (\bibinfo {year} {2014})}\BibitemShut
  {NoStop}%
\bibitem [{\citenamefont {Nishino}(2004)}]{Nishino2004}%
  \BibitemOpen
  \bibfield  {author} {\bibinfo {author} {\bibfnamefont {Y.}~\bibnamefont
  {Nishino}},\ }\href
  {https://doi.org/10.4028/www.scientific.net/msf.449-452.909} {\bibfield
  {journal} {\bibinfo  {journal} {Materials Science Forum}\ }\textbf {\bibinfo
  {volume} {449-452}},\ \bibinfo {pages} {909} (\bibinfo {year}
  {2004})}\BibitemShut {NoStop}%
\bibitem [{\citenamefont {Jiang}\ \emph {et~al.}(2020)\citenamefont {Jiang},
  \citenamefont {Xia}, \citenamefont {Zhou}, \citenamefont {Niu}, \citenamefont
  {Chen}, \citenamefont {Luo}, \citenamefont {Liu}, \citenamefont {Zhou},
  \citenamefont {Fan},\ and\ \citenamefont {Wang}}]{Jiang2020}%
  \BibitemOpen
  \bibfield  {author} {\bibinfo {author} {\bibfnamefont {J.}~\bibnamefont
  {Jiang}}, \bibinfo {author} {\bibfnamefont {J.}~\bibnamefont {Xia}}, \bibinfo
  {author} {\bibfnamefont {T.}~\bibnamefont {Zhou}}, \bibinfo {author}
  {\bibfnamefont {Y.}~\bibnamefont {Niu}}, \bibinfo {author} {\bibfnamefont
  {Y.}~\bibnamefont {Chen}}, \bibinfo {author} {\bibfnamefont {J.}~\bibnamefont
  {Luo}}, \bibinfo {author} {\bibfnamefont {J.}~\bibnamefont {Liu}}, \bibinfo
  {author} {\bibfnamefont {J.}~\bibnamefont {Zhou}}, \bibinfo {author}
  {\bibfnamefont {J.}~\bibnamefont {Fan}},\ and\ \bibinfo {author}
  {\bibfnamefont {C.}~\bibnamefont {Wang}},\ }\href
  {https://doi.org/10.1080/09500839.2020.1732492} {\bibfield  {journal}
  {\bibinfo  {journal} {Philosophical Magazine Letters}\ }\textbf {\bibinfo
  {volume} {100}},\ \bibinfo {pages} {128} (\bibinfo {year}
  {2020})}\BibitemShut {NoStop}%
\bibitem [{\citenamefont {Disalvo}(1999)}]{Disalvo1999}%
  \BibitemOpen
  \bibfield  {author} {\bibinfo {author} {\bibfnamefont {F.~J.}\ \bibnamefont
  {Disalvo}},\ }\href {https://doi.org/10.1126/science.285.5428.703} {\bibfield
   {journal} {\bibinfo  {journal} {Science}\ }\textbf {\bibinfo {volume}
  {285}},\ \bibinfo {pages} {703} (\bibinfo {year} {1999})}\BibitemShut
  {NoStop}%
\bibitem [{\citenamefont {Jonson}\ and\ \citenamefont
  {Mahan}(1980)}]{Jonson1980}%
  \BibitemOpen
  \bibfield  {author} {\bibinfo {author} {\bibfnamefont {M.}~\bibnamefont
  {Jonson}}\ and\ \bibinfo {author} {\bibfnamefont {G.~D.}\ \bibnamefont
  {Mahan}},\ }\href {https://doi.org/10.1103/PhysRevB.21.4223} {\bibfield
  {journal} {\bibinfo  {journal} {Physical Review B}\ }\textbf {\bibinfo
  {volume} {21}},\ \bibinfo {pages} {4223} (\bibinfo {year}
  {1980})}\BibitemShut {NoStop}%
\bibitem [{\citenamefont {Quintela}\ \emph {et~al.}(2009)\citenamefont
  {Quintela}, \citenamefont {Rivadulla},\ and\ \citenamefont
  {Rivas}}]{Quintela2009}%
  \BibitemOpen
  \bibfield  {author} {\bibinfo {author} {\bibfnamefont {C.~X.}\ \bibnamefont
  {Quintela}}, \bibinfo {author} {\bibfnamefont {F.}~\bibnamefont
  {Rivadulla}},\ and\ \bibinfo {author} {\bibfnamefont {J.}~\bibnamefont
  {Rivas}},\ }\bibfield  {journal} {\bibinfo  {journal} {Applied Physics
  Letters}\ }\textbf {\bibinfo {volume} {94}},\ \href
  {https://doi.org/10.1063/1.3120280} {10.1063/1.3120280} (\bibinfo {year}
  {2009})\BibitemShut {NoStop}%
\end{thebibliography}%

\end{document}